\documentclass[twocolumn,rmp,showpacs,preprintnumbers,amsmath,amssymb]{revtex4}
\usepackage{graphicx}% Include figure files
\usepackage{dcolumn}% Align table columns on decimal point\cite{Dooling:2002js}
\usepackage{bm}% bold math
\usepackage{longtable}% bold math
\newcommand{\sm}{{Standard~Model~}}

\newcommand{\beq}{\begin{equation}}

\newcommand{\eeq}{\end{equation}}
\newcommand\beqn{\begin{eqnarray}}
\newcommand\eeqn{\end{eqnarray}}
\newcommand{\bea}{\begin{eqnarray}}
\newcommand{\eea}{\end{eqnarray}}
\def\r2{\sqrt 2}
\def\beq{\begin{equation}}
\def\eeq{\end{equation}}
\def\beqn{\begin{eqnarray}}
\def\eeqn{\end{eqnarray}}

\def\sinW2{\sin^2\theta_W}
\def\AEM{\alpha_{EM}}

\def\mz2{M_{z}^2}
\def\c2b{\cos 2\beta}

\def\m#1{{\tilde m}_#1}

\def\mc#1{{\tilde m}_{\chi^{+}_#1}}

\def\mxi{{\tilde m}_{\chi^{0}_i}}
\def\mci{{\tilde m}_{\chi^{+}_i}}
\def\mz{M_z}

\def\cb{\cos\beta}
\def\sb{\sin\beta}

\def\Fq2{F_{2}(q^2)}
\begin{document}

\preprint{APS/123-QED}

\title{CP Violation From  Standard Model  to Strings}% Force line breaks with \\

\author{Tarek Ibrahim}

\affiliation{
Department of Physics,  Faculty of Science, University of Alexandria, Alexandria, Egypt\\
}%

\author{Pran Nath}

\affiliation{Department of Physics, Northeastern University, Boston, Massachusetts 02115, USA\\
}%

\date{\today}% It is always \today, today,

\begin{abstract}

A  review  of CP violation from  the Standard Model to strings  is given  which
  includes  a broad landscape of particle physics models, encompassing  the
 non-supersymmetric 4D extensions of the standard model, and models based on
supersymmetry,   on extra dimensions, on strings and on branes.  The supersymmetric models
discussed  include complex mSUGRA and its extensions, while the models based on extra dimensions include
5D  models including  models based on warped geometry.
CP  violation beyond the standard model is central to achieving the desired amount of baryon asymmetry
in the universe via baryogenesis and leptogenesis. They also  affect a  variety of particle
physics phenomena: electric dipole moments, $g-2$, relic density and detection rates for
neutralino dark matter in supersymmetric theories,  Yukawa unification in
grand unified and string based models, and sparticle  production cross sections,
and their decays patterns and signatures at hadron colliders. Additionally CP violations can
generate CP even-CP odd Higgs mixings,  affect the neutral  Higgs spectrum and  lead  to phenomena detectable
at colliders. Prominent among  these are  the CP violation effects  in decays of neutral and charged Higgs bosons.
 Neutrino masses introduce new sources of CP violation  which may be explored in neutrino
 factories  in the future. Such phases  can also enter in proton stability in unified  models of particle interactions.
The current experimental  status of CP violation is discussed and  possibilities for the
future  outlined.
\end{abstract}
% Valid PACS numbers may be entered using the \verb+\pacs{#1}+ command.
\pacs{Valid PACS appear here}% PACS, the Physics and Astronomy
%                              %display desired
\maketitle

\tableofcontents
\section{Introduction}

We begin with a brief  history of the  considerations that led to question the validity of CP symmetry as 
an exact  symmetry for elementary particles.  The history is tied to the  issue of electric dipole moments 
and  we  need to retrace  the steps back  to 1950 when it was generally accepted that
the particle electric  dipole moments vanished due to parity symmetry.  However, in 1950 it was first 
observed  by Purcell and Ramsey\cite{purcell1950}, that there was no experimental evidence
for the parity symmetry for  nuclear  forces and for elementary particles, and thus the
possible existence of an electric dipole moment for these needed to be tested  experimentally. 
They and their graduate student James Smith then carried  
out the first
such test by showing experimentally in 1951 that the magnitude of the  
electric
dipole moment of the neutron was less than $3\times 10^{-20}$ e.cm where e is the
charge of the proton
\footnote{The experimental results of Purcell, Ramsey and Smith while completed in 1951
were not published till much later\cite{Smith:1957ht}. However, they were quoted in
other publications\cite{Smith1951,Ramsey1956,Lee:1956qn}.}.
After the violation of parity symmetry proposed by
T.D. Lee and C.N. Yang \cite{Lee:1956qn} was confirmed \cite{Wu:1957my},
it was  argued by many that the elementary electric dipole moments would
vanish due to the combined charge conjugation and parity symmetry, i.e., 
 CP symmetry  (or equivalently under a time reversal  symmetry under the assumption 
 of  CPT invariance).  However, it was then  pointed out by 
Ramsey\cite{Ramsey1958} and independently by 
 Jackson and  collaborators\cite{Jackson1957} 
that T invariance was also
an assumption and needed to be checked experimentally
(A brief review of early history can be found in \cite{Ramsey1998}). 
%The first experiment
%performed at Oak Ridge by Purcell, Ramsey and Smith\cite{Smith:1957ht}
%  yielded no 
%evidence for the neutron electric dipole moment but produced a limit of 
%$|d_n|<5\times 10^{-20}$ ecm. 
Since then the search for CP violations
 has been  vigorously pursued.  The CP violation was eventually discovered in the Kaon
system by Val Fitch, James Cronin and collaborators in 1964\cite{Christenson:1964fg}.
Shortly thereafter it was pointed
out by  Andre Sakharov\cite{Sakharov:1967dj}
that CP violations play an important role in generating the
baryon asymmetry in the universe.  However, it has recently been realized that 
sources of CP violation beyond what exist in the Standard Model are needed
for this purpose.  In this context over the past decade  a very
significant body of work on CP violation beyond the Standard Model has appeared.
It encompasses non-supersymmetric models, supersymmetric models, models 
based on extra dimensions and warped dimensions,  and string models.   
There is currently no review which encompasses these developments.
The purpose of this review is to bridge this gap. 
Thus in this review we present a broad overview of CP violation starting from the Standard Model
and ending with strings. CP violation is central to understanding the phenomena in particle physics 
as well as in cosmology. Thus CP violation enters in K and B physics, and as mentioned 
above CP violation 
beyond the Standard Model is deemed necessary to explain the desired baryon asymmetry
in the universe. Further,  new sources of CP violation beyond the Standard Model could also show up
in sparticle production at the LHC, and in the new generation of experiments
underway on neutrino physics.  In view of the importance of CP violation in particle physics
and in cosmology it is also important to explore the possible origins of such violations.
These topics are the focus of this review.  We give now a  brief outline of the contents of 
this review. \\

In  Sec.(\ref{b}) we give a discussion of CP violation in the Standard Model and of the strong 
CP problem.  The electroweak sector of the Standard Model contains one phase which appears
in the Cabibbo-Kobayashi-Maskawa (CKM) matrix. The CKM matrix satisfies unitarity constraints
including the well known unitarity triangle constraint where the three angles $\alpha, \beta,
\gamma$ defined in terms of ratios involving the products of  CKM matrix elements and their complex
conjugates sum to $\pi$. 
 In addition the quantum chromo dynamic 
(QCD) sector of the Standard Model brings in another source of CP violation - the strong CP 
phase $\theta_{QCD}$. The natural size of this phase is  $O(1)$  which would produce a huge 
contribution to the electric dipole moment (EDM) of the neutron in violation of the existing 
experimental bounds.  A brief discussion  of these issues is given in Sec.(\ref{b}). 
 A review of the experimental evidence for CP violation and of the  searches for 
 evidence of  other CP 
 violation such as in the electric dipole moment of elementary particles and of atoms 
  is given in Sec.(\ref{c}).  Here we discuss the current experimental situation
in  the K and B system.  In the Kaon system two parameters $\epsilon$ (indirect CP violation) and 
$\epsilon'$ (direct CP violation)  have played an important role in the discussion of CP violation
in this system. Specifically the measurement  of $\epsilon'/\epsilon$ rules out the so called
superweak theory of CP violation while the measurement is consistent with the Standard Model 
prediction. In this section we also give an analysis of experimental  constraints 
 on the angles $\alpha, \beta, \gamma$   of the unitarity triangle discussed in  Sec.(\ref{b}).
 The  current experimental limits of  the EDMs of the electron, of the neutron and of 
 $^{199}Hg$ are also discussed. 
 \\

In Sec.(\ref{d}) we  give  a discussion of the CP violation in some non-supersymmetric extensions
of the Standard Model. These  include  the Left-Right (LR)  extensions,   the two Higgs  doublet  model  
and extensions with more than two Higgs doublets. It is shown that such extensions contain more 
sources of CP violation. For example, the LR extensions with the gauge  group $SU(2)_L\times SU(2)_R\times U(1)_Y$
and three  generations contains seven CP phases instead of one phase that the Standard Model has. 
Similarly it is shown that the number of allowed  CP phases increases with the number of Higgs doublets.
 Further, new sources of CP violation arise as one increases the number of allowed generations.
 CP violation in the context of supersymmetric extensions of the Standard Model are discussed in
 Sec.(\ref{e}). Here one finds  that the minimal supersymmetric standard model (MSSM) has a large
 number (i.e., 46) of phases which, however, is reduced to two phases in the minimal supergravity
 unified model (mSUGRA).  However, more phases are  allowed if one considers 
 supergravity unified models with non-univesal soft  breaking at the grand unified  (GUT) scale
 consistent with flavor  changing neutral current (FCNC) constraints.  A discussion of CP violation
 in extra  dimension models is given in Sec.(\ref{f}).  In this section we give an exhibition of 
 the phenomena of spontaneous vs  explicit  CP violation.  In this section we also discuss 
 CP violation in the context of warped extra dimensions. \\

 A discussion of CP violation in strings is given in   Sec.(\ref{g}).  It is shown that soft breaking in
 string models is parametrized by vacuum expectation values (VEVs) of the dilaton ($S$) and of the 
 moduli fields ($T_i$) which carry CP violating phases.  Additionally CP phases can occur
 in the Yukawa couplings. Thus CP violation is quite generic in string models. 
 We give specific illustration of this in a Calabi-Yau compactification of an $E_8\times E_8$ 
 heterotic string and in orbifold compactifications. Here we also discuss CP violation in D brane 
 models. Finally in this section we discuss the possible connection of SUSY CP phases with
 the CKM phase. 

A  discussion of the computation of the EDM of an elementary Dirac fermion is given in
  Sec.(\ref{h}) while that of a charged lepton in supersymmetric models is given in  Sec.(\ref{i}).
  In Sec.(\ref{j}) we give an analysis of the  EDM of quarks in supersymmetry. The supersymmetric contributions to the EDM of a  quark 
involve  three  different pieces which include the electric  dipole, the chromo electric
dipole and the purely gluonic  dimension six operators. The contributions 
of each of these are discussed in Sec.(\ref{j}). Typically for low lying
sparticle masses the supersymmetrtic contribution to the EDM of the 
electron and of the neutron  is generally  in excess of the current experimental bounds.
 This poses a  serious difficulty for supersymmetric models. Some ways to overcome 
 these are also discussed in Sec.(\ref{j}). Two  prominent ways to accomplish this include
 either a heavy sparticle spectrum with sparticle masses lying in the TeV region, or the
 cancellation mechanism where contributions arising from  the electric  dipole, the chromo electric
dipole and the purely gluonic dimension six operators largely cancel. \\
 
If the large SUSY CP phases can be made consistent with the EDM constraints, then such large
phases  can affect a variety of supersymmetric phenomena. We discuss several such 
phenomena in  Sec.(\ref{k}). These include analyses of the effect of CP phases on 
$g_{\mu}-2$, on CP even-CP odd Higgs mixing in the neutral Higgs sector, 
and  on the b quark mass. 
Further, CP phases can affect significantly the neutral Higgs decays into $b\bar b$ and 
$\tau\bar \tau$ and the decays of the charged Higgs  into $\bar t b, \bar \nu_{\tau} \tau$
and the decays $H^{\pm}\to \chi^{\pm}\chi^0$.  These phenomena are also discussed 
in Sec.(\ref{k}). Some of the other phenomena affected by  CP phases include 
the relic density  of neutralino dark matter, proton decay via
dimension six operators,  the decay  $B^0_s \to \mu^+\mu^-$, decays of the 
sfermions,  and the decay $B\to \phi K$.  These are all discussed in some detail
in   Sec.(\ref{k}).  Finally in this section we discuss the $T$ and $CP$ odd operators
and their observability at colliders.
An analysis of the interplay between CP violation and flavor is given Sec.(\ref{m}).
Here we first discuss the mechanisms which may allow the muon EDM to be much
larger than the electron EDM, and accessible to a  new proposed experiment on the
muon EDM which may extend the sensitivity of this measurement by several orders 
of magnitude and thus make it potentially observable.  In this section  an analysis of the
effect of CP phases on $B\to X_s\gamma$ is also given. This FCNC process is of 
importance as it constrains the parameter space of MSSM and also constrains the
 analyses of dark matter. 
  Sec.(\ref{n}) is devoted to a study of CP violation in neutrino physics. Here a discussion
  of CP violation and leptogenesis is given, as well as a discussion on the observability  of
  Majorana phases.  \\

 Future prospects  for improved measurement of CP violation in experiments are
 discussed in  Sec.(\ref{o}). These include improved experiments for the 
 measurements of the EDMs,  B physics experiments at the LHCb which is 
 dedicated to the study of B physics,   Super Belle proposal, as well as 
 superbeams  which  include  the study of possible CP violation
 in neutrino physics.  Conclusions are given in Sec.(\ref{p}).  
 Some further mathematical details are given in the Appendices in 
 Sec.(\ref{q}).    \\

\section{CP violation in the Standard Model and the strong CP problem }\label{b}
 The  electroweak sector of the \sm  with three generations of quarks and leptons  has one
 CP violating phase which enters via the  Cabbibo-Kobayashi-Maskawa (CKM) matrix V.
Thus the electroweak interactions  contain the CKM matrix in the
charged current sector
 \beqn
 g_2 \bar u_i \gamma_{\mu}V_{ij} (1-\gamma_5)d_j W^{\mu}  + H.c.
 \eeqn
 where $u_i=u,c,t$ and $d_j=d,s,b$ quarks.
  The CKM matrix obeys the unitarity constraint  $(VV^{\dagger})_{ij}=\delta_{ij}$ and can
 be parameterized in terms of  three mixing angles and  one CP violating
 phase.  For the  case  $i\neq j$ the unitarity constraint can be  displayed as a
 unitariy triangle, and there are six such unitarity triangles.
Thus the unitarity of the CKM matrix for the first and the third column gives 
 \beqn
 V_{ud}V^*_{ub} + V_{cd}V^*_{cb} + V_{td} V^*_{tb}=0.
\eeqn
One can display this  constraint as  a unitarity triangle by defining the angles
 $\alpha, \beta, \gamma$   so that
\beqn
 \alpha =arg(-{V_{td}V^*_{tb}}/{V_{ud}V^*_{ub}}), ~
\beta =arg(-{V_{cd}V^*_{cb}}/{V_{td}V^*_{tb}}),\nonumber\\
\gamma =arg(-{V_{ud}V^*_{ub}}/{V_{cd}V^*_{cb}})
~~~~~~~~~~~~~~~~~~~~~ \eeqn which satisfy the constraint
$\alpha+\beta+\gamma=\pi$. One can parameterize CP violation in  a
way which is independent of the phase conventions. This is the so
called Jarlskog
  invariant~\cite{Jarlskog:1985ht}
  J  which can be defined in nine different ways,   and one of which is given by
\beqn
J= Im(V_{us}V_{ub}^*V_{cb}V_{cs}^*).
\eeqn
An interesting observation is that the CKM is hierarchical and allows for expansion in $\lambda \simeq 0.226$
so one may write $V$ as a perturbative expansion in $\lambda$ which up $O(\lambda^3)$  is given by
\beqn
 \left(
\begin{array}{ccc}
 1-{\lambda^2\over 2} & \lambda & A\lambda^3 (\rho -i \eta)\\
  -\lambda     & 1-{\lambda^2\over 2}
  & A\lambda^2\\
 A\lambda^3 (1-\rho -i\eta) & -A\lambda^2 & 1\end{array}\right)
\eeqn In this representation the Jarlskog invariant is given by
$J\simeq A^2 \lambda^6 \eta$, and the CP violation enters via
$\eta$.

The \sm has  another source of CP violation in addition to the one that appears in the CKM matrix.
This source  of CP violation arises  in the strong interaction sector of the theory from  the term
$ \theta \frac{\alpha_s}{8\pi}G\tilde G$,
  which is of topological origin.  It gives  a large contribution to the EDM of the
  neutron and consistency with current experiment requires
  $\bar\theta =\theta+  ArgDet(M_uM_d)$    to be small
  $\bar\theta <O(10^{-10})$.
 One solution to the strong CP problem is the vanishing of the up quark mass.
However, analyses based on chiral  perturbation theory and on lattice gauge
theory appear to indicate a non-vanishing mass for the up quark.
Thus a resolution to the strong CP problem appears to require beyond the Standard Model
physics.  For example, 
one proposed solution is the Peccei-Quinn mechanism~\cite{Peccei:1977ur}
and  its refinements\cite{Dine:1981rt,Zhitnitskii:1980, Kim:1979if}
which leads to  axions. But currently severe limits exist
on the corridor in which axions can exist.
There is much work in the  literature  regarding  how  one may  suppress the
strong CP violation effects (for a review see \cite{Dine:2000cj}).
 In addition to the  use of axions or  a massless up
quark one also has the possibility of using  a symmetry to suppress the  strong
CP effects \cite{Barr:1984qx,Nelson:1983zb}.\\

  The solution to the strong CP in the framework of Left-Right symmetric models is discussed in
  \cite{Mohapatra:1997su,Babu:2001se}.
  Specifically in the analysis of  \cite{Babu:2001se}  the strong CP parameter
$\bar\theta$ is zero  at the tree level, due to parity (P),
but is induced due to P -violating effects
below the unification scale.
In the analysis of  \cite{Hiller:2001qg} a solution to the strong CP problem
using supersymmetry  is proposed.  Here one  envisions a solution to the
strong CP problem based on supersymmetric non-renormalization
theorem. In this scenario CP is broken spontaneously and its breaking
is communicated to the MSSM by radiative corrections. The strong CP phase is
protected by a SUSY non-renormalization theorem and remains
exactly zero while the loops can generate a large CKM phase from
wave function renormalization.
Another idea advocates promoting the $U(1)$ CP violating phases of the
supersymmetric standard model to dynamical variables, and then allowing
the  vacuum to relax near a CP conserving point \cite{Dimopoulos:1995kn}.
In the analysis of ~\cite{Demir:2000ng}
 an axionic solution of the strong CP problem with a
Peccei-Quinn mechanism  using
the gluino rather than the quarks is given and the spontaneous breaking
of the new U(1) global symmetry is connected to the supersymmetry breaking
with  a solution to the $\mu$ problem~\cite{Demir:2000ng}.
 Finally, in the analysis of \cite{Aldazabal:2002py}
a solution based on gauging away the strong CP problem is proposed.
Thus the work of  \cite{Aldazabal:2002py} proposes a  solution that
involves the existence of an unbroken gauged $U(1)_X$ symmetry whose
gauge boson gets a Stueckelberg mass term by combining with a
pseudoscalar field $\eta (x)$ which has an axion like coupling to
$G\tilde G$. Thus the $\theta$ parameter can be gauged away by a
$U(1)_X$ transformation. The additional $U(1)_X$ generates mixed
gauge anomalies which are canceled by the addition of an appropriate
Wess-Zumino term.  We will  assume from here on that the strong CP
problem is solved by one or the other of the techniques outlined
above.

%%%%%%%%%%%%%%%%%%%%%%%%%%%%%%%%%%%%%%%%
\section{~Review of  experimental evidence  on CP violation and searches for other  evidence}
\label{c}
There  are currently four  pieces  of  experimental evidence for CP violation.
These  consist  of  (i) the  observation of indirect  CP violation ($\epsilon$),  and (ii) of direct CP violation
($\epsilon'/\epsilon$) in the Kaon system,  (iii) the observation of  CP violation in B physics,  and (iv) an indirect
evidence for CP violation due to the existence of baryon asymmetry in the universe.
 Thus far the experimental evidence indicates  that
the CP violation in the K and B physics  can be  understood within the framework  of the standard model.
However,  an understanding of  baryon asymmetry in the universe requires  a new  source of CP violation
We briefly review these below.

\noindent
A.  CP violations in the Kaon system\footnote{For a  review  of this topic
see  \cite{Bertolini:1998vd,Winstein:1992sx}.}\\
Historically the first indication for CP violation came from the
observation of the decay $K_L\to \pi^+\pi^-$.  In  order to
understand this phenomenon  we begin with the states $K^0$ (with
strangeness $S=+1$) and $\bar K^0$ (with strangeness $S=-1$). From
the above  one can construct CP even and  CP odd  eigenstates,
\beqn K_{1,2}={1\over \sqrt 2} (K^0\pm\bar K^0). \eeqn One can
arrange $\bar K^0$ to be the CP conjugate of $K^0$, i.e.,
$CP|K^0>=|\bar K^0>$,  and in that  case
 $K_1$ is the CP even and $K_2$ is the CP odd state.  The decay of  neutral K's
come in two varieties:  $K_S (K_L)$ with lifetimes $\tau_S= 0.89\times 10^{10} s
(\tau_L= 5.2\times 10^{-8})$ with dominant decays $K_S\to \pi^+\pi^-,\pi^0\pi^0 (K_L\to 3\pi, \pi l\nu)$.
 If these were the only decays one would identify $K_S$ with $K_1$ and $K_L$ with $K_2$.
 However,  the decay of the $K_L\to \pi^+\pi^-$ provided  the first  experimental evidence for
 the existence of CP violation~\cite{Christenson:1964fg}.  This experiment indicates that the
 $K_S(K_L)$ are mixtures of CP even and CP odd  states  and one may write
 \beqn
 K_S= \frac{K_1+\bar \epsilon K_2} {(1+|\bar \epsilon|^2)^{1\over 2}},
 ~~K_L= \frac{K_2+\bar \epsilon K_1} {(1+|\bar \epsilon|^2)^{1\over 2}}.
 \eeqn
 Experimentally one attempts to measure two independent CP violating parameters $\epsilon$
 and $\epsilon'$ which are defined by
 \beqn
 \epsilon = \frac{ <(\pi\pi)_{I=0}|{\cal L_W}| K_L>}{ <(\pi\pi)_{I=0}|{\cal L_W}| K_S> },
 \eeqn
where $\cal{L_W}$ is the  Lagrangian for the weak  $\Delta S=1$ interactions,
 and
 \beqn
 \epsilon' = \frac{ <(\pi\pi)_{I=2}|{\cal L_W}| K_L>}{ <(\pi\pi)_{I=0}|{\cal L_W}| K_L> }
 -  \frac{ <(\pi\pi)_{I=2}|{\cal L_W}| K_S>}{ <(\pi\pi)_{I=0}|{\cal L_W}| K_S> }.
 \eeqn
The parameter $\epsilon'$ is  often referred to as a measure of
direct CP violation while $\epsilon$ is referred to as a measure of
indirect CP violation in the Kaon system. An  accurate determination
of  $\epsilon$ has existed  for many years so that

\beqn
|\epsilon| = (2.266\pm 0.017)\times 10^{-3}.
\eeqn
The determination of direct CP violation is more recent and here one
has~\cite{Burkhardt:1988yh,Fanti:1999nm,Alavi-Harati:1999xp}
\beqn
\epsilon'/\epsilon = (1.72\pm 0.018) \times 10^{-3}.
\eeqn
The  above  result  rules  out the so called  superweak theory of CP violation~\cite{Wolfenstein:1964ks}
but is consistent with the predictions of the Standard Model.
A detailed discussion of direct CP violation can be  found in \cite{Bertolini:1998vd}.
There are other Kaon processes where CP violation effects can, in principle,  be discerned.
The most prominent among these is the decay $K_L\to \pi^0\nu\bar \nu$. This process is
fairly clean in that it provides a direct determination of the quantity $V_{td}V_{ts}^*$.
The Standard Model prediction for the branching ratio  is~\cite{Buras:2004uu}
 $BR(K_L\to \pi^0\nu\bar \nu) =
(3.0\pm 0.6)\times 10^{-11}$ while the current experimental limit is~\cite{Anisimovsky:2004hr}
$BR(K_L\to \pi^0\nu\bar \nu) < 1.7\times 10^{-9}$.  Thus an improvement in experiment
by a factor of  around  $10^2$ is needed to test the \sm prediction. On the other hand
significantly larger  contribution to this branching ratio can arise in
 beyond  the \sm physics~\cite{Grossman:1997sk,Buras:2004uu,Buras:2004qb,Colangelo:1998pm}.
A new experiment, 391a, is underway at KEK which would have a significantly
improved sensitivity for the measurement of this branching ratio and its  results could
provide a window to testing new physics in this channel.

We turn now to  B physics. There is considerable literature in this area to which the reader
is directed for details~(\cite{Carter:1980hr, Bigi:1983cj,Bigi:1981qs,Dunietz:1986vi}.
For reviews see
\cite{ Peruzzi:2004tn,Sanda:2004qd,Barberio:1998rm,Quinn:1998jt,Nakada:1994pk,Nardulli:1993ud,Hitlin:1991uc,
Stone:2006rc,Harrison:1998yr}).
CP violations can occur in charged B or neutral B decays such
as $B_d=\bar b d$ and $B_s=\bar bs$.
 In the  $B^0-\bar B^0$ system the
mass eigenstates can be labeled as $B_H$ and $B_L$
with
\beqn
|B_L>= p |B^0> +q|\bar B^0>,\nonumber\\
|B_H>= p |B^0>- q|\bar B^0>, \eeqn where $p(q)$ may be parameterized
by \beqn
p=\frac{1+\epsilon_B}{\sqrt{2(1+|\epsilon_B|^2)}},\nonumber\\
 ~~q= \frac{1-\epsilon_B}{\sqrt{2(1+|\epsilon_B|^2)}}.
\eeqn
A quantity of interest is the  mass difference between these
states, i.e., $\Delta m_s=m_{B_H}-m_{B_L}$.
 Next let us consider a state f which is accessible to both $B^0$ and $\bar B^0$.
A quantity sensitive to CP violation is the asymmetry which is defined by
\beqn
a_f(t) = \frac{ \Gamma(B^{0}(t)\to f)-\Gamma(\bar B^{0}(t) \to  f)}
{ \Gamma(B^{0}(t)\to f)+\Gamma(\bar B^{0}(t)\to  f)}
\eeqn
where $B^{0}(t)$ ($\bar B^{0}(t)$) denote the states which were  initially $B^0$($\bar B^{0}$).
The analysis of the asymmetry becomes specially simple if the final state is an eigen state of CP.
$A_f(t)$ may  be written in the form
\beqn
A_f(t)=  A_f^c \cos(\Delta m t) + A_f^s \sin(\Delta m t)
\eeqn
where
\beqn
A_f^c = \frac{1-|\lambda|^2}{1+|\lambda |^2}, ~~A_f^s = \frac{-2Im \lambda}{1+|\lambda |^2}.
\eeqn
Here   $\lambda\equiv {q\bar A_f}/{p A_f}$, where $A_f=<f|H|B^0>$, and $\bar A_f=<f|H|\bar B^0>$.
An interesting  aspect of
 $a_f$ is that it is free of hadronic uncertainties and for  the \sm case it is determined
fully in terms of the CKM parameters.  This would be the case  if only one amplitude contributes
to the decay $B^0(\bar B^0)\to f$.  More generally one has more than one diagram contributing
with different CKM  phase dependence which make the extraction of CKM phases less transparent.
Specifically $B^0(\bar B^0)$ decays may in general involve penguin diagrams which tend to
contaminate the simple analysis outlined above. Gronau  and London
have  proposed an  isospin
analysis which can disentangle the effect of the tree and penguin contributions when the final
states in $B^0(\bar B^0)$ are  $\pi^+\pi^-$ and $\pi^0\pi^0$ which is useful in the analysis of all
the CKM angles~\cite{Gronau:1990ka,Gronau:1990ra}.
The decay final states $J/\Psi K_S$  is  interesting in that it is a CP   eigen state  and
it has  a large branching ratio and to leading order  is dominated by a single CKM phase.
Specifically, the relation $\bar A_{J/\Psi K_S}/A_{J/\Psi K_S}= 1$ holds
 to within a percent~\cite{Boos:2004xp},
$A_{J/\Psi K_S}^s =  \sin (2\beta)$  and  $A_{J/\Psi K_S}^c = 0$.
Thus $B^0(\bar B^0)$ decay into this mode gives  a rather clean
measurement of $\sin 2\beta$. BaBar and Belle have both measured CP
asymmetries utilizing the charm decays. Using the decays $B^0(\bar
B^0)\to J/\Psi K_S$ and $B^0(\bar B^0)\to J/\Psi K_L$ BaBar and
Belle have obtained  a determination of the CP asymmetry $\sin
(2\beta)$ and the world average for this is~\cite{Barberio:2006bi}
\beqn \sin (2\beta)= 0.685 \pm 0.032 \eeqn While the analysis of CP
asymmetries  in the $J/\Psi K_S$  system is the cleanest way to
determine $\sin (2\beta)$ there are additional constraints on
$\beta$ that are  indirect such as from $\Delta m_d$,  and  $\Delta
m_s$. These lead to a constraint on $\beta$ with  $\beta$ lying in
the range $(13^0, 31^0)$
at $95 \%$ C.L.~\cite{Long:2004gd,Charles:2004jd}.\\

\begin{figure}
\vspace{0.0cm}
 \hspace{-1.0cm}
\includegraphics*[angle=0, scale=0.9]{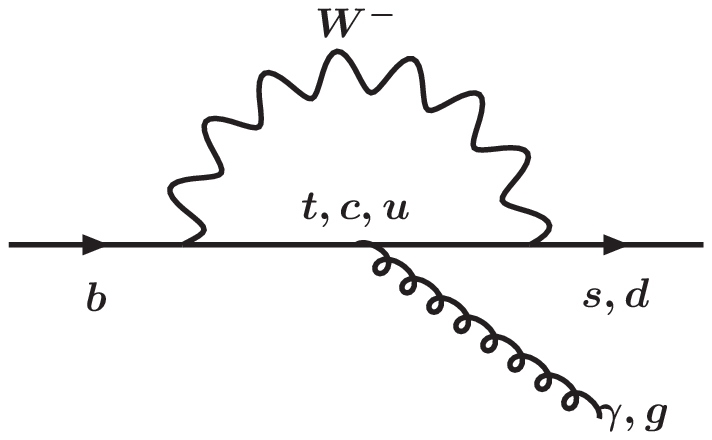}
\vspace{-1.0cm}
\caption{The penquin diagram that contributes to B decays.}
\label{penguin}
\end{figure}

The determination of $\alpha$ comes from the measurement of processes of type
$B^0\to \pi^+\pi^-, ~\rho^+\rho^-$ since the  combinations of phases that enter here are
via $\sin(2(\beta+\gamma))=-\sin(2\alpha)$.  One problem arises  due to the contribution of
the penguin diagram Fig.(\ref{penguin})   which does not contain any  weak phase. The penguin
diagram can thus contaminate the otherwise  neat weak phase dependence of this  process.
A possible  cure come from the fact that one can use the analysis of ~\cite{Grossman:1997jr} to
put an upper limit on the branching ratio for $B^0\to \rho^0\rho^0$.
 The current determination of $\alpha$ gives
$\alpha =(96\pm 13\pm 11)^0$ ~\cite{Stone:2006rc}.
The determination of $\gamma$ comes from the charged decays $B^{\pm}\to D^0K^{\pm}$.
The current  experimental values from BaBar and Belle are
$\gamma=(67\pm 28\pm 13\pm 11)^0$, and  $\gamma =(67^{+14}_{-13}\pm 13\pm 11)^0$~\cite{Stone:2006rc,Asner:2005wf}.
A detailed analysis of global fits  to the CKM matrix can be  found in ~\cite{Charles:2004jd,Charles:2006yw}. \\

We discuss now $D^0-\overline{D^0}$ system. In analogy with the neutral B system we
introduce the two neutral mass
eigen states $D_1,D_2$ defined by
\begin{eqnarray}
|D_1>= p |D^0>+ q |\overline{D^0}>, \nonumber\\
|D_2>= p |D^0>   - q |\overline{D^0}>.
\end{eqnarray}
The D mesons are produced as flavor eigen states but they evolve as admixtures of the mass
eigen states which govern their decays.   The analysis of $D^0$ and $\overline{D^0}$ decays  
by BaBar\cite{Aubert:2007wf}
and by Belle\cite{Staric:2007dt}
  finds no evidence of CP violation.   
  For further details the reader is 
directed to \cite{Nir:2007ac}.\\

The fourth piece of experimental evidence  for CP violation in nature is indirect. It arises from the existence
of a baryon asymmetry in the universe which is generally expressed by the ratio
\beqn
n_B/n_{\gamma}= (6.1^{0.3}_{-0.2})\times 10^{-10}
\eeqn
An attractive picture for the understanding of the baryon asymmetry is that the asymmetry was
generated in the very early history of the universe  within the context of an inflationary universe
starting with no  initial baryon asymmetry (for a recent review on matter-antimatter
asymmetry see ~\cite{Dine:2003ax}).  The basic mechanism how  this can come about
was  already enunciated a long time ago by  ~\cite{Sakharov:1967dj}.
According to Sakharov
there are three  basic  ingredients that govern the generation of baryon asymmetry.
(i) One needs a source of baryon number  violating interactions if one starts out
 with a universe which  initially has  no net baryon  number.  Such interactions arise quite
 naturally in grand unified models  and in string models. (ii)  One needs CP violating interactions
 since otherwise  would be a balance between processes   producing
 particles vs processes producing anti-particle leading to a vanishing net baryon asymmetry.
 (iii) Finally, even with baryon number  and CP violating interactions the production of a net
 baryon asymmetry  would require  a departure from thermal equilibrium.
 Thus one finds  that one of the essential ingredients
  for the generation of the  baryon asymmetry  in the early universe   is the
existence of CP violation.  However, the CP violation in the Standard Model is not
sufficient to generate the desire amount of baryon asymmetry and one needs a source
of CP violation above and beyond what is present in the Standard Model. Such sources
of CP violation are abundant in supersymmetric theories. \\

In addition to the baryon asymmetry in the universe there are other avenues which may reveal
the existence  of new sources  of CP violation beyond what exists  in the Standard  Model.
The EDMs of elementary particles  and of atoms  are prime candidates  for these.
The largest values of EDMs in the framework of the Standard Model SM are very small. SM predicts for the case of the electron the value of $d_e\simeq 10^{-38} ecm$ and for the case of the neutron the value that ranges from $10^{-31}$ to
$10^{-33} ecm$  \cite{Bernreuther:1990jx,Booth:1993af,Shabalin:1982sg,Gavela:1981sk,Bigi:1991rh,Khriplovich:1981ca}.\\

So far no electric dipole moment for the electron or for  the neutron has been detected,
and thus strong bounds on these quantities exist.
For the  electron the current experimental  limit is ~\cite{ Regan:2002ta},
\beqn
|d_e|<1.6\times 10^{-27} ecm ~~~~~~~(90\% ~CL).
\label{EDMe}
\eeqn
For the neutron the Standard Model gives $d_n \sim 10^{-32\pm  1}$
ecm while the current experimental limit is~\cite{Baker:2006ts}
\beqn
|d_n|<2.9\times 10^{-26} ecm ~~~~~~~(90\% ~CL).
\label{EDMn}
\eeqn
In each case one finds that the \sm prediction for the EDM is several orders of magnitude
smaller than the
current experimental limit and thus far beyond the reach of experiment even with
improvement in sensitivity by one to two orders  of magnitude.
 On the other hand
many models of new physics beyond  the \sm  generate  much larger EDMs
and such models are already being  constrained by the EDM experiment. Indeed
improved sensitivities  in  future experiment may lead to a detection of such
effects or put even more stringent constraints on the new physics models.
The EDM of the atoms also provides a sensitive test of CP violation.
An example is  Hg-199 for which the  current limits are    \cite{Romalis:2000mg},
\beqn
|d_{H_g}|<2\times 10^{-28} ecm.
\label{EDMhg}
\eeqn

\section{CP violation in some non-susy extensions of the Standard Model}\label{d}

While the \sm contains just one CP phase  more  phases can appear  in extensions
of the Standard Model.   In general the  violations of CP can be either  explicit or spontaneous.
The CP violation is called explicit if  redefinitions of fields cannot  make all the
couplings  real in the interaction structure of the theory. The remaining phases
provide  an explicit source of CP violation.
CP violation is  called  spontaneous if the
model starts out with all the couplings  being real but spontaneous breaking in the
Higgs  sector  generates a non-removable phase  in one of the vacuum expectation values
in the Higgs fields at the minimum of the potential.  Returning to CP violation in the extension of the
Standard Model, such extensions could be based on an extended gauge  group,
on an extended  Higgs sector,  or on an extended fermionic  content (see, for example,
\cite{Accomando:2006ga}).
An example of a   model with an  extended gauge sector is the left-right (LR) symmetric
model based on the gauge group
$SU(2)_L\times SU(2)_R\times U(1)$~\cite{Mohapatra:1974hk}.
For $n_g$
number of generations the number of phases is given by $N_L+N_R$ where
$N_L=(n_g-1)(n_g-2)/2$ is exactly what one has in $SU(2)_L\times U(1)_Y$ model
and $N_R=n_g(n_g+1)/2$ are  additional set  of phases that arise  in the LR model.
For the case  of three generations this leads to 7 CP  phases instead of just one CP phase
that  one has  in the  Standard Model.
 An analysis of EDM   in LR models for the
electron and for the neutron is given in ~\cite{Frank:1999sy,Frank:1999qk}.\\

The simplest extension of the \sm with an extended Higgs sector is the so called  two
Higgs doublet model~\cite{Lee:1973iz,Lee:1974jb}
 (2HDM) which contains two $SU(2)$ doublets which have exactly the same quantum
 numbers
$\Phi_i=(\phi^+_i, \phi^0_i)$, i=1,2.  One problem with the model is that it
leads to flavor changing neutral currents (FCNC) at the tree level if one allows
couplings of both $\Phi_i$ to the up and down quarks. The  FCNC can be suppressed
by imposing a discrete  ${\cal Z}_2$ symmetry~\cite{Glashow:1976nt}
 such that under  ${\cal Z}_2$ one has
$\Phi_2\to -\Phi_2$ and $u_{iR}\to -u_{iR}$ and the remaining fields are unaffected.
Under the above symmetry the most general  renormalizable scalar potential one can
write  is
\beqn
V_0= -\mu_1^2 \Phi_1^{\dagger}\Phi_1
-\mu_2^2  \Phi_2^{\dagger}\Phi_2
+
\lambda_1 (\Phi_1^{\dagger}\Phi_1)^2 +  \lambda_2 (\Phi_2^{\dagger}\Phi_2)^2+\nonumber\\
+\lambda_3 (\Phi_1^{\dagger}\Phi_1) (\Phi_2^{\dagger}\Phi_2)^2
+\lambda_4  |\Phi_1^{\dagger}\Phi_2|^2+\nonumber\\
+ ( \lambda_5  (\Phi_1^{\dagger}\Phi_2)^2 +H.c.) ~~~~~~~~~~~~
 \eeqn
However, with an exact  ${\cal Z}_2$ discrete symmetry CP cannot be broken either explicitly
or spontaneously in a  2HDM model~\cite{Branco:1980sz,Branco:1979pv,Mendez:1991gp}.
Thus to have CP in the 2HDM model one must allow for violations of the discrete symmetry,
but arrange for suppression of FCNC. If the couplings allow for FCNC at the tree level, then
they must be suppressed  either by heavy Higgs
 masses~\cite{Branco:1985pf,Lahanas:1977my} or by adjustment of couplings or
fine  tunings  so that FCNC are  suppressed but  CP violation is allowed~\cite{Liu:1987ng}. \\

However, the hard breaking of the ${\cal Z}_2$ discrete symmetry is generally
considered not acceptable.  A more desirable possibility is violation of the discrete
symmetry only via  soft terms~\cite{Branco:1985aq}.
 Here the FCNC are not allowed at the  tree level but the inclusion of the soft terms
allows for CP violation.  Such a term is of the form
\beqn
V_{soft} = -\mu_{3}^2 \Phi_1^{\dagger}\Phi_2+ H.c.
\eeqn
Soft breaking of the ${\cal Z}_2$ symmetry can allow  both explicit and
spontaneous CP violation.  Thus explicit CP violation can occur in $V=V_0+V_{soft}$
if one has\cite {Grzadkowski:1999ye}
$Im(\mu_3^{*4} \lambda_5)\neq 0$.
For  the case when $Im(\mu_3^{*4} \lambda_5)= 0$  a spontaneous violation of $CP$ can
arise. Specifically, in this case one can choose phases  so that
$<\Phi_1>=v_1/\sqrt 2$ ($v_1>0$)  and $<\Phi_2>=e^{i\theta}v_2/\sqrt 2$ ($v_2>0)$
with the normalization
\beqn
\sqrt{v_1^2+v_2^2}=2m_W/g_2=246 {\rm GeV.}
\eeqn
 The  conditions for CP violation in a 2HDM model,  both explicit and spontaneous, have more
 recently been studied using  basis independent potentially complex invariants which are combinations
 of mass  and coupling
 parameters.
 These invariants also are helpful in distinguishing between
 explicit and spontaneous CP violation in the Higgs sector.
 For further discussion, the reader is refereed to the works of
 \cite{ Branco:2005em,Gunion:2005ja,Ginzburg:2004vp,Botella:1994cs,Lavoura:1994fv, Davidson:2005cw}.
While the spontaneous breaking of CP discussed above involves  $SU(2)$ Higgs
doublets which may enter in the  spontaneous breaking of the electro-weak symmetry, 
similar spontaneous violations of CP  can occur in sectors not related to 
electro-weak symmetry breaking.\\

 In the  absence of CP violation, the Higgs sector of the theory after spontaneous breaking of the
 $SU(2)_L\times U(1)_Y$ symmetry gives two CP even, and one CP odd  Higgs in the neutral
 sector.  In the  presence of CP violation, either explicit or spontaneous,  the CP eigenstates mix
 and the mass eigenstates  are admixtures of CP even and CP odd states.
The above leads to interesting phenomenology  which is discussed in detail in
\cite{Grzadkowski:1999ye,Mendez:1991gp}.
The number of independent  CP phases  increases  very rapidly with increasing
number of Higgs doublets. Thus, suppose we consider an $n_D$ number of Higgs
doublets. In this case the number of independent  CP phases  that can appear
in the unconstrained  Higgs potential is~\cite{Branco:2005em}
$N_p =n_D^2(n_D^2-1)/4 -(n_D-1)$.  For $n_D=1,2,3$ one gets $N_p=0,2,16$,
and thus the number  of independent CP phases  rises rather rapidly as the number
of Higgs  doublets increases.  An analysis of the EDMs in the two Higgs model is given in
~\cite{Barger:1996jc,Hayashi:1994xf}.
 Finally, one may consider
extending the fermionic sector of theory with inclusion of
additional generations. Such an extension brings in more possible
sources of CP violation. Thus, for example, with four generation of
quarks the extended  CKM matrix will be $4\times 4$. Such a matrix
can be parameterized in terms of six angles and  three
phases~\cite{Barger:1981du,Oakes:1982ey}. Thus generically
extensions of  the \sm will in general have more  sources of CP violation
than the \sm. We discuss CP violation in supersymmetric
theories next.
While the spontaneous breaking of CP discussed above involves  $SU(2)$ Higgs
doublets which may enter in the  spontaneous breaking of the electro-weak symmetry, 
similar spontaneous violations of CP  can occur in sectors not related to 
electro-weak symmetry breaking.

\section{CP violation in supersymmetric theories}\label{e}
 Supersymmetric models are one of the leading candidates for new physics
 (for review see~\cite{Nath:1983fp,Nilles:1983ge,Haber:1984rc,Martin:1997ns})
 since they allow for a technically natural solution to the gauge hierarchy problem.
 However, supersymmtetry is not  an exact symmetry of nature, Thus one must
 allow for breaking of supersymmetry in a way that does not violate the
 ultraviolet behavior of the theory and destabilize the hierarchy. This can be
 accomplished by the introduction of soft breaking.  However,
the soft breaking sector in the  minimal supersymmetric standard model (MSSM)
allows for a large number of arbitrary
parameters ~\cite{Girardello:1981wz,Dimopoulos:1981zb}.
Indeed in softly broken supersymmetry with the particle content of MSSM
additionally 21 masses, 36 mixing angles and 40 phases \cite{Dimopoulos:1995ju}.
 which makes the model rather unpredictive. \\

The number of parameters is significantly reduced  in the minimal supergravity unified
models under the assumptions of a flat Kahler metric as explained below.
The minimal supergravity model and supergravity model in general are
constructed using techniques of applied N=1 supergravity, where one couples
chiral matter multiplets and a vector multiplet  belonging to the adjoint representation
of a gauge group to each other and to supergravity.  The supergravity couplings
can then be described in terms of three arbitrary functions: the superpotential $W(z_i)$
which is a holomorphic function of the chiral fields $z_i$, the K\"ahler  potential
$K(z_i, z_i^{\dagger})$ and the gauge kinetic energy function
$f_{\alpha\beta}(z_i, z_i^{\dagger})$ which transforms  like the symmetric product
of two adjoint representations. In supergravity models
  supersymmetry is broken in a so called hidden sector and is  communicated
to the physical sector where quarks and lepton live via gravitational interactions.
The size of the soft breaking mass, typically the  gravitino mass $m_{3\over 2}$,
 is $\sim \kappa^2  |<W_h>|$, where $W_h$ is the
superpotential in the hidden sector where supersymmetry breaks and
$\kappa= 1/M_{Pl}$, where $M_{Pl}$ is the Planck mass.
The simplest
model  where supersymmetry breaks in the hidden sector via  a super Higgs  effect
is given by
  $W_h=m^2 z$ where $z$ is the Standard Model singlet super Higgs field.
The breaking of supersymmetry by supergravity interactions in the hidden sector
gives $z$ a VEV of size $\sim \kappa^{-1}$, and thus with $m\sim 10^{10-11}$ GeV, the
soft breaking mass is of size $\sim 10^{3}$ GeV. \\

In the minimal supergravity model one assumes that the K\"ahler  potential has
no generational dependence and is flat and further  that the gauge kinetic
energy function is diagonal and has no field dependence, i.e., one has
effectively $f_{\alpha\beta}\sim \delta_{\alpha\beta}$. In this case one
finds that the low energy theory obtained after integrating the GUT scale
masses has the following soft breaking
potential ~\cite{Chamseddine:1982jx,Hall:1983iz,Nath:1983aw}

\beqn
{\cal V_{SB}} =m_{\frac{1}{2}} \bar \lambda^{\alpha} \lambda^{\alpha}
+ m_0^2 z_az_a^{\dagger} + (A_0W^{(3)}+B_0W^{(2)} +H.c.)\nonumber\\
\label{softlag}
\eeqn
 where $W^{(2)}$ is the quadratic   and $W^{(3)}$ is cubic in the fields.

 The physical sector  of supergravity models consist of the MSSM fields,
 which include the three generations of quarks and leptons and their
 superpartners, and a pair of  $SU(2)_L$ Higgs doublets $H_1$ and $H_2$ and
 their superpartners which are the corresponding Higgsino  fields
 $\tilde H_1$ and $\tilde H_2$.
 For the case of MSSM one has
 \beqn
 W^{(2)}= \mu_0 H_1H_2,\nonumber\\
 W^{(3)}= \tilde Q Y_{U}  H_2 \tilde u^c + \tilde Q Y_{D}  H_1 \tilde d^c +
 \tilde L Y_{E}  H_2 \tilde e^c
\label{w23}
 \eeqn
Here $H_1$ is Higgs doublet that gives mass to the bottom quark and the lepton,
and $H_2$ gives mass to the up quark.
As is evident from Eqs(\ref{softlag}) and (\ref{w23}) the minimal supergravity theory
is characterized by the parameters
: $m_0, m_{\frac{1}{2}}, A_0, B_0$ and $\mu_0$.
An interesting aspect of supergravity models is that they
allow for spontaneous breaking of the $SU(2)_L\times U(1)_Y$  electroweak symmetry~\cite{Chamseddine:1982jx}.
 This can be accomplished in an efficient  manner by radiative breaking using renormalization group
  effects~\cite{Inoue:1982pi,Ibanez:1982fr,Alvarez-Gaume:1983gj,Ellis:1983bp,Ibanez:1983di,Ibanez:1984vq,Ibanez:2007pf}. \\

 To exhibit spontaneous breaking one considers
  the scalar potential of the Higgs  fields by evolving the potential to
  low energies by renormalization group effects such that
   \beqn
V=V_0+\Delta V
\label{veff1}
\eeqn
where $V_0$ is the tree level potential \cite{Nath:1983fp,Haber:1984rc,Nilles:1983ge}
\beqn
V_0=m_1^2 |H_1|^2+m_2^2 |H_2|^2+(m^2_3 H_1.H_2+H.c.)\nonumber\\
+\frac{g_2^2+g_1^2}{8}|H_1|^4+\frac{g_2^2+g_1^2}{8}|H_2|^4-\frac{g_2^2}{2}|H_1.H_2|^2\nonumber\\
+\frac{g_2^2-g_1^2}{8}|H_1|^2|H_2|^2.
\label{veff2}
\eeqn
and  $\Delta V$ is the one loop correction to the effective potential and is given by \cite{Coleman:1973jx,Weinberg:1973ua,Arnowitt:1992qp,Carena:2000yi}
\beqn
\Delta V=\frac{1}{64\pi^2}Str(M^4(H_1,H_2)(\log{\frac{M^2(H_1,H_2)}{Q^2}}-\frac{3}{2})).\nonumber\\
\label{veff3}
\eeqn
  Here $Str=\Sigma_iC_i(2J_i+1)(-1)^{2J_i}$,   where the sum runs over all particles with spin $J_i$ and $C_i(2J_i+1)$ counts the degrees of freedom of the particle $i$ and $Q$ is the running scale which is to be in the electroweak region.
 The  gauge coupling constants and the soft parameters are subject to  the supergravity boundary
conditions:  $\alpha_2(0)= \alpha_G= \frac{5}{3} \alpha_Y(0);$
$m_i^2(0)=m_0^2+\mu_0^2, ~i=1,2;$ and $m_3^2(0)=  B_0\mu_0$.
As one evolves the potential downwards from the GUT scale using renormalization group
equations\cite{Machacek:1984zw,Machacek:1983fi,Machacek:1983tz,Martin:1993zk,Jack:1994rk},
 a breaking of the electro-weak
symmetry occurs when the   determinant  of the Higgs mass$^2$ matrix
turns negative so that $(i) ~~m_1^2 m_2^2 -2m_3^4<0,$ and further
for a stable  minimum to  exist one  requires  that the potential be
bounded from below  so that $(ii)  ~m_1^2+m_2^2 -2|m_3^2|>0$.
Additionally one must impose the constraint that there be  color and
charge  conservation. Defining $v_i =<H_i>$ as  the VEV of the
neutral component of the Higgs $H_i$,  the necessary conditions for
the minimization of the potential, i.e.,  $\partial V/\partial
v_i=0$ , gives two constraints. One of these can be  used to
determine the magnitude  $|\mu_0|$ and the other can be used to
replace $B_0$ by $\tan\beta\equiv <H_2>/<H_1>$.   In this case  the
low energy supergravity model or mSUGRA can be parameterized by
$m_0, m_{\frac{1}{2}}, A_0,  \tan\beta$ and $\rm{sign}(\mu_0)$. 
It  should be noted that fixing the value $|\mu|$ using radiative breaking does
entail fine tuning but a measure of this is model dependent (see, for example,
 ({Chan:1997bi}) and the references therein).
The above discussion is for the case  when there are no CP violating phases
in the theory. In the presence of CP phases $m_{\frac{1}{2}}, A_0,
\mu_0$  become complex and one may parameterize them so that 
\beqn
m_{\frac{1}{2}} =|m_{\frac{1}{2}}|e^{i\xi_{1\over 2}}, ~~A_0 =|A_0|
e^{i\alpha_0},  \mu_0=|\mu_0|e^{i\theta_{\mu_0}}. \label{soft1} 
\eeqn
Now not all the phases are independent. Indeed, in this case only
two phase combinations are independent, and in the analysis of the
EDMs  one finds these to be $\xi_{1\over 2}+\theta_{\mu_0}$ and
$\alpha_0+\theta_{\mu_0}$. Often one rotates away the phase of the
gauginos which is equivalent to setting $\xi_{1\over 2}=0$, and thus
one typical choice of parameters for the complex mSUGRA  (cmSUGRA)
case is

\beqn
m_0,  ~~|m_{\frac{1}{2}}|, ~ \tan\beta,  ~~|A_0|; ~~ \alpha_0, ~~\theta_{\mu_0}  ~~~({\rm  cmSUGRA}).
\label{soft2}
\eeqn
  However, other choices are  equally valid: thus, for example, the independent  soft
breaking parameters  can be chosen to be
$m_0,  ~~|m_{\frac{1}{2}}|, \tan\beta,  ~~|A_0|, ~~ \alpha_0, ~~\xi_{\frac{1}{2}}$.
mSUGRA model was derived using a  super Higgs  effect which breaks supersymmetry in the hidden
sector by VEV formation of a  scalar super Higgs field.  
Alternately one can view breaking of
supersymmetry as  arising from   gaugino condensation where  in analogy with QCD where  one forms
the condensate  $q\bar q$ one has that the strong dynamics of an asymptotically free gauge theory 
 in the hidden sector
produces a gaugino condensate with $<\lambda \gamma^0 \lambda>=\Lambda^3$. The above  can 
lead  typically to  supersymmetry breaking 
   and a gaugino mass  of size $m_{\frac{3}{2}} \sim \kappa^2  \Lambda^3$.   With  
       $ |\Lambda >| \sim (10^{12-13})$ GeV     one   will have
   an  $m_{\frac{3}{2}} $  again in the electro-weak region \cite{Nilles:1982ik,Ferrara:1982qs,Dine:1985rz,Taylor:1990wr}.
   \\

   The assumption of  a flat K\"ahler  potential and of a flat kinetic energy function in supergravity unified
   models is essentially a  simplification, and in general  the nature  of the physics at the Planck scale
   is  largely unknown. For this reason one  must also consider more general K\"ahler 
    potentials ~\cite{Soni:1983rm,Kaplunovsky:1993rd}   and
   also allow for the non-universality of the gauge kinetic energy function. In this case the number
    of soft parameters grows, as also do the number  of CP phases. Thus, for example, the gaugino
    masses will be complex and  non-universal, and the trilinear parameter $A_0$, which is in general
    a matrix in the generation space,  will also be  in general
   non-diagonal and complex. A simplicity assumption to maintain the appropriate constraints on
   flavor changing neutral currents is  to assume  a diagonal form for $A_0$ at the GUT scale.
   Additionally,  the Higgs masses for $H_1$ and $H_2$  at the GUT scale could also be non-universal.
   Thus in general for the non-universal supergravity unification a canonical set of soft parameters
   at the GUT scale will consist of ~\cite{Matalliotakis:1994ft,Olechowski:1994gm,Polonsky:1994rz,Nath:1997qm}
   \beqn
  m_{H_i}= m_0(1+\delta_i), ~i=1,2\nonumber\\ m_{\alpha}=|m_{\alpha}| e^{i\xi_{\alpha}} ,~~ \alpha=1,2,3\nonumber\\
  A_a =|A_a| e^{i\alpha_a}, ~a=1,2,3
  \label{soft3}
   \eeqn
   which contain  several additional CP phases beyond the two phases in complex mSUGRA.
    However, not all the
    phases are independent, as some phases can be eliminated by field redefinitions. Indeed in physical
    computations only a certain set of phases appear, as  discussed  in detail in \cite{Ibrahim:1999aj}
    (also see Appendix \ref{qE}).
It should  be  kept in mind that for  the case  of non-universalities the renormalization group evolution
gives an additional correction  term   at low energies   \cite{Martin:1993zk}.\\

As is apparent from the preceding discussion  radiative breaking of
the electroweak symmetry plays a central role in the supergravity
unified models.  An interesting phenomena here is the  existence of
two branches of radiative  breaking:  one is the conventional branch
known since the early eighties (we call this the ellipsoidal branch
(EB)) and the other was  more  recently discovered, i..e, it is the
so called hyperbolic branch (HB).  The two branches can be
understood simply by examining the condition of radiative breaking
which is a constraint on the soft parameters $m_0, m'_{1/2},  A_0$
of the form  \cite{Chan:1997bi}
  \beqn
C_1m_0^2+C_3m'^2_{1/2}+C_2'A_0^2+\Delta \mu^2_{loop}=
     {M_Z^2\over 2} + \mu^2.
\label{hb1} \eeqn Here $\Delta\mu^2_{loop}$ is the loop
correction~\cite{Arnowitt:1992qp,Carena:2000yi}, and
$m_{1/2}'=m_{1/2}+\frac{1}{2}A_0C_4/C_3$, where $C_i$ are determined
purely in terms of the gauge and the Yukawa couplings but depend on
the renormalization group scale $Q$. The behavior of radiative
breaking is controlled in a significant way by the loop correction
$\Delta\mu^2_{loop}$  especially for moderate to large values of
$\tan\beta$. For small values of $\tan\beta$ the loop correction
$\Delta\mu^2$ is small around $Q\sim M_Z$, and the $C_i$  are
positive and thus Eq.(\ref{hb1}) is  an ellipsoidal constraint on
the soft parameters. For a given value of $\mu$,  Eq.(\ref{hb1})
then puts an upper limit on the sparticle masses.  However, for
moderate to large values of $\tan\beta$,
 $\Delta\mu^2$ becomes sizable. Additionally $C_i$ develop a significant
 $Q$  dependence.  It is then possible to choose a point $Q=Q_0$ where
  $\Delta\mu^2$  vanishes and quite interestingly here one finds that
  one of the $C_i$ (specifically $C_1$)  turns negative, drastically changing
  the nature of the symmetry breaking constraint Eq.(\ref{hb1}) on the
  soft parameters. Thus in this case the soft parameters in Eq.(\ref{hb1})
  lie on the surface of a  hyperboloid and thus for a fixed  value of $\mu$ the
  soft  parameters can get very large with $m_0$ getting as large as 10 TeV or
  larger.   The direct observation of  squarks and sleptons may be difficult on this branch,
  although charginos, neutralinos and even gluino may be accessible. However,
  the HB  does have other desirable features such as suppression of
  flavor changing neutral currents, and suppression of the SUSY EDM contributions.
  Further, HB still allows for satisfaction of relic density constraints  with R parity
  conservation if the lightest neutralino is the lightest
  supersymmetric particle (LSP).  We note in passing that the   so called focus
  point region~\cite{Feng:1999zg} is included in the hyperbolic
  branch~\cite{Chan:1997bi,Baer:2003ru,Lahanas:2003bh}.  \\

   There is a potential danger in supergravity theories in that the hierarchy could be destabilized
 by non-renormalizable couplings in supergravity models
since they can lead to power law divergences. This issue has been investigated by several authors:
at one loop by~\cite{Bagger:1993ji,Gaillard:1994sf}
and at two loop by ~\cite{Bagger:1995ay}.  The analysis shows that at the one loop  level the
minimal supersymmetric standard model appears to be  safe  from divergences~\cite{Bagger:1993ji}.
In addition to the breaking of supersymmetry by gravitational interactions, there are a variety
of other scenarios for supersymmetry breaking. These include  gauge mediated and anomaly
mediated breaking for which  reviews can be found in  \cite{Giudice:1998bp,Luty:2005sn}.
 Finally as  is clear from the preceding discussion in supergravity models and  in MSSM there is no
 CP violation at the tree level in the Higgs sector of the theory. However, this situation changes when
 one includes the loop correction to the Higgs potential. This leads to  the generation of CP violating
 phase for one the Higgs VEVs and leads to mixings between the CP even and the CP odd Higgs fields.
   This phenomenon is very interesting from the experimental view point and  will be discussed
 in greater  detail later.

While the Standard Model contribution to the EDMs of the electron and of the neutron is small and beyond
  the pale of observation of the current or the future experiment, the situation in supersymmetric models is quite
  the opposite. Here the new sources  of CP violation can generate large contributions to the  EDMs even
  significantly above the current experimental limits. Here one needs special mechanisms to suppress the
  EDMs such as mass suppression~\cite{Nath:1991dn,Kizukuri:1992nj}
     or the cancelation mechanism to control the effect of large CP phases
  on the EDMs. ~\cite{Ibrahim:1997nc,  Ibrahim:1997gj,Ibrahim:1998je,Ibrahim:1999af,  Chattopadhyay:2000fj,  Ibrahim:2001ht}.
Specifically for the cancelation mechanism the phases can be large
and thus affect  a variety of CP phenomena which can be observed in
low energy experiments and at accelerators. The literature on this topic is quite large. A sample of these  analyses can be found in 
~\cite{Chattopadhyay:1998wb,Ibrahim:1999hh,Ibrahim:1999aj,Ibrahim:2000tx,Ibrahim:2000qj,Ibrahim:2001ym,Ibrahim:2001jz,Ibrahim:2002zk,Ibrahim:2002fx,Ibrahim:2003ca,Ibrahim:2003jm,Ibrahim:2003tq,Gomez:2004ek,Ibrahim:2004cf,Gomez:2004ef,Ibrahim:2004gb,Gomez:2005nr,Gomez:2006uv,Alan:2007rp,Falk:1998pu,Huang:1999an,Huang:2000ha,Huang:2000tz,Akeroyd:2001kt,Boz:2000wr,Demir:1999hj,Bartl:2006hh,Bartl:2003pd}.

  \section{ CP violation in extra dimension models}\label{f}
    Recently there has been significant activity in the physics of extra
    dimensions~\cite{Arkani-Hamed:1998rs,Gogberashvili:1998vx,Antoniadis:1998ig,Antoniadis:1990ew,Randall:1999vf,Randall:1999ee}.
          One might speculate on the
  possibility of generating CP violation in a natural way from models derived from extra dimensions (For an early
  work see ~\cite{Thirring:1972de}).
  It turns out that it is indeed possible to do
  so ~\cite{Khlebnikov:1987zg,Sakamura:1999fa,Branco:2000rb,Huang:2001np,Chang:2001yn,Chaichian:2001nx,Chang:2001uk,Dienes:2001zz,Burdman:2003nt,Grzadkowski:2004jv}.
   The idea is to utilize properties  of the hidden compact dimensions in extra  dimension models. Thus in
  extra  dimension models  after compactification the physical four dimensional space is a slice of the
  higher dimensiional space and  such a slice   can be placed  in different locations in extra  dimensions.
  In the discussion below we will label such a slice  as a brane.  We consider now  a
  simple argument  which illustrates how CP violation in extra dimension models 
   can arise ~\cite{Chang:2001yn}. Thus consider a $U(1)$ gauge theory with left-handed fermions $\Psi_i$ (i=1-4),
  where $i=1,2$ have charges +1 and  $i=3,4$ have charges  $-1$, and also consider a real scalar field $\Phi$ which is neutral.
  We assume that the fermion fields  are in the bulk and the scalar field is confined to the $y=0$ brane.
 The fields  $\Psi_{1}, \Psi_{2}$  and $\Phi$ are   assumed to be even and $\Psi_{3L}, \Psi_{4L}$ are assumed to be odd under $y\to -y$
 transformation. Further, under CP symmetry define  the fields to
  transform so that $\Psi_{1L}\to (\Psi_{3L})^c, \Psi_{2L}\to (\Psi_{4L})^c$, and $\Phi\to - \Phi$ where $(\Psi_L)^c$ has  the meaning of a 4D charge
   conjugate of $\Psi$.  One constructs  a 5D Lagrangian invariant under $y\to -y$ transformation of the form
  \beqn
  M_5^{-1}\lambda_5  \delta(y)\Phi  [ \Psi^T_{iL}C^{-1} \Psi_{2L} -(\Psi_{3L})^{cT}C^{-1} (\Psi_{4L})^c]\nonumber\\
  +\mu [ \Psi^T_{iL}C^{-1} \Psi_{2L} -(\Psi_{3L})^{cT}C^{-1} (\Psi_{4L})^c]  + H.c.
  \eeqn
  On integration over the y co-ordinate the interaction terms in 4D arise from the couplings on the y=0 brane and thus
  the zero modes of the fields odd in y are absent, which means that the effective interaction at low energy in
  $( \lambda   \Phi +\mu)  \Psi_{1L}^{0T} \Psi_{2L}^{(0)}$ which violates CP provided $Im(\lambda^*\mu)\neq 0$.
    Next we discuss a  more detailed illustration of this CP violation arising from extra dimensions.
   This  illustration is an explicit exhibition of  how  violations of CP  invariance can occur
   in the compactification of a 5D QED~\cite{Grzadkowski:2004jv}.
    Thus  consider the Lagrangian in 5D of the form
   \beqn
   {\cal L}_5 = -\frac{1}{4} V_{MN}^2  +\bar \Psi (i\gamma^MD_M-m_i) \Psi +{\cal L}_{gh}.
   \eeqn
  Here $V_M$ is the vector potential in 5d space with co-ordinates $z^M$, where $M=0,1,2,3,5$ so that
  $z^M=(x^{\mu}, y)$, where $\mu=0,1,2,3$, and where $D_M= \partial_M+ig_5q V_M$ is the gauge covariant derivative,
  with $g_5$ the $U(1)$ gauge coupling constant, and $q$ the charge of fermion field.
  The theory is invariant under the following gauge transformations

  \beqn
  \psi(z) \to e^{-ig_5 q \lambda} \psi(z)\nonumber\\
  V_M(z)\to V_M(z) + \partial_M\lambda(z),
  \eeqn
  and additionally under the CP transformations in 5D
  \beqn
  z^m\to \eta^M z^M, ~~V^M\to \eta^M V^M, ~~\psi\to P \gamma^0\gamma^2 \psi^*
  \eeqn
 where $\eta^{1,2,3}=-1= -\eta^{0,5}$ and $P=1$.
  We compactify the theory
 in the fifth dimension on a circle with radius R assuming periodic boundary conditions  for the gauge fields but
 assuming  the twisted  boundary condition for the fermion field
  \beqn
  \psi(x,y+R)= e^{i\alpha} \psi(x,y).
  \eeqn
 One can now carry out a mode expansion in 4D and recovers a massless zero mode $V_{\mu}(x)$  for the vector field
 (the photon). One also finds in addition a  massless field  $\phi(x)$ which is the zero mode of the $V_5(x,y)$  expansion.
 This is so because while  $V_5^n, n\neq 0$ modes can be eliminated by an appropriate gauge choice,
 while  $\phi$ is a gauge
 singlet and remains in the spectrum.  We note in passing that the presence of the zero mode is a consequence of
 the specific compactification chosen. Thus compactification, on $S^1/Z_2$, rather than on the circle will remove the
 field $\phi$.  Now while $\phi$ is massless at the tree level, it can develop a mass when loops contributions are
 included. Thus an analysis of one loop effective potential gives~\cite{Grzadkowski:2004jv}.
 \beqn
 V_{eff}= \frac{1}{2\pi^4R^4}\sum_i [ \beta^2_i L_{i_3}(\gamma_i) +3\beta_i L_{i_4}(\gamma_i) + 3L_{i_5}(\gamma_i)]
 \eeqn
  where $\beta_i =mR$,  $\gamma_i=exp(i\omega_i R-\beta_i)$, and  where $\omega_i=(\alpha_i+g_5q_iR\phi_0)$,
  and $\phi_0=<\phi>$,  and $L_{i_n}$ is the polylogarithm function.

 Now it turns out that for the case when one has a single fermion, there is no CP violation, but CP violation is possible when
  there are two fermions and one can assume the boundary conditions in this case so that
 $  \psi_1(x,y+R)=  \psi_1(x,y)$ and       $ \psi_2(x,y+R)= e^{i\alpha} \psi_2(x,y)$.  In this situation the Yukawa couplings
 for the fermions  violate CP. An interesting phenomenon  here is  that  the above mechanism exhibits examples of
 both spontaneous CP violation as  well as explicit CP violation. Thus for the case $\alpha =0, \pi$ one finds that
 the effective  potential is symmetric in $\phi_0$ and one has two degenerate minima  away from $\phi_0=0$ and thus
 here one has spontaneous breaking of CP. For other choices  of $\alpha$, the effective potential is not symmetric in
 $\phi_0$ and one has explicit violation of CP.  The fact that CP is indeed violated in this example can be tested
 by an explicit computation of the EDM of the fermions which is non-vanishing and suppressed   by the inverse size
 of the extra dimension.

 \begin{figure}
% \hspace{-.5cm}
\includegraphics*[angle=0, scale=0.5]{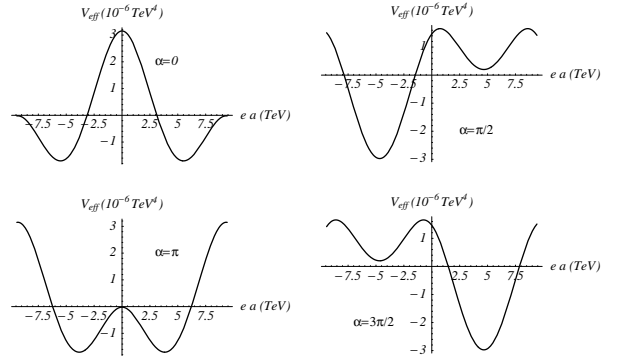}
\vspace{.5cm}
\caption{An exhibition of the phenomena of spontaneous vs explicit breaking in a 5D compactification
model \cite{Grzadkowski:2004jv}.  The figure gives the effective potential $V_{eff}$ for four cases
of twist angles with $\alpha=0$,  $\pi/2$, $\pi$, $3\pi/2$. The cases $\alpha=0, \pi$ correspond to spontaneous breaking
and $\alpha =\pi/2, 3\pi/2$ correspond to explicit breaking.}
\label{effpot}
\end{figure}

 We turn now  to another mechanism for the generation of CP violation in extra dimensional theories.  This scenario is
 that of split fermions where the hierarchies of fermion masses and couplings are  proposed to arise from a fermion location
 mechanism under  a kink background  wherein   the quark and
 leptons of different generations being confined to different points in a  fat
 brane~\cite{Arkani-Hamed:1999dc,Mirabelli:1999ks,Kaplan:2001ga,Kaplan:2000av}.
  To illustrate the fat  brane paradigm  consider the 4+1 dimensional action of two fermions
   \beqn
 S_5 = \int d^4x dy [ \bar Q[ i\gamma_M \partial^M +\Phi_Q(y)]Q +\nonumber\\
 + \bar U [i\gamma_M\partial^M +\Phi_U(y)] U +
 \kappa H Q^c U ].
 \eeqn
 The quantities $\Phi_{Q,U}$ are potentials which confine the quarks at different points in the extra dimension.
 As a model one may consider these  as  Gaussian functions centered around points  ${\it l}_q$ ( i.e.,
 functions of the form $exp(-\mu^2(y-{\it l}_q)^2)$) and ${\it l}_u$
  where $1/2\sqrt \mu$ is the width of the Gaussian.
 After expanding the  fields in their normal modes and integrating over the extra dimension the Yukawa interaction
 in 4D including the generation index will take the form
 \beqn
 {\cal L}_Y= \lambda^u_{ij} Q_iU_jH+ \lambda_{ij}^d  Q_i D_j H^*,
 \eeqn
      where $\lambda_{ij}^u$  is defined by
      \beqn
      \lambda_{ij}^u = \kappa_{ij} e^{-{1\over 2}\mu^2({\it l}_{q_i}-  {\it l}_{u_i})},
            \eeqn
and $\lambda_{ij}^d$ is similarly defined.  The above structure indicates that the Yukawa textures  are governed by
the location of the quarks in the extra dimension. Detailed analyses, however, indicate that this  scenario leads
to an insufficient amount of CP violation to explain the value of $\epsilon_K$ in Kaon decay. Thus the scenario above
gives a value of the Jarskog invariant $J\leq 5\times 10^{-9}$ while one needs $J\sim 10^{-5}$ to get the proper value
of $\epsilon_K$.   The above shortcoming can be corrected by extending the analysis to two extra
 dimensions~\cite{Branco:2000rb}.
In this case one finds the   Jarlskog invariant $J\simeq 2.2\times 10^{-5}$  which is of desired strength to explain CP violation in the
Kaon decay.  An extension to include  masses for  the charged leptons and neutrinos has been carried out
in  \cite{Barenboim:2001wy}. \\

An analysis  using the fermion localization mechanism for generating
quark-lepton textures within a supersymmetric  SU(5) GUT theory is
carried out in the analysis of  \cite{Kakizaki:2001ue} where the
different $SU(5)$ chiral multiplets are localized along different
points in the extra dimension. The analysis allows one to generate a
realistic pattern of quark masses and mixings  and lepton masses.
The CP violation is of sufficient strength here since $J\sim
O(10^{-5})$.  An additional feature of this model is that dimension
5 proton decay operators are also naturally suppressed due to the
fact that these operators contain an overlap of wavefunctions  of
different chiral multiplets and are thus exponentially suppressed.

Similar analyses can be  carried out in the framework of a non-factorizable
geometry~\cite{Huber:2000ie,Grossman:1999ra,Abe:2001ax,Chang:1999nh}
 based on the metric
\beqn
ds^2=e^{-2\sigma(y)} (dx)^2 -dy^2,
\eeqn
where $\sigma(y)=k|y|$.  Under the $Z_2$ orbifold symmetry the 5D fermion transform as
$\Psi(-y)_{\pm}= \pm \gamma_5 \Psi(y)_{\pm}$. The $\Psi_{\pm}$ have the mode  expansion
\beqn
\Psi(x,y)_{\pm} = \frac{1}{\sqrt{2\pi r_c}} \sum_{n=0}^{\infty} \psi_{n\pm} (x) f_{\pm}^{(n)} (y).
\eeqn
The zero modes of $\Psi_{\pm}$ are the left-handed and the right handed Weyl spinors. Masses for these  are  generated  by
the 5D Higgs couplings which are of the form

\beqn
\int d^4x dy \sqrt{-g} \lambda_{ij} H \bar \Psi_{i+} \Psi_{j-}.
\eeqn
For the zero mode they give rise to a Dirac mass term of the form~\cite{Huber:2000ie}
\beqn
m_{ij}= (2\pi r_c)^{-1}  \int_{-\pi r_c}^{\pi r_c} dy \lambda_{ij} H(y) f_{i+}^{(0)}(y)  f_{i-}^{(0)}(y)
\eeqn
where
\beqn
f^{(0)}= ( { {e^{2\pi r_c ( {1\over 2}-c)} -1}  \over {2\pi k r_c( {1\over 2}-c)}})^{-\frac{1}{2}} e^{(2-c)\sigma}
\eeqn
where c  is a parameter that characterizes the location of the fermion in the extra  dimension.
For $c<1/2$ the fermion is localized near the $y=0$ brane while for $r=\pi r_c$ it is localized near $y=\pi r_c$
brane.  With the appropriate choice  of the c's  one may generate  a realistic  pattern of quark masses and mixings
and a realistic CKM matrix. However, an explicit determination of the Jarlskog invariant  appears not
to have  been carried out.  The texture models using  extra dimensions do generally require a high level of
fine tuning in the selection of locations where the  fermions are placed. Thus models of this type
do not  appear very natural.  For  related works on CP violation and extra dimensions
see \cite{Ichinose:2002hb,Huang:2001np,Sakamura:1999fa,Dooling:2002js}.

\section{CP violation in strings}\label{g}

We discuss now the possible origins of CP violation in
 SUSY, string and brane models (for review of string theory see \cite{Green:1987sp,Green:1987mn,Polchinski:1998rq,Polchinski:1998rr}).
  One possible origin is string
 compactification\cite{Wu:1990ay,Kobayashi:1994ks,Bailin:2000ra,Bailin:1998xx,Bailin:1997iz,Dent:2001cc,Dent:2001ut,Faraggi:2002vx,Witten:1985xc}.
   One may call this hard CP violation since this
 type of CP violations can exist even without soft terms.
Now Yukawa couplings which are formed via
 string compactification will carry this type of CP violation
 and the CKM phase $\delta_{CKM}$ which arises from the Yukawas
 is therefore  a probe of CP violation arising from string compactification
 (assuming there is  no CP violation arising from the Higgs  sector).
 A second source of  CP violation is via soft breaking.
 If SUSY contributions to K and B physics turn out to be small,
then one has a plausible bifurcation, i.e.,
the CP violations in K and B physics are probe of string compactification,
and baryogenesis and other CP phenomena that may be seen in sparticle
decays etc become a probe of   soft  breaking.

  Regarding soft breaking in string theory,  such an analysis would entail specifying the K\"ahler  potential, the superpotential,
  and the gauge kinetic energy function on the one hand and the mechanism of breaking on the other. Each of these are
  model dependent. However, it is possible to parameterize the breaking as in  gravity mediated breaking in
  supergravity. Thus one can write  the general form of the soft terms in the  form
\beqn
V_{\rm soft} =  m_{\alpha}^2 C_\alpha \bar C_{\bar \alpha}
               + A_{\alpha\beta\gamma} Y_{\alpha\beta\gamma} C_\alpha C_\beta C_\gamma\nonumber\\
               +\frac12 ( B_{\alpha\beta} \mu_{\alpha\beta} C_\alpha C_\beta + \, {\rm H.c.}\, )
               + \, \cdots,
\eeqn
where the general expressions for the  scalar masses $m_\alpha$,  trilinear couplings
$A_{\alpha\beta\gamma}$ and  the bilinear  term $B$  can be given. For the case when
$K_{\alpha\bar\beta} = \delta_{\alpha\bar\beta} K_\alpha$, one has  \cite{Kaplunovsky:1993rd,Brignole:1993dj}
\beqn
m_\alpha^2 = m^2_{3/2} + V_0 - F^I \bar F^{\bar J} \partial_I \partial_{\bar J} \ln ( {K}_\alpha) \ , \nonumber\\
A_{\alpha\beta\gamma} = c F^I \left( \partial_I {K} + \partial_I \ln (Y_{\alpha\beta\gamma})
  - \partial_I \ln (  {K}_\alpha  {K}_\beta  {K}_\gamma ) \right)  \ , \nonumber\\
B_{\alpha\beta} = c F^I \left( \partial_I {K} + \partial_I \ln(\mu_{\alpha\beta}) -
 \partial_I \ln ( {K}_{\alpha} {K}_{\beta} )  \right) + \ \cdots, \ .
\eeqn
while the gaugino masses are given by
\beqn
m_a = \frac{1}{2\Re(f_a)} F^I \partial_I f_a \ .
\eeqn
An efficient way to parameterize $F^I$ is given by
\cite{Brignole:1993dj} \beqn
F^S=\sqrt{3} m_{\frac{3}{2}} (S+S^*)\sin\theta e^{-i\gamma_S},\nonumber\\
F^i=\sqrt{3} m_{\frac{3}{2}} (T+T^*)\cos\theta \Theta_i
e^{-i\gamma_i}, \eeqn where $\theta$, $\Theta_i$ parameterize the
Goldstino direction in the
 S, $T_i$ field space and $\gamma_S$ and $\gamma_i$ are the $F^S$ and
$F^i$ phases, and $\Theta^2_1+\Theta^2_2+\Theta^2_3=1$.

 \subsection{Complex Yukawa couplings  in string compactifications}\label{gA}
 The Yukawa couplings arise at the point of string compactification, and it is interesting to
 ask how the Yukawa couplings   develop CP phases.  It is also interesting to determine if
 such phases are small or large. Consider, for example, the compactification of the $E_8\times E_8$
 heterotic   string
 on  a six dimensional Calabi-Yau (CY) manifold.  In this case the massless families  are
 either (1,1) or (2,1) harmonic forms. For the case when hodge number $h_{11}>h_{21}$,
 the massless mirror  families are (1,1) forms while if $h_{21}>h_{11}$ the massless families are
 (2,1) forms.  For the case when the families are (1,1) the cubic couplings among the
 families  have been discussed in \cite{Strominger:1985ks}.
  The analysis for the case when $h_{21}>h_{11}$  is more involved.
One specific model of interest that can lead to complex Yukawas corresponds to compactification
on the manifold $K_0'$\cite{Schimmrigk:1987ke,Gepner:1987qi}

\beqn
P^1\equiv  \sum_{i=0}^{3} z_i^3 +a_0 (z_1z_2z_3) =0\nonumber\\
P^2\equiv \sum_{i=0}^{3}  z_ix_i^3  =0
\eeqn
which
 is deformed from the manifold $K_0$ (corresponding to the case  $a_0=0$) in the ambient space
 $CP^3\times CP^2$ by a single (2,1) form $(z_1z_2z_3)$.  The $K_0$ has 35  $h_{21}$ forms and
 8 $h_{11}$ forms, giving an Euler characteric $\chi =2(h_{21}-h_{11})$ and the number of net  mass less
 families is $|\chi|/2$ \cite{Sotkov:1988xw}.

 By modding out by two discrete groups $Z_3$ and $Z_3'$ one gets a three generation
 model. The discrete symmetries  are $Z_3$ and $Z_3'$ where
 \beqn
 Z_3:~~~g:~~(z_0,z_1,z_2,z_3:x_1,x_2,x_3)\to \nonumber\\(z_0,z_2,z_3,z_1; x_2,x_3,x_3,x1), \nonumber\\
  Z_3':~~~h:~~(z_0,z_1,z_2,z_3:x_1,x_2,x_3)\to \nonumber\\ (z_0,z_1,z_2,z_3; x_1,\alpha x_2, \alpha^2 x_3).
 \eeqn
 where $\alpha^3=1, \alpha \neq 1$.  The group $Z_3'$ is not freely acting and leaves three tori invariant.
 These invariant tori have to be blown up in order to obtain a smooth CY manifold. Such a blowing up
 procedure produces six additional (2,1) and (1,1) forms which,  however, leave the net number
 of generations unchanged.
One considers now the flux breaking of $E_6$ on this manifold. If one embeds a single factor, $Z_3$ or $Z_3'$
in the $E_6$, then $E_6$  can break to $SU(3)^3$ or $SU(6)\times U(1)$ each of which leave the Standard Model
gauge group unbroken. However, the case $SU(6)\times U(1)$ cannot be easily broken further since an adjoint
representation does not arise in the massless spectrum. Thus typically one considers the $SU(3)^3$ possibility.
In this case there are  two possibilities : Case(A), where $Z_3$ is embedded trivially and $Z_3'$ is embedded
non-trivially, and case (B) where  $Z_3'$ is embedded trivially and $Z_3$ is embedded non-trivially.
Now for case (A) one may choose $U_g= (id)_C\times (id)_L\times (id)_R$,  $U_h= (id)_C\times \alpha (id)_L\times
\alpha (id)_R$,
where  $U_g$ is defined  so that
$g \to U_g$  is  a homomorphism of $Z_3$ into $E_6\ni U_g$\cite{Witten:1985xc},
and similarly for $U_h$, where (id) stands
for an identity matrix, and $C_L, R$ stand for color, left and right -handed subgroups of $SU(3)^3$.
The analysis of Yukawa couplings in this case has been carried out and the  couplings can be made
all  real. Thus in this case there is no CP  violation arising in the Yukawa sector at the compactification scale.

We consider next case (B) where essentially one has an interchange in the definitions of $U_g$ and $U_f$ so that
\beqn
  U_g= (id)_C\times \alpha (id)_L\times\alpha (id)_R,\nonumber\\
 U_h= (id)_C\times (id)_L\times (id)_R
\eeqn
In this case  the massless states that survive flux breaking of $E_6$ transform under $Z_3$ as follows

\beqn
Z_3 L=L,~ Z_3 Q=\alpha Q,~ Z_3Q^c =\alpha^2 Q^c\nonumber\\
Z_3 \bar L=\bar L,~ Z_3 \bar Q=\alpha^2 \bar Q,~ Z_3\bar Q^c =\alpha \bar Q^c\
\eeqn
where the leptons transform as  $L(1,3,\bar 3)$, quarks  as $Q(3,\bar 3, 1)$, and conjugate quarks as $Q^c(\bar 3,1,3)$.
 The barred quantities represent the mirrors,  so that $\bar L(1, \bar 3, 3)$, $\bar Q (\bar 3, 3,1)$, and $\bar Q^c(3,1,\bar 3)$.
 In this  model the number of generations and mirror generation are identical to that of the Tian-Yau
 model\cite{Greene:1986jb,Greene:1986bm}  so that
 there are  9 lepton generations and 6 mirror generations, 7 quark generations and 4 mirror  quark generations,  7 conjugate
 quark generations and 4 mirror conjugate quark generations, providing us with three net families of quarks and leptons.
 The analysis of Yukawa couplings has  been carried out on the manifold  $K_0$ by many author.\\

  Our focus here is
 the  $(27)^3$ couplings which  are unaffected by the instantons \cite{Distler:1988ms}
  and here one can use the techniques of \cite{Candelas:1987se} to determine
 the couplings.    An analysis for case (B) was carried out in \cite{Wu:1990ay}.
     The Yukawa couplings  determined in this fashion have unknown normalizations for the kinetic energy.
 However, symmetries can be used to obtain constraints on the normalizations. Including these normalization constraints
 into account it is found  that Yukawas  depend  on $\alpha$ in a non-trivial manner, and thus CP  is violated in
 an  intrinsic  manner. Further, the  CP phase entering in the coupling is large. 
 The CP violation on the $K_0'$ manifold
 persists even when the modulus $a_0$ is real, so in this sense CP violation is intrinsic.

\subsection{CP violation in orbifold models}\label{gB}
Next we discuss the possibility of spontaneous CP violation in some heterotic string models.
What we consider are  field point  limits of such models so we are essentially discussing
supergravity models with the added constraint of modular invariance (T duality).
The duality constraints have been utilized quite extensively in the analysis of
gaugino condensation and SUSY breaking
\cite{Ferrara:1990ei,Binetruy:1990ck,Font:1990nt,Cvetic:1991qm,Nilles:1990jv,Gaillard:2007jr}
and have also been utilized recently in the analysis of spontaneous breaking of
CP \cite{Dent:2001ut,Dent:2001cc,Giedt:2002ns,Acharya:1995ag,Bailin:1997fh}.\\

The scalar potential in supergravity and string theory is given by \cite{Chamseddine:1982jx,Cremmer:1982wb}
\beqn
V=e^{K}[(K^{-1})^i_j D_iW D_j^{\dagger}W^{\dagger}
 -3 WW^{\dagger}] +V_{D},
\eeqn
where $K$ is the K\"ahler potential, $W$ is superpotential  and
 $D_iW=W_i + K_iW$, with the subscripts denoting
derivatives with respect to 
the corresponding fields.
As noted above we now use the added constraint  of  $T$-duality symmetry.
Specifically we assume that the scalar potential in the effective four dimensional theory depends
on the  dilaton field $S$ and on the (K\"ahler) moduli fields
$T_i$ (i=1,2,3), and it is
invariant under the modular transformations (to keep matters  simple,
we do not include  here the dependence on the  so called complex structure $U$-moduli)
 \beqn T_i\rightarrow
T'_i=\frac{a_iT_i-ib_i}{ic_iT_i+d_i},
 (a_id_i-b_ic_i)=1,
\eeqn
where  $a_i,b_i,c_i,d_i \in Z$.
Under the modular transformations, $K$ and $W$ undergo a K\" ahler transformation
while the scalar potential $V$ is invariant.
For the K\"ahler  potential we assume   essentially a no scale form \cite{Lahanas:1986uc}
\beqn
 K = D(z) -\sum_ilog(T_i+\bar T_i) + K_{IJ} Q_I^{\dagger}Q_J +H_{IJ} Q_IQ_J,\nonumber
 \eeqn
where   $D(z)=-log(z)$, and for $z$ one may consider
\beqn
 z= (S+\bar S +\frac{1}{4\pi^2} \sum_i^3 \delta_i^{GS}
 log(T_i+\bar T_i)),
 \eeqn
 where  $\delta_i^{GS}$ is the one  loop correction to the K\"ahler
potential from the Greene-Schwarz mechanism,
and $Q$ are the matter fields  consisting of the quarks, the leptons
and the Higgs. For the superpotential in the visible sector  one may consider
\beqn
W_v= \tilde \mu_{IJ}
Q_IQ_J + \lambda_{IJK}Q_IQ_JQ_K.
\eeqn
 Under $T$-duality, $Q$'s transform as
 \beqn
Q_I\rightarrow Q_I\Pi_i (ic_iT_i+d_i)^{n^i_{Q_I}}.
\eeqn
In general, $K_{IJ}, H_{IJ}, \mu_{IJ}$ and $\lambda_{IJK}$ are
functions of the moduli. The constraints on $n^i_{Q_I}$ are such
that $V$ is modular invariant.
Analyses of soft  SUSY breaking terms using modular invariance of the type
above has been extensively discussed in the literature assuming moduli
stabilization. In such analyses one generically finds that CP is indeed violated
if one assumes that the moduli are in general complex.

 However, minimization of the potential and stabilization of the dilaton VEV
 is a generic problem in such models and requires additional improvements.
 Often this is accomplished  by
 non-perturbative corrections  to the potential.  Thus one might
 consider non-perturbative contributions to the superpotential so  that
 \beqn
 W_{np}= \Omega(\sigma) \eta(T)^{-6}.
 \eeqn
Here $\eta(T)$ is the Dedekind function, and
 we have assumed a single overall modulus $T$,
and  $\sigma=S+ 2 \tilde \delta^{GS} log \eta (T)$ and
  $\tilde \delta^{GS} = -(3/4\pi) \delta^{GS}$. Additionally  one can assume non-perturbative corrections
  to the K\"ahler  potential and treat $D(z)$ as a function to be determined by non-perturbative effects.
  The analysis shows that for a wide array of parameters minima typically occur at the self-dual points
  of the modular group, i.e., $T=1$ and $T=e^{i\pi/6}$.  However, for some choices
  of the parameters $T$ can take complex values away from the fixed point. Nonetheless CP phases
  arising from such points are  very small since in the soft parameters they come multiplied by
  the function $G(T,\bar T) =(T+\bar T)^{-1} +2dlog(\eta(T)/dT$ the imaginary part of which varies
  very  rapidly as the real part changes. Thus large CP phases do not appear to arise using the
  moduli stabilization of the type above\cite{Bailin:1997fh}.  \\

The situation changes  significantly if   $W_{np}$ contains an additional factor $H(T)$ where
\beqn
H(T)=   \left (\frac{G_6(T) }{\eta(T)^{12}}  \right )^m  \left (\frac{G_4(T) }{\eta(T)^{8}}  \right )^m P(j),
\eeqn
where $G_4(T)$ and  $G_6(T)$ are Eisenstein functions of modular weight 4 and 6, m, n are  positive
integers and  $P(j)$ is a polynomial of $j(T)$ which is an absolute modular invariant.  Alternately
 $H$ can be
expressed in the form
\beqn
H(T) =(j-1728)^{m/2} j^{n/2} P(j)
\eeqn
The form on $H(T)$ is
dictated by the  condition that no singularities  appear in the fundamental domain.  In this case
to achieve dilaton stabilization
with  $T$  modulus not only on the boundary of the fundamental domain
but also inside the fundamental domain and thus $T$ has  a substantial imaginary part.
 In this  case it is possible to get CP  phases for the soft parameters
which   can lie in the range $10^{-4}-10^{-1}$\cite{Bailin:1997fh}.
Thus with the absolute modular invariant in the
superpotential large  CP phases  can appear in the  soft  breaking in orbifold compactifications
of the type discussed above. \\

In the analysis of \cite{Faraggi:2002vx}
the issue of CP violation and FCNC in string models
with  anomalous $U(1)_A$-dilaton  supersymmetry breaking mechanism was investigated.
Here
scalar masses arise dominantly from the $U(1)_A$ contribution while the
dilaton generates the main contribution to the gaugino masses. Further,  the dilaton contributions
to the trilinear terms and to the gaugino masses have the same phase. In this class of models
the nonuniversal components of the trilinear soft SUSY breaking parameter are typically small
and one has suppression of FCNC and of  CP in this class of models.

\subsection{CP violation on D brane models}\label{gC}

Considerable progress has occurred over the recent past in the development of Type I and Type II string theory.
Specifically D branes have provided a new  and better understanding of Type I string theory
and connection with  Type IIB orientifolds.  Further, the advent of  D branes open up the possibility of a new
class of model building
(for recent reviews  on D branes see \cite{Blumenhagen:2005mu,Blumenhagen:2006ci,Polchinski:1996na}).
Thus a stack of N D branes  can produce generally an $SU(N)$ gauge group or
a subgroup of it,   and open strings with both ends terminating 
 on the same stack give rise  to a vector multiplet
corresponding to the gauge group of the stack.  Further,  open strings beginning on one end and  ending on
another transform like the bifundamental representations and can  be chiral. Thus these are possible
candidates for massless quarks, leptons, and Higgs fields.
  A simple possibility for  model    building occurs with  compactification on $T^6/Z_2\times Z_2$.
  In addition to the axion-dilaton field s  the   moduli space  consists in this case
  of the K\"ahler  ($t_m$) and  the complex structure ($u_m$) moduli  (m=1,2,3).  For the moduli fields one has the
  K\"ahler  potential of the form
  \beqn
  K_0 = -ln(s+\bar s) -\sum_{m=1}^3 ln(t_m+\bar t_m) -\nonumber\\
  \sum_{m=1}^3 ln(u_m+\bar u_m).
  \eeqn
  Consider now complex scalars $C^{[99]}_i$
  along the direction i with ends of the open string ending in each case  on a D9-brane.
In this case  one can obtain the K\"ahler  potential including the complex scalar field by  the
translation $t_m+\bar  t_m\to t_m+\bar t_m -|c^{[99]}_m|^2$.   For the case of strings  with both ending
on the same $D5_i$ brane one can show using either T-duality\cite{Ibanez:1998rf}  or by use of
Born-Infeld action\cite{Kors:2003wf,kors}
 that
the K\"ahler  potential  is modified by making the replacement $s+\bar s\to s+\bar s -|C_m^{[5_m,5_m]}|^2$.
For the case when one has both D9- branes and $D5_m$-branes  the  modified K\"ahler  potential reads

\beqn
K^{[99+55]} =
-\ln\Big( s+ \bar s - \sum_{m=1}^3 |C_m^{[5_m5_m]}|^2 \Big)-
\nonumber\\
-\sum_{m=1}^3 \ln\Big( t_m + \bar t_{\bar m} - |C_m^{[99]}|^2 - \frac12 \sum_{n,p=1}^3
 \gamma_{mnp} |C_n^{[5_p5_p]}|^2 \Big).~
\eeqn
To construct the K\"ahler  potential for the case when one has open strings with one  end on D9-branes
and the other end on $D5_m-$ branes, or for  the case when open strings end on two different $D5$ branes,
 one can use  the analogy to heterotic strings with
$\mathbb Z_2$-twisted matter fields \cite{Ibanez:1998rf,Kors:2003wf}.
Alternately one can use
string perturbation theory \cite{Lust:2004cx,Lust:2004fi,Bertolini:2005qh}.
 The result is
\beqn \label{n2}
K^{[95]} = \frac12 \sum_{m,n,p=1}^3 \gamma_{mnp}
\frac{|C^{[95_m]}|^2}{(t_n+\bar t_{\bar n})^{1/2}(t_p+\bar t_{\bar p})^{1/2}} \nonumber\\
 +
\frac12 \sum_{m,n,p=1}^3 \gamma_{mnp}
\frac{|C^{[5_m5_n]}|^2}{(t_p+\bar t_{\bar p})^{1/2}(s+\bar s)^{1/2}} \ .
\eeqn
Explicit formulae for the soft parameters using these  results are given in the literature.
However, one needs to keep in mind the configurations of the type discussed above
are  the so called $\frac{1}{2} BPS$ states, and in this case the spectrum of open states
falls  into $N=2$ multiplets, which implies that the spectrum is not chiral.   Similar considerations
apply to open strings which start and end on $D_3$ and $D_7$ branes, and results for
these can be obtained by using T dualities.

For realistic model building one needs to work with intersecting  D branes.
Thus in Calabi-Yau orientifolds of Type IIA one has  D6-branes that intersect
on the compactified  6 dimensional manifold.  Sometimes it is convenient to work
in the T-dual picture of Type IIB strings where the geometrical picture of branes
intersecting is replaced by internal world volume gauge field backgrounds, called
fluxes on the D9 and D5 branes.  The fluxes ${\cal F}_a^m$ where a labels the set  of
branes, are rational numbers, i.e.,
${\cal F}_a^m$ = ${m}_a^m$ $/ $  ${n}_a^m$, in order to satisfy charge quantization constrains.
The fluxes determine the number of chiral  families.  Further, the condition that $N=1$ supersymmetry
be valid is  a further constraint on the moduli and the fluxes and may be expressed
in the form \cite{Kors:2003wf,Bachas:1995ik,Berkooz:1996km}
\beqn
\sum_{m=1}^3 \frac{s+\bar s}{ t_m+\bar t_m} {\cal F}_a^m  =
=\prod_{m=1}^3
{\cal F}_a^m
\eeqn
In the presence of fluxes the gauge kinetic energy function $f_a$ is given by
\beqn \label{gcfu}
f_a = \prod_{m=1}^3 {\bf n}_a^{(m)}
 \left( s  - \frac12 \sum_{m,n,p=1}^3 \gamma_{mnp} {\bf F}_a^{(n)} {\bf F}_a^{(p)} t_m  \right) \ .
\eeqn
The computation of the  K\"ahler metric for the case of  an open string  with both ending  on some given stack $a$, $C_m^{[aa]}$,
can be computed by dimensional reduction\cite{Kors:2003wf} or string perturbation
theory\cite{Lust:2004cx} and is given by
\beqn
K^{[aa]} &=& \sum_{m=1}^3
\frac{|C_m^{[aa]}|^2}{(s+ \bar s)(t_m + \bar t_{\bar m})(u_m + \bar u_{\bar m})} \frac{4\Re (f_a)}{1 + \Delta_a^{(m)}}\ ,
\nonumber\\
\Delta_a^{(m)} &=& \frac12 \sum_{n,p=1}^3 \gamma_{mnp}
\frac{ ( t_n +\bar t_{\bar n} )( t_p +\bar t_{\bar p} )}
  {(s+\bar s)( t_m +\bar t_{\bar m} )} \left( {\bf F}_a^{(m)} \right)^2 \ .
\eeqn
Now  the technique above using the heterotic dual or Born-Infeld works  for $\frac{1}{2}$BPS brane
configurations.   However, for  the bifundamental fields $C^{[ab]}$
that connect the different stacks of branes with different world volume gauge flux one needs  an actual
string perturbation calculation  and here the result for the K\"ahler  potential is \cite{Lust:2004cx}
\beqn
K^{[ab]} = \frac{|C^{[ab]}|^2}{\prod_{m=1}^3 (u_m + \bar u_{\bar m})^{\theta_{ab}^{(m)}}}
 \frac{\Gamma(\theta_{ab}^{(m)})^{1/2}}{\Gamma(1-\theta_{ab}^{(m)})^{1/2} }\ ,\nonumber\\
\theta_{ab}^{(m)} = {\rm arctan}\Big( \frac{ {\bf F}_a^{(m)} }{\Re(t_m)} \Big) \ .
\eeqn
Using the above  one can obtain explicit expressions for the soft parameters.  These  have
been worked out in detail in  several papers. One can count the number of CP phases
that enter  in the analysis. They are  the phases arising from $s, t_m, u_m$ (m=1,2,3).
These can be reduced with extra  restrictions such as, for example,  dilation dominance
 which would imply only one  CP phase $\gamma_s$.

\subsection{SUSY CP phases and the CKM matrix}\label{gD}
 A natural question is if  there is a connection between the soft  SUSY CP phases
 and the CKM phase $\delta_{CKM}$.   A priori it would  appear that there is no connection
 between these two since they arise from two very different sources.
 Thus the $\delta_{CKM}$  arises from the Yukawa interactions
 (assuming there is no CP violation in the Higgs  sector)
 which
  from the string view  point originates at the point when the  string compactifies from
 10 dimensions to four dimensions. This is the point where we begin to identify
 various  species of quarks and leptons and their couplings to the Higgs bosons.
 On the other hand soft SUSY phases  arise from the spontaneous breaking
 of supersymmetry and enter only in the dimension $\leq 3$ operators.
 Thus it  would appear that  they are  disconnected.
 While this conclusion is largely true it is not  entirely so. The reason for this is
 that in SUGRA models
the trilinear soft term $A_{\alpha\beta\gamma}$ contains a dependence
 on Yukawas so that\cite{Kaplunovsky:1993rd,Nath:1983aw}
\beq
 A_{\alpha\beta\gamma} =F^i\partial_i Y_{\alpha\beta\gamma} +..
 \eeq
 Thus the phase  of the Yukawa couplings  enters in the  phase  of
 the  trilinear  coupling.  However, the phase  relationship between
 $A$ and  $Y$ is not rigid, since even for  the case  when there is no phase in the Yukawas one
 can generate  a  phase of $A$, and  conversely even for the case
 when $\delta_{CKM}$ is maximal one  may constrain $A$ to have
 zero phase.
Further, it is entirely possible that the Yukawa couplings are  all real and $\delta_{CKM}$ arises
from CP violation in the Higgs  sector as  originally
conjectured \cite{Lee:1973iz,Lee:1974jb,Weinberg:1976hu}. A more recent analysis of this
possibility is given in  \cite{Chen:2007nx}.

On a more theoretical level it was initially thought that CP violation could occur in string theory
in either of the two ways: spontaneously or explicitly \cite {Strominger:1985it}.
 However, it  was conjectured later that CP symmetry in string
theory is a gauge theory and it is  not violated explicitly\cite{Choi:1992xp,Dine:1992ya}.
We do not address this issue further here.

\section{The  EDM of  an elementary  Dirac fermion}\label{h}

If the spin-1/2 particle has electric dipole moment EDM $d_f$, it would interact with the electromagnetic tensor $F_{\mu \nu}$ through
\beqn
   {\cal L} = -\frac{i}{2} d_f \bar{\psi} \sigma_{\mu \nu} \gamma_5 \psi F^{\mu \nu}
   \eeqn
which in the non-relativistic limit reads \beqn {\cal L} = d_f
\psi^{\dagger}_A \vec{\sigma}.\vec{E} \psi_A \eeqn where $\psi_A$ is
the large component of   the Dirac field. The above Lagrangian is
not renormalizable, so it does not exist at the tree level of a
renormalizable quantum field theory. However, it could be induced at
the loop level if this theory contains sources of CP violation at
the tree level. Thus suppose we wish  to determine the EDM of a
particle with the  field $\psi_f$ due to the exchange of  two other
heavy fields:  a spinor $\psi_i$ and a scalar $\phi_k$. The
interaction that contains CP violation is given by \beqn {\cal L} =
L_{ik} \bar{\psi_f} P_L \psi_i \phi_k +R_{ik} \bar{\psi_f} P_R
\psi_i \phi_k +H.c. \eeqn Here ${\cal L}$ violates CP invariance iff
$Im(L_{ik} R^*_{ik})\neq 0$. A direct analysis shows that the
fermion $\psi_f$  acquires a one loop EDM $d_f$ which is given by \beqn d_f =
\frac{m_i}{16 \pi^2 m^2_k} Im(L_{ik} R^*_{ik}) (Q_i
A(\frac{m^2_i}{m^2_k})+Q_k B(\frac{m^2_i}{m^2_k})), \eeqn where
\beqn
A(r) = \frac{1}{2(1-r)^2} (3-r+\frac{2 ln r}{1-r})\nonumber\\
B(r) = \frac{1}{2(1-r)^2} (1+r+\frac{2r ln r}{1-r}).
\label{ABr}
\eeqn
We will utilize this result in  EDM analyses in the following discussion.

 \begin{figure}
  \vspace{-1cm}
   \hspace{-1.0cm}
\includegraphics*[width=9cm, height=12cm]{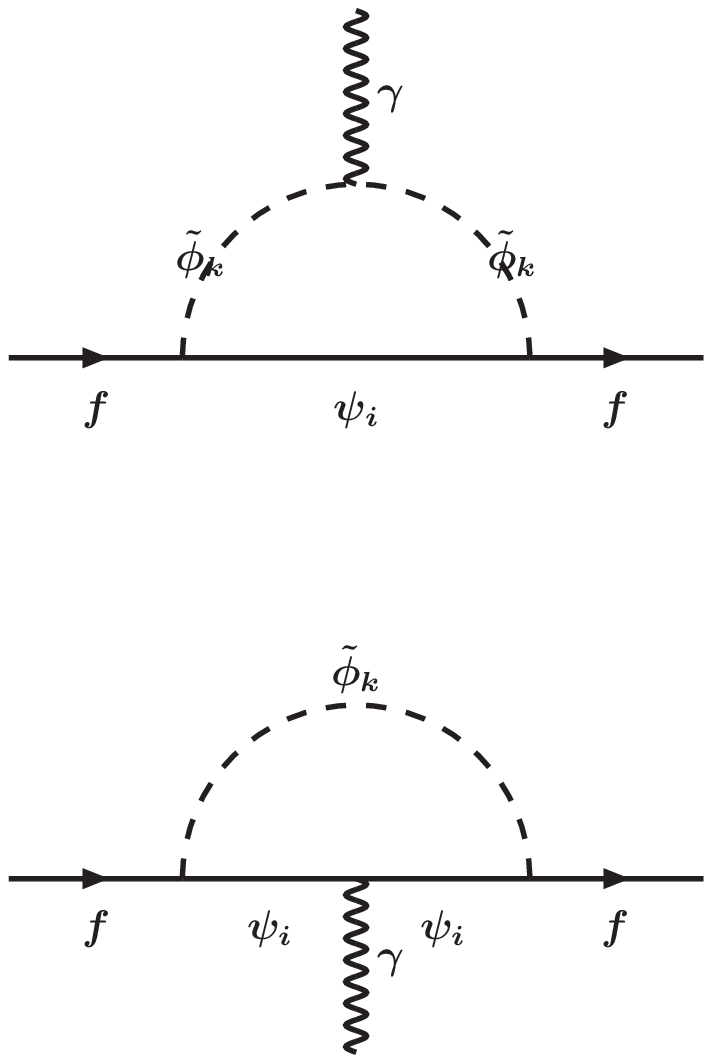}
\vspace{-.5cm}\caption{Contributions to the electric dipole moment of
a lepton or of a quark from the exchange of the charginos, the
neutralinos and the gluino.
The internal dashed line in the loop is the scalar
field $\phi_k$,  the solid line is the fermion
field $\psi_i$ and the external wiggly line is the  external photon line. 
} \label{EDMdipole}
\end{figure}

\section{EDM of a charged lepton  in SUSY}\label{i}

We discuss now the EDM of a charged lepton in MSSM using the results of
the previous section.
As mentioned in Sec.IV, in softly broken supersymmetric models as many as 40 additional phases
can appear.
However, only certain combinations of phases appear  in a  given process  and the number of
such combinations depends on the process.  We discuss now the details.

In the computations here we use the  Lagrangian of applied $N=1$ supergravity for the
case of MSSM fields with inclusion of soft breaking\cite{Haber:1984rc,Nath:1983fp,Nilles:1983ge}.
The EDM of a charged lepton receives contributions from chargino, neutralino, and slepton 
exchanges.
 A discussion of the chargino and neutralino masses is given in Sec.(\ref{qA}) while 
 a discussion of the slepton and squark masses is given in Sec.(\ref{qB}). 
 For the case of the charged lepton we find
\beq
{d_{e-chargino}^{E}}/{e}=\frac{\AEM}{4\pi\sinW2}
    {m_{\tilde{\nu}e}^2} \sum_{i=1}^{2} \mci {\rm Im}
     (\Gamma_{ei})
    {\rm A}(\frac{\mci^2}{m_{\tilde{\nu}e}^2})
\eeq
where $U$ and $V$ are as defined in Appendix \ref{qA} and where
$\Gamma_{ei}=
 (\kappa_e U_{i2}^* V_{i1}^*)=|\kappa_e| U_{R2i}^*U_{L1i}$.
A direct inspection of $\Gamma_{ei}$ shows that it depends on only
one combination,
i.e.,
$\xi_2+\theta_{\mu}+\theta_H$ where the phase
$\theta_H$ comes from the Higgs sector and as discussed later 
is generated at the loop level.
\\

  The neutralino exchange contribution to the EDM of the fermion is  as follows:
\beqn
{d_{f-neutralino}^E}/{e}=\frac{\AEM}{4\pi\sinW2}\sum_{k=1}^{2}\sum_{i=1}^{4}
{\rm Im}(\eta_{fik})
               \frac{\mxi}{M_{\tilde{f}k}^2}\nonumber\\
     \times    Q_{\tilde{f}}
{\rm B}(\frac{\mxi^2}{M_{\tilde{f}k}^2})
\label{EDMneutralino}
\eeqn
where
\beqn
\eta_{fik} = (a_0 X_{1i} D_{f1k}^*
  + b_0 X_{2i}D_{f1k}^*+
     \kappa_{f} X_{bi} D_{f2k}^*)\nonumber\\
     ( c_0 X_{1i} D_{f2k}
     -\kappa_{f} X_{bi} D_{f1k})
\eeqn
where
$a_0=-\r2 \tan\theta_W (Q_f-T_{3f})$, $b_0=-\r2 T_{3f}$,
$c_0=\r2 \tan\theta_W Q_f$, and in $X_{bi}$
b=3(4) for $T_{3q}=-\frac{1}{2}(\frac{1}{2})$.
The following three combinations of phases appear in $\eta_{fik}$:
$\xi_1+\theta_{\mu}+\theta_H$, $\xi_2+\theta_{\mu}+\theta_H$
and
 $\alpha_f+\theta_{\mu}+\theta_H$.
 We note in passing that
the contribution from the neutrino Yukawa couplings to the lepton electric dipole moment
is computed in \cite{Farzan:2004qu}, and the
  charged Higgs contributions to the lepton EDM in a two-Higgs doublet model
is discussed in   \cite{Kao:1992jv}.

\section{EDM of quarks in SUSY}\label{j}
The quarks receive contribution from the electric dipole operator  ($d_q^E$),
from the chromoelectric dipole operator ($d_q^C$),  and from the purely
gluonic dimension six operator of Weinberg ($d_q^G$).  Thus
\beqn
d_q=d_q^E +d_q^C+ d_q^G
\eeqn
We discuss  these in further detail below.

\subsection{The electric dipole moment operator contribution to EDM
of quarks}\label{jA}
The electric dipole moment operator receives contributions from the gluino,
chargino and neutralino exchanges.  The gluino exchange contributes
to the EDM of the quarks as follows
\beqn
{d_{q-gluino}}/{e}=\frac{-2 \alpha_{s}}{3 \pi}  m_{\tilde{g}}Q_{\tilde{q}}
{\rm Im}(\Gamma_{q}^{11}) \nonumber\\
 \times \left(\frac{1}{M_{\tilde{q}1}^2}
 {\rm B}(\frac{m_{\tilde{g}}^2}{M_{\tilde{q}1}^2}) -\frac{1}{M_{\tilde{q}2}^2}
{\rm B}(\frac{m_{\tilde{g}}^2}{M_{\tilde{q}2}^2}) \right),
\eeqn
where  $\tilde{q}_1$ and  $\tilde{q}_2$ are the mass eigenstates,  and
 $\Gamma_{q}^{1k}=e^{-i\xi_3} D_{q2k} D_{q1k}^*$,
  $\alpha_s$=${g_{s}^2}\over {4\pi}$, $m_{\tilde{g}}$ is the gluino mass,
   and  $B(r)$ is as defined by Eq.(\ref{ABr}).
    An explicit
   analysis gives  $\Gamma_{q}^{12}=-\Gamma_{q}^{11}$ where

\beqn
{\rm Im}(\Gamma_{q}^{11})=\frac{m_q}{M_{\tilde{q}1}^2-M_{\tilde{q}2}^2}
        (m_0 |A_q| \sin (\alpha_q -\xi_3)+\nonumber\\
         |\mu| \sin
        (\theta_{\mu}+\theta_H+\xi_3) |R_q|),
\eeqn
which holds for both signs of
 $M_{\tilde{q}1}^2-M_{\tilde{q}2}^2$.
 It is easy to see that the
 combinations of phases that enter  are ($\alpha_q$-$\xi_3)$ and
 $\xi_3+\theta_{\mu}+\theta_H$, or alternately one can choose
 them to be $\alpha_q+\theta_{\mu}+\theta_H$ and
 $\xi_3+\theta_{\mu}+\theta_H$.

The  chargino contribution to the EDM for the up quark is  as follows
\beqn
{d_{u-chargino}}/{e}=\frac{-\AEM}{4\pi\sinW2}\sum_{k=1}^{2}\sum_{i=1}^{2}
      {\rm Im}(\Gamma_{uik})
               \frac{\mci}{M_{\tilde{d}k}^2}\times\nonumber\\
                ~[Q_{\tilde{d}}
                {\rm B}(\frac{\mci^2}{M_{\tilde{d}k}^2})+
    (Q_u-Q_{\tilde{d}}) {\rm A}(\frac{\mci^2}{M_{\tilde{d}k}^2})].
\eeqn
Here  $A(r)$ is as defined by Eq.(\ref{ABr}) and
\beq
\Gamma_{uik}=\kappa_u V_{i2}^* D_{d1k} (U_{i1}^* D_{d1k}^*-
        \kappa_d U_{i2}^* D_{d2k}^*)
\eeq
and
\beqn
\kappa_u=\frac{m_ue^{-i\theta_H}}{\r2 m_W \sb},
 ~~\kappa_{d,e}=\frac{m_{d,e}}{\r2 m_W \cb}
\eeqn
and explicitly
\beqn
\Gamma_{ui1(2)}=|\kappa_u| (cos^2 \theta_d /2) [U_{L2i} U^*_{R1i}]\nonumber\\
-(+) \frac{1}{2} |\kappa_u \kappa_d| (sin \theta_d)
[U_{L2i}U^*_{R2i}] e^{i\{\xi_2 -\phi_d\}}
\eeqn
The EDM here depends only on two combinations of phases:
 $\alpha_d+\theta_{\mu}+\theta_H$ and
$\xi_2+\theta_{\mu}+\theta_H$ with $\xi_2-\alpha_d$ being just a
linear combination of the first two.
A similar analyses hold for the chargino
contributions to the down quark and one gets only two phase combinations
which are identical to the case above with $\alpha_d$ replaced by
$\alpha_u$.
The neutralino exchange contribution to the EDM of quarks  is given by
Eq.(\ref{EDMneutralino}).
The sum of the gluino, the chargino and the neutralino exchanges
discussed above gives the total contribution from the electric dipole
operator to the quark EDM.

\subsection{The chromoelectric dipole moment contribution to the EDM  of
quarks}\label{jB}
For the case of the quarks one has two more operators that contribute. These are the
 quark chromoelectric dipole moment ($\tilde d^C$)  and the purely gluonic dimension six operator.
 For the operator $\tilde d^C$ we  have  the effective dimension five operator
\begin{equation}
{\cal L}_I=-\frac{i}{2}\tilde d^C \bar{q} \sigma_{\mu\nu} \gamma_5 T^{a} q
 G^{\mu\nu a},
 \end{equation}
 where $T^a$ are the $SU(3)$ generators.
   Contributions to $\tilde d^C$
of the quarks from the gluino,
 the chargino and from the neutralino exchange are given by
\beq
\tilde d_{q-gluino}^C=\frac{g_s\alpha_s}{4\pi} \sum_{k=1}^{2}
     {\rm Im}(\Gamma_{q}^{1k}) \frac{m_{\tilde{g}}}{M_{\tilde{q}_k}^2}
      {\rm C}(\frac{m_{\tilde{g}}^2}{M_{\tilde{q}_k}^2}),
\eeq

\beq
\tilde d_{q-chargino}^C=\frac{-g^2 g_s}{16\pi^2}\sum_{k=1}^{2}\sum_{i=1}^{2}
      {\rm Im}(\Gamma_{qik})
               \frac{\mci}{M_{\tilde{q}k}^2}
                {\rm B}(\frac{\mci^2}{M_{\tilde{q}k}^2}),
\eeq
and
\beq
\tilde d_{q-neutralino}^C=\frac{g_s g^2}{16\pi^2}\sum_{k=1}^{2}\sum_{i=1}^{4}
{\rm Im}(\eta_{qik})
               \frac{\mxi}{M_{\tilde{q}k}^2}
                {\rm B}(\frac{\mxi^2}{M_{\tilde{q}k}^2}),
\eeq
where B(r) is defined by Eq.(\ref{ABr}) and C(r) is given by
\beq
C(r)=\frac{1}{6(r-1)^2}(10r-26+\frac{2rlnr}{1-r}-\frac{18lnr}{1-r}).
\label{Cr}
\eeq
We note that
all of the CP violating phases are contained in the factors
${\rm Im}(\Gamma_{q}^{1k})$, ${\rm Im}(\Gamma_{qik})$, and
 ${\rm Im}(\eta_{qik})$. But these are precisely the same factors
that appear in the gluino, the chargino and the neutralino contributions to
the electric dipole operator.

  \begin{figure}
  \vspace{0.0cm}
 \hspace{-1.0cm}
\includegraphics*[width=9cm, height=9cm]{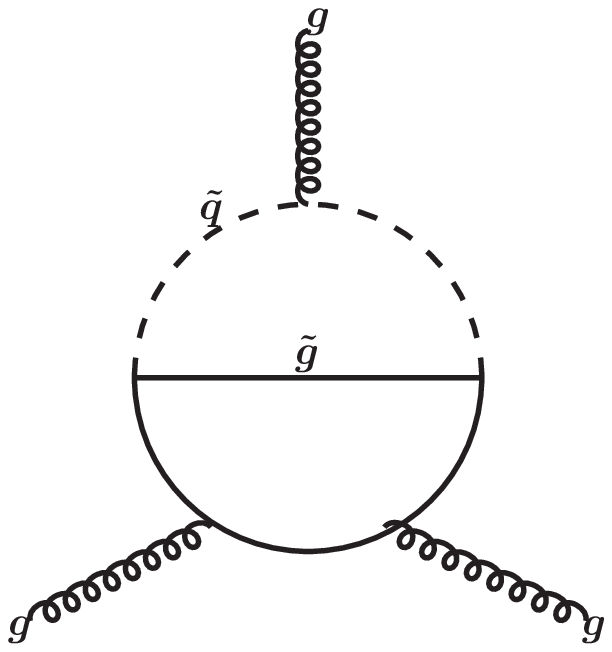}
\vspace{0.0cm}\caption{The quark-squark-gluino exchange contribution to the purely
gluonic dimension six operator.
The  dashed line in upper semi circle in the loop is the squark $\tilde{q}$,
  the internal horizontal solid line is the gluino  $\tilde g$, the solid line on the
  lower semicircle in the loop is the quark $q$, while the external wiggly lines
  are the gluons. 
}
\label{dim6EDM}
\end{figure}

 \subsection{The contribution of the purely gluonic operator to the EDM of
 quarks}\label{jC}
    The purely  gluonic dimension six operator
 which contributes to the dipole moment is  \cite{Weinberg:1989dx}
 \begin{equation}
{\cal L}_I=-\frac{1}{6}\tilde d^G f_{\alpha\beta\gamma}
G_{\alpha\mu\rho}G_{\beta\nu}^{\rho}G_{\gamma\lambda\sigma}
\epsilon^{\mu\nu\lambda\sigma},
\end{equation}
where $G_{\alpha\mu\nu}$ is the
 gluon field strength tensor, $f_{\alpha\beta\gamma}$
 are the Gell-Mann coefficients, and $\epsilon^{\mu\nu\lambda\sigma}$
is the totally antisymmetric tensor with $\epsilon^{0123}=+1$. An
analysis of $d^G$ including the quark-squark-gluino exchange (see
Fig.(\ref{dim6EDM})  where one of the loops contributing to this operator is shown) with gluino phase  $\xi_3$ but with squark
$mass^2$ matrix treated real is given in \cite{Dai:1990xh}.
Including the phases  from $A_t$ and $\mu$ in the squark mass$^2$
matrix the analysis  of $\tilde d^G$ gives \cite{Dai:1990xh,Ibrahim:1997nc} 
\beqn
\tilde d^G=-3\alpha_s(\frac{g_s}{4\pi m_{\tilde g}})^3
(m_t(z_1^t-z_2^t)Im(\Gamma^{12}_t)H(z_1^t,z_2^t,z_t)\nonumber\\
+m_b(z_1^b-z_2^b)Im(\Gamma^{12}_b)H(z_1^b,z_2^b,z_b)).\nonumber\\
\eeqn
\noindent
Here
\beqn
\Gamma_q^{1k}=e^{-i\xi_3}D_{q2k}D_{q1k}^*,
z^q_{\alpha}=(\frac{M_{\tilde{q}\alpha}}{m_{\tilde{g}}})^2,
z_q=(\frac{m_q}{m_{\tilde{g}}})^2,\nonumber\\
\eeqn
and $H(z_1,z_2,z_3)$ is defined by
\beqn
H(z_1,z_2,z_3) = \frac{1}{2} \int_0^1 dx \int_0^1 du \int_0^1 dy x(1-x)u \frac{N_1N_2}{D^4},
\eeqn
where
\beqn
N_1= u(1-x) +z_3 x(1-x) (1-u) -\nonumber\\
-2 ux [z_1 y +z_2 (1-y)],\nonumber\\
N_2= (1-x)^2 (1-u)^2 +u^2 -\frac{1}{9} x^2 (1-u)^2, \nonumber\\
D=u(1-x) + z_3 x(1-x) (1-u) +ux[z_1 y +z_2 (1-y)].~~
\eeqn

For the case when $m_{\tilde q}, m_{\tilde g} >> m_q$ one obtains for H the following expression
\beqn
H \simeq -\frac{m_{\tilde g}^2}{m_{q}^2} I(z^q_2),
\eeqn
where $I(z)$ is defined by
\beqn
I(z)= \frac{1}{6(z-1)^2} [2(z-1) (11z -1) +\nonumber\\
+ (1-16z -9z^2)ln z]. \eeqn The contribution  of the last two
operators  to the EDM of the quarks can be computed using the naive
dimensional analysis\cite{Manohar:1983md}. This technique can be
expressed in terms of a rule using the  `reduced' coupling
constants.  Thus for example, for a coupling constant g appearing in
an interaction of dimensionality (mass)$^D$ and  containing N field
operators the reduced coupling is $(4\pi)^{2-N} M^{D-4} g$ where
$M$ is the chiral-symmetry breaking scale and has the value $M=1.19$
GeV. Thus the rule means that the reduced coupling of any term in
the effective  hadronic theory at energies below  $M$ is given by a
product of the reduced couplings of the operators appearing in the
effective  Lagrangian at energies below M, that produces this term.
Using this rule for the chromoelectric  and  purely gluonic
dimension six operators one finds  there contribution to the EDM of
the quarks  are given as follows

\beqn
d^E_q= d_q \eta^E, ~~
d^C_q=\frac{e}{4\pi}\tilde{d^C_q} \eta^C,~~
d^G_q=\frac{eM}{4\pi}\tilde{d^G_q} \eta^G, \eeqn where $\eta^E$,
 $\eta^C$ and
$\eta^G$ are renormalization group evolution of $d_q$, 
$\tilde{d^C_q}$ and $\tilde{d^G_q}$ from the electroweak scale to
the hadronic scale. A discussion of how these renormalization group
factors  are computed is discussed in Sec.(\ref{qC}). Their
numerical value is estimated to be 
$\eta^E\approx 0.61$ \cite{Degrassi:2005zd}, $\eta^C\approx \eta^G\sim 3.4$.
The alternate technique to estimate the contributions of the
chromoelectric operator is to use the QCD sum
rules\cite{Khriplovich:1996gk}. To obtain the neutron EDM, we use
the non-relativistic $SU(6)$ quark model which gives $
d_n=\frac{4}{3} d_d- \frac{1}{3} d_u$.

\subsection{The cancelation mechanism  and other remedies for the  CP problem
in   SUSY, in strings and in branes}\label{jD}

Thus MSSM contains new sources of CP violation and these phases
would induce EDMs of the fermions in the theory. Taking the values
of the parameters of the model at their phenomenologically favorable
range $(m_{1/2}\sim m_0\sim 100 GeV, \tan\beta \sim 10, \theta_{\mu}
\sim \alpha_0\sim 1)$ one finds that the EDMs of the electron and
neutron exceed the experimental bounds by several orders of
magnitude.  This problem is certainly a weakness of the low energy
SUSY and needs to be  corrected to make the theory viable. 
Various remedies have been suggested in the literature to
overcome this problem. The first of these is the suggestion that the
first generation of sleptons and the first two generations of
squarks are very heavy  \cite{Nath:1991dn} (see also
\cite{Kizukuri:1992nj}). This means the production and study of
these particles at LHC will be difficult if not impossible.  There
is another reason that this possibility is not attractive is that,
the annihilation rate of the lightest supersymmetric particle LSP
may be too low in this range of masses and as a result the relic
density of the LSP may be larger than the observed dark matter
density. Another suggestion is that the phases are small
$O(10^{-2})$
  \cite{Dugan:1984qf,
   Ellis:1982tk,Polchinski:1983zd,Franco:1983xm,Garisto:1996dj,Weinberg:1989dx,Braaten:1990gq,Braaten:1990zt,Dai:1990xh,arnowitt:1990eh,Gunion:1990iv,Arnowitt:1990je}.
 However, a  small phase constitute a fine tuning and there will not be any interesting display of CP violation in colliders. Moreover, electroweak baryogenesis cannot take place in this case\cite{Kuzmin:1985mm}.
   A third possibility first proposed in \cite{Ibrahim:1997nc,Ibrahim:1997gj,Ibrahim:1998je}
    is that there are internal cancelations among the various contributions to the neutron EDM
    and to the electron EDM, leading to compatibility with experiment with large phases and a SUSY spectrum that is still within the reach of the accelerators.

    This is the most interesting solution because it leaves room for a host of non trivial CP violating as well as CP conserving phenomena to be discovered at colliders and elsewhere.
By CP violating properties we mean those properties that vanish in
the limit of CP conservation like the  EDMs and the neutral Higgs
bosons mixing. By CP conserving phenomena, we mean those properties
that exist in the absence of CP violation but they differ if CP
violation is included like $g_{\mu}-2$.  Following the work of
\cite{Ibrahim:1997nc,Ibrahim:1997gj,Ibrahim:1998je} there is  much
further work on the cancelation mechanism in the
 literature\cite{Falk:1998pu,Falk:1999tm,Brhlik:1999qr,Brhlik:1998zn,Brhlik:1999ub,Brhlik:1999pw, Brhlik:1999hs,Pokorski:1999hz,Accomando:1999zf,  Accomando:1999uj,Abel:2001cv,Bartl:1999bc,Bartl:2001wc,Chattopadhyay:2000fj,Ibrahim:1998je,Ibrahim:1999af,Barger:2001nu}.

 As was shown above, the quark and the lepton EDMs in general depend on ten independent phases providing one with considerable freedom for the satisfaction of the EDM constraints. Numerical analyses show the existence of significant regions of the parameter space where the cancelation mechanism holds.
 We describe here a straightforward technique for accomplishing the satisfaction of the electron EDM and the
 neutron EDM constraints. For the case of the electron one finds that the chargino component of the electron is independent of $\xi_1$ and the electron EDM as a whole is independent of $\xi_3$. Thus the algorithm to discover a point of simultaneous cancelation for the electron EDM and for the neutron EDM is a straightforward one. For a given set of parameters we vary $\xi_1$ until we reach the cancelation for the electron EDM since only one of its components (the neutralino) is affected by that parameter. Once the electric dipole moment constraint on the electron is satisfied we vary $\xi_3$ which affects only the neutron EDM keeping all other parameters fixed. By using this simple algorithm one can generate any number of simultaneous cancelations.  The EDM of the atoms also provide a sensitive test of CP violation.
An example is the EDM of  Hg-199 for which the  current limits are given by Eq.(\ref{EDMhg}).
Among the phases that enter  the EDM of
 Hg-199  is  the phase  $\alpha_s$.  We note that $\alpha_s$ enters only in $d_{H_g}$ to one loop
 order,  and thus it can be varied to achieve  a simultaneous cancelation in  $d_{H_g}$ and a
 consistency with the experimental limits.
 Illustrative examples of points in the parameter space where cancelations occur and
 all the EDM constraints are satisfied are given in Tables 1 and 2 in  Sec.(\ref{qD}).
 It needs to be emphasized that while the cancellations among the various
contributions to the EDMs  are pretty 
generic the suppression of the EDMs for the electron and for the neutron
do require fine tuning.  On the positive side the above, of course, leads to a narrowing of 
the parameter
space of the theory. 
  \begin{figure}
 \hspace{-0.25cm}
\includegraphics*[width=9cm, height=6cm]{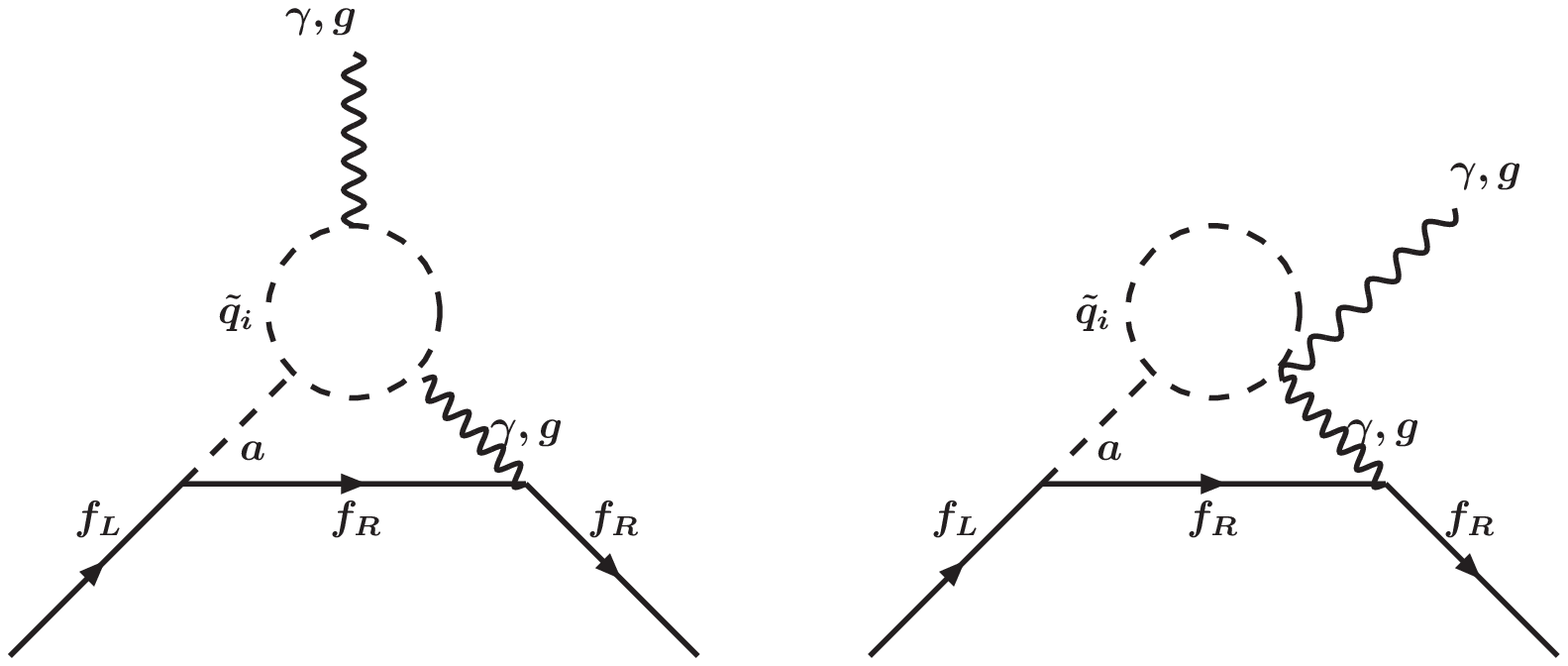}
\vspace{-0.5cm}\caption{Two loop Barr-Zee type diagrams that contribute to the EDMs
in supersymmetry \cite{Chang:1998uc}.}
\label{loop2}
\end{figure}

In theories where the Higgs mixing parameter $\mu$ obeys the simple
scaling behavior as the rest of the SUSY masses the EDMs exhibit a
simple scaling behavior under the simultaneous scaling on $m_0$ and
$m_{1/2}$. In the scaling region the  knowledge of a single point in
the MSSM parameter space where the cancelation in the EDMs occurs
allows one to generate a trajectory in the $m_0-m_{1/2}$ plane where
the cancelation mechanism holds and the EDMs are small. Thus under
the transformation $m_0\rightarrow \lambda m_0$, $m_{1/2}\rightarrow
\lambda m_{1/2}$
 $\mu$ itself obeys the same scaling, i.e., $\mu \rightarrow \lambda \mu$ in the large $\mu$ region.
 In this case $d_e$ exhibits the scaling behavior
\beqn d_e\rightarrow \lambda^{-2} d_e \eeqn The same scaling
relation holds for the electric and for the  chromoelectric
operators of the quarks \beqn d^E_q\rightarrow  \lambda^{-2} d^E_q,
~~ d^C_q\rightarrow  \lambda^{-2} d^C_q \eeqn For the gluonic
dimension 6 operator we find the following scaling \beqn
d^G_q\rightarrow  \lambda^{-4} d^G_q \eeqn Thus the scaling property
of $d_q$ will be more complicated. However,  as $\lambda$ gets large
the contribution of $d^G_q$ will fall off faster than $d^E_q$ and
$d^C_q$ and in this case one will have the scaling $d_q\rightarrow
\lambda^{-2} d_q$ and so $d_n\rightarrow \lambda^{-2} d_n$. Thus
scaling property of EDMs allows one to promote a single point in the
SUSY parameter space where cancelation occurs to a trajectory in the
parameter space. 
With the scaling property one can arrange the cancellation mechanism to work for
  the EDMs over  a much larger region of the parameter space \cite{Ibrahim:1999af}
    than would otherwise be possible
\cite{Pospelov:2005pr}.
The scaling phenomenon also has implications for
the satisfaction of the EDM constraints in string and D-brane
models\cite{Ibrahim:1999af}. As stated already in general only
certain phase combinations  appear in the analysis of a given
physical quantity. Some examples of such combinations 
are given in  Table 3 in Sec.(\ref{qE}).
 For other
solutions to the SUSY CP problem see
\cite{Nir:1996am,Dimopoulos:1995kn,Abel:2001vy,Babu:1999xf}.
  \begin{figure}
 \hspace{-1.5cm}
\includegraphics*[width=10cm, height=13cm]{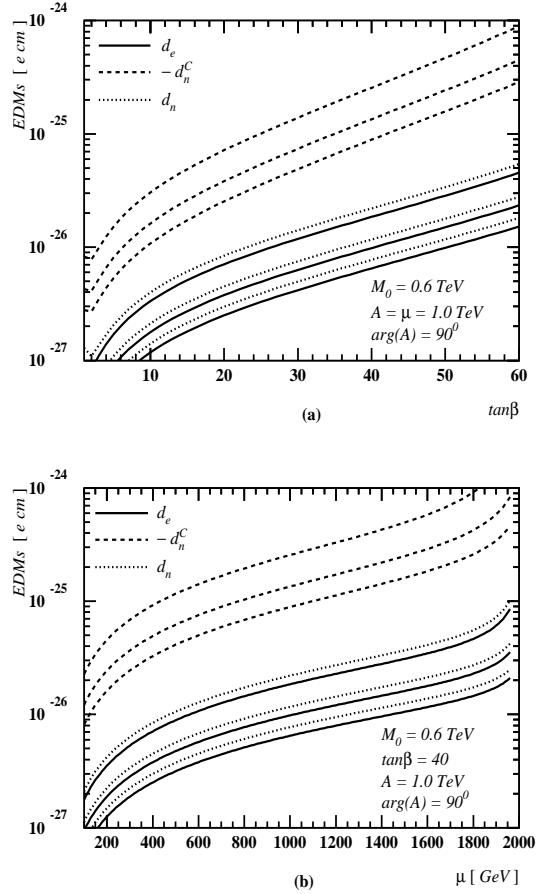}
\vspace{-1cm}\caption{Estimate of the size of two loop contribution to the EDMs
in supersymmetry with phases only in the third generation \cite{Chang:1998uc}.}
\label{2loopEDM}
\end{figure}

 \subsection{Two loop contribution to EDMs}\label{jE}
   Two loop contributions to the  EDMs can be quite significant.  Thus the analysis of \cite{Barr:1990vd,Gunion:1990iv}
  showed that significant
 contributions to the EDM of the electron and of the neutron can result if the  Higgs-boson exchange mediates CP violation.
 A more recent analysis in the same spirit is given by \cite{Chang:1998uc}  for the MSSM case. Here the CP phases
  arising from the Higgs boson couplings  to the stop and the sbottom enter and these are not stringently constrained by data.
 Thus CP phases in the third generation could be quite substantial
  consistent with the EDM constraints.
  We discuss now the two loop analysis in further detail.
 We assume that the large CP phases  arise only   in the third
generation  trilinear   soft parameters $A_{\tau, t, b}$ and the relevant two loop interactions
arise via the CP-odd Higgs  $a(x)$ (see Fig.(\ref{loop2})) whose interactions with fermions and sfermions are
given by

\beqn
{\cal L}_a = \frac{g  m_f }{2 M_W} R_f ia \bar f \gamma_5 f + v \xi_f a (-\tilde f_1^*\tilde  f_1 +\tilde f_2^*
\tilde f_2),
\eeqn
where $g$ is related to the  W boson mass by $M_W=g v/2$, $R_f=\cot\beta (\tan\beta)$ for
$T_3^f=\frac{1}{2}(-\frac{1}{2})$. The   diagrams of Fig.(\ref{loop2}) give the following
contribution to the EDM of a fermion at the electroweak scale
\beqn
d_f/e= \frac{3\alpha_{em}}{64 \pi^3} \frac{R_f m_f}{ m_a^2} \sum_{q=t,b} \xi_q Q_f Q_q^2
[F(x_{1a})- F(x_{2a})],
\eeqn
where
$x_{ia} =(m_{\tilde q_i}/m_a)^2$ (i=1,2), $\xi_q$ (q=t,b)  are defined  by
\beqn
\xi_b =\frac{2m_b\sin 2\theta_b Im(A_b e^{i\delta_b})} { v^2\sin 2\beta}, \nonumber\\
\xi_t =\frac{m_t\sin 2\theta_t Im( \mu e^{i\delta_t})} {v^2 \sin^2\beta },
\eeqn
and
where $\delta_q= arg(A_q+ R_q \mu^*)$.  The function $F(x)$ is given by the loop integral
\beqn
F(x) =  \int_0^1 dy \frac{y(1-y)}{x-y(1-y)} ln\left( \frac{y(1-y)}{x}\right).
\eeqn
Similarly the contribution to CEDM at the electroweak scale is given by
\beqn
d_f^C/e= \frac{\alpha_s}{128 \pi^3} \frac{R_f m_f}{ m_a^2} \sum_{q=t,b}\xi_q [F(x_{1a})- F(x_{2a})].~
\eeqn
A numerical analysis of the EDM is given in Fig.(\ref{2loopEDM}) and  indicates that one can satisfy the EDM constraints in certain ranges  of    the parameter space. However, it remains to be seen how one can naturally
suppress  phases in the first two generations while allowing them only in the third generation.
The reader is also directed to several other works on two loop analyses of EDMs: \cite{Feng:2004vu,Feng:2006ei,Degrassi:2005zd,Chang:1990tw,Chang:1990sf,Pilaftsis:2002fe}.
Specifically, a complete account of all dominant 2-loop Barr-Zee type graphs in the
CP  violating MSSM is given in  in \cite{Pilaftsis:2002fe}. 
The analyses of EDMs given in this section were based on the assumption of R parity conservation. For analyses of EDMs
without R parity see \cite{Keum:2000ak,Keum:2000ea,Faessler:2006at,Hall:1983id}.

\section{CP effects and SUSY phenomena}\label{k}

As noted earlier with the cancelation mechanism the phases can be
large, and thus their effects could be visible in many
supersymmetric
phenomena\cite{Choi:1998nv,Baek:1998yn,Choi:1999mv,Goto:1998iy,Aoki:1999xg,Barr:1999hf,Asatrian:1999sg,Kneur:1999nx,Ma:1999im,Choi:1999cc,Dedes:1999sj,Dedes:1999zh,Choi:1999kn,Kribs:1999df,Mrenna:1999ai,Okada:1999zk,Huang:1999an,Huang:2000ha,Huang:2000tz}.
Below we discuss several  of these phenomena and refer to the
literature above for others.

 \begin{figure}
 \vspace{0.0cm}
 \hspace{-0.5cm}
%\vspace{-4.0}
%\includegraphics*[angle=0, scale=0.5]{f.ps}
\includegraphics*[width=9cm, height=12cm]{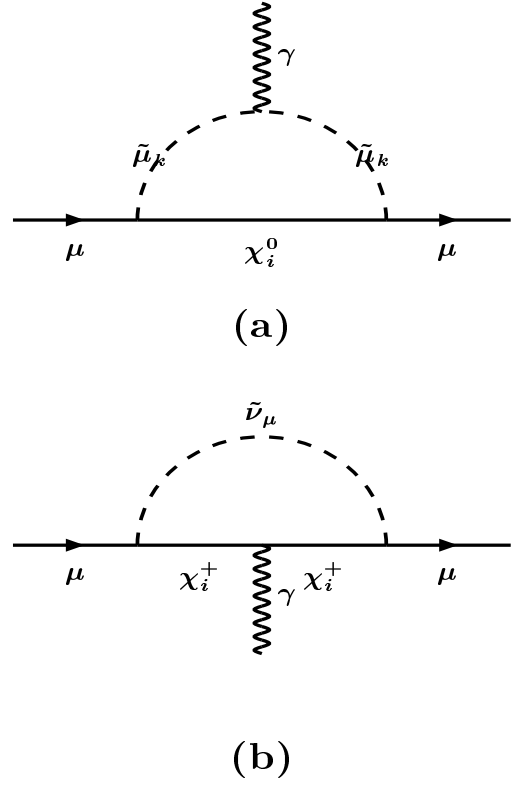}
\vspace{-0.5cm}
\caption{Chargino and neutralino exchanges contributing to the muon g-2 which generate
dependence of $g_{\mu}-2$ on phases.}
\label{mug2}
\end{figure}

\subsection{SUSY phases and $g_{\mu}-2$}\label{kA}

The effects of CP violating phases on the supersymmetric electroweak contributions to $g_{\mu}-2$ have been
investigated\cite{Ibrahim:1999hh,Ibrahim:1999aj,Ibrahim:2001ym}.
The parameter $a_{\mu}\equiv \frac{g_{\mu}-2}{2}$ is induced by loop corrections to the muon vertex with the photon field. In MSSM the muon interacts with other  fermions $\psi_i$ and  scalars $\phi_k$ through
\beqn
{\cal L} = L_{ik} \bar{\mu} P_L \psi_i \phi_k +R_{ik} \bar{\mu} P_R \psi_i \phi_k +H.c.
\eeqn
 where $\psi_i$ stands for the neutralino  (chargino) and $\phi_k$ stands for  the smuon (scalar neutrino).
 The one loop contribution to $a_{\mu}$ is given by
\beqn
a_{\mu}=a^1_{\mu}+a^2_{\mu}
\eeqn
   Here $a^1_{\mu}$ comes from the neutralino exchange contribution and
$a^2_{\mu}$ comes from the chargino exchange contribution so that

\beqn
a^1_{\mu}=\frac{m_{\mu}}{8\pi^2 m_i}Re(L_{ik}R^{*}_{ik})I_1(\frac{m^2_{\mu}}{m^2_i},\frac{m^2_k}{m^2_i})\nonumber\\
+\frac{m^2_{\mu}}{16\pi^2 m^2_i}(|L_{ik}|^2+|R_{ik}|^2)I_2(\frac{m^2_{\mu}}{m^2_i},\frac{m^2_k}{m^2_i}),
\label{a1}
\eeqn
and
\beqn
a^2_{\mu}=\frac{m_{\mu}}{8\pi^2 m_i}Re(L_{ik}R^{*}_{ik})I_3(\frac{m^2_{\mu}}{m^2_i},\frac{m^2_k}{m^2_i})\nonumber\\
-\frac{m^2_{\mu}}{16\pi^2 m^2_i}(|L_{ik}|^2+|R_{ik}|^2)I_4(\frac{m^2_{\mu}}{m^2_i},\frac{m^2_k}{m^2_i}).
\label{a2}
\eeqn
Here
\beqn
I_1(\alpha,\beta)=-\int_{0}^{1} dx \int_{0}^{1-x} dz \frac{z}{\alpha z^2 +(1-\alpha - \beta)z +\beta},\nonumber\\
I_2(\alpha,\beta)=\int_{0}^{1} dx \int_{0}^{1-x} dz \frac{z^2-z}{\alpha z^2 +(1-\alpha - \beta)z +\beta},\nonumber\\
I_3(\alpha,\beta)=\int_{0}^{1} dx \int_{0}^{1-x} dz \frac{1-z}{\alpha z^2 +(\beta-\alpha-1)z +1},\nonumber\\
I_4(\alpha,\beta)=\int_{0}^{1} dx \int_{0}^{1-x} dz \frac{z^2-z}{\alpha z^2 +(\beta-\alpha-1)z +1}~~~~
\label{i1234}
\eeqn
In the supersymmetric limit the soft  breaking terms vanish
and $a_{\mu}$ should
vanish as well\cite{Ferrara:1974wb,Barbieri:1993av}.
A careful limit of Eqs.(\ref{a1}) and (\ref{a2}) shows that in the
 supersymmetric  limit  the sum of the $W$ exchange contribution, in the standard model part, and of the chargino exchange contributions, in the supersymmetric counterpart, cancel. Thus
\beqn
a^W_{\mu}+a^{\chi^+}_{\mu}=0
\eeqn
Similarly one can show that the $Z$ boson exchange and the contribution of the massive modes of the neutralino sector in the supersymmetric limit cancel.
\beqn
a^Z_{\mu}+a^{\chi^0}_{\mu}(massive)=0
\eeqn
One can show that the massless part of the neutralino spectrum in the supersymmetric limit gives the value of $-\alpha_{em}/2\pi$. Thus it gives the same magnitude but is opposite in sign to  the famous photon exchange result.

\begin{figure}
 \hspace{-1.0cm}
\includegraphics*[angle=90, scale=0.3]{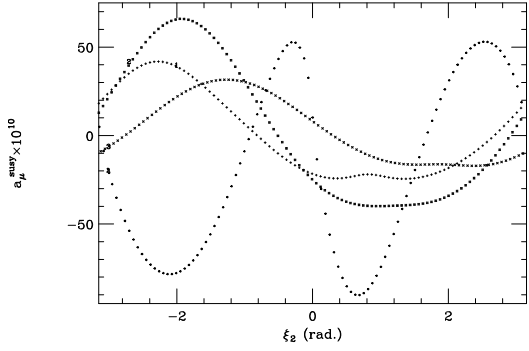}
%\includegraphics*[angle=270,width=3cm, height=4cm]{gmu_phae.ps}
%\vspace{-.5cm}
\caption{ An exhibition of the dependence of $a_{\mu}$ on  a SUSY CP phase.
The curves  correspond to the four cases below
 \cite{Ibrahim:1999aj}
: (1) $m_0$=70, $m_{1/2}$=99, $tan\beta$=3 , $|A_0|$=5.6,
$\xi_1$=$-1$,  $\xi_3$=0.62;  $\theta_{\mu}$= 2.35, $\alpha_{A_0}$=.4;
(2) $m_0$=80, $m_{1/2}$=99, $tan\beta$=5 , $|A_0|$=5.5,
$\xi_1$=$-0.8$, $\xi_3$=0.95;
$\theta_{\mu}$= 1.98, $\alpha_{A_0}$= 0.4;
(3) $m_0$=75, $m_{1/2}$=132, $tan\beta$=4 , $|A_0|$=6.6,
$\xi_1$=$-1$,  $\xi_3$=2.74;
$\theta_{\mu}$= 1.2, $\alpha_{A_0}= -1.5$;
(4) $m_0$=70, $m_{1/2}$=99, $tan\beta$=6 , $|A_0|$=3.2,
$\xi_1$=0.63,    $\xi_3$=0.47,
$\theta_{\mu}$= 2.7, $\alpha_{A_0}=-.4$
where all masses are in GeV units and all phases are in rad.
}
\label{amucp}
\end{figure}

The CP dependence of $a_{\mu}$ arises  from the effect of the phases on the  sparticle  masses, and on their
effects on $L_{ik}$ and $R_{ik}$ and significant variations can arise in $a_{\mu}$ as the phases are varied.
An illustration of this phenomenon is given in Fig.(\ref{amucp}).  Because of the significant dependence
of $a_{\mu}$ on the phases it is possible to constrain the CP phases using  the current data on $a_{\mu}$.
This is done in \cite{Ibrahim:2001ym}. Further details on the analysis of this section are given
in Sec.(\ref{qF}). 

\subsection{SUSY CP phases and CP even -CP odd   mixing in the neutral Higgs 
 boson sector}\label{kB}

Another important effect of CP violating phases is their role in determining the spectrum and CP properties of the
neutral Higgs fields arising due to mixings of the CP even-CP odd
Higgs\cite{Pilaftsis:1998dd,Pilaftsis:1998pe,Pilaftsis:1999qt}.

Such mixings between CP even and CP odd Higgs  bosons cannot occur at the tree level, but are possible when
loop corrections  to the effective potential are included.  To calculate such mixings  we use the
one loop effective potential as given by Eq.(\ref{veff1}).
We assume that the   $SU(2)$ Higgs doublets
 $H_{1,2}$ have non-vanishing vacuum expectation $v_1$ and $v_2$ so that we can write
\beqn
(H_1)
 =\frac{1}{\sqrt 2}
\left(\begin{array}{c} v_1+\phi_1+i\psi_1\\
             H_1^- \end{array} \right),\nonumber\\
(H_2)
=\frac{e^{i\theta_H}}{\sqrt 2} \left(\begin{array}{c} H_2^+ \\
             v_2+\phi_2+i\psi_2 \end{array} \right).
\eeqn
For the present case with the inclusion of CP violating effects,
  the variations with respect to
the fields $\phi_1, \phi_2, \psi_1,\psi_2$ give the following
\beq
-\frac{1}{v_1}(\frac{\partial \Delta V}{\partial \phi_1})_0=
m_1^2+\frac{g_2^2+g_1^2}{8}(v_1^2-v_2^2)+m_3^2 \tan\beta \cos\theta_H,
\label{min1}
\eeq
\beq
-\frac{1}{v_2}(\frac{\partial \Delta V}{\partial \phi_2})_0=
m_2^2-\frac{g_2^2+g_1^2}{8}(v_1^2-v_2^2)+m_3^2 cot\beta \cos\theta_H,
\label{min2}
\eeq

\beq
\frac{1}{v_1}(\frac{\partial \Delta V}{\partial \psi_2})_0=
m_3^2 \sin\theta_H= \frac{1}{v_2}
(\frac{\partial \Delta V}{\partial \psi_1})_0,
\label{min3}
\eeq
where the subscript 0 means that the quantities are evaluated
at the point $\phi_1=\phi_2=\psi_1=\psi_2=0$.
As noted in \cite{Demir:1999hj}
only one of the two equations
in Eq.(\ref{min3}) is independent. \\

One can have sizable contributions to the potential corrections from top-stop,
bottom-sbottom \cite{Pilaftsis:1999qt,Choi:2000wz,Demir:1999qm,Ibrahim:2000qj},
$W-H^+-\chi^+$ sector \cite{Ibrahim:2000qj}
 and from the $\chi^0-Z-h^0-H^0$ sector \cite{Ibrahim:2002zk,Ham:2002ps}.
  The mass-squared matrix of the neutral Higgs bosons is defined by
\beqn
M^2_{ab}=(\frac{\partial^2 V}{\partial \Phi_a \Phi_b})_0,
\eeqn
where $\Phi_a=(a=1-4)$ are defined by
\beqn
\{\Phi_a\}=\{\phi_1,\phi_2,\psi_1,\psi_2\},
\label{phibasis}
\eeqn
and the subscript $0$ means that we set $\phi_1=\phi_2=\psi_1=\psi_2=0$.
The dominant contributions come from
the stop, sbottom and chargino.
With the inclusion of the stop, the sbottom,   and of the chargino
contributions
one finds that $\theta_H$ is determined by

\beqn
m_3^2 \sin\theta_H =\frac{1}{2} \beta_{h_t} |\mu| |A_t| \sin\gamma_t
f_1(m_{\tilde t_1}^2, m_{\tilde t_2}^2)
+\frac{1}{2}  \beta_{h_b}
|\mu|\nonumber\\
 |A_b| \sin\gamma_b
f_1(m_{\tilde b_1}^2, m_{\tilde b_2}^2)
-\frac{g^2_2}{16\pi^2}|\mu| |\tilde m_2| \sin\gamma_2
f_1(m_{\tilde \chi_1}^2, m_{\tilde \chi_2}^2),\nonumber\\~
\label{defthetah}
\eeqn
where
\beqn
 \beta_{h_t}=\frac{3 h_t^2}{16\pi^2},
~~ \beta_{h_b}=\frac{3 h_b^2}{16\pi^2};\nonumber\\
\gamma_t=\alpha_{A_t} + \theta_{\mu}, ~~
\gamma_b=\alpha_{A_b} + \theta_{\mu},~~
\gamma_2=\xi_2 + \theta_{\mu},
\label{phasecom1}
\eeqn
and $f_1(x,y)$ is defined by
\beq
f_1(x,y)=-2+log\frac{xy}{Q^4} + \frac{y+x}{y-x}log\frac{y}{x}.
\eeq
The inclusion of the stau and the neutralino sectors  in the analysis
would contribute extra terms to Eq.(\ref{defthetah}) that are dependent 
on the phase $\gamma_{\tau}= \alpha_{A_{\tau}} +\theta_{\mu}$ and 
$\gamma_1=\xi_1+\theta_{\mu}$.
The tree and loop contributions to $M_{ab}^2$ are given by
\beq
M_{ab}^2= M_{ab}^{2(0)}+ \Delta M_{ab}^2,
\eeq
where  $M_{ab}^{2(0)}$ are the contributions at the tree level and
 $\Delta M_{ab}^2$ are the loop contributions where

\beqn
\Delta M_{ab}^2=
\frac{1}{32\pi^2}
Str(\frac{\partial M^2}{\partial \Phi_a}\frac{\partial M^2}{\partial\Phi_b}
log\frac{M^2}{Q^2}+\nonumber\\
+ M^2 \frac{\partial^2 M^2}{\partial \Phi_a\partial \Phi_b}
log\frac{M^2}{eQ^2})_0,~~
\eeqn
 and where e=2.718.
Computation of the $4\times 4$ Higgs  mass$^2$  matrix in the basis
of Eq.(\ref{phibasis}) gives
\beqn
\left(\begin{array}{cccc} M_{11}+\Delta_{11} &
-M_{12}+\Delta_{12} &\Delta_{13}s_{\beta} & \Delta_{13} C_{\beta}\\
-M_{12}+\Delta_{12} &
M_{22}+\Delta_{22} & \Delta_{23}  s_{\beta} & \Delta_{23} c_{\beta} \\
\Delta_{13} s_{\beta} & \Delta_{23}s_{\beta}  & M_{33} s_{\beta}^2 &  M_{33}s_{\beta}c_{\beta}\\
\Delta_{13} c_{\beta} & \Delta_{23}c_{\beta}  & M_{33}  s_{\beta}c_{\beta} &  M_{33}c_{\beta}^2
 \end{array} \right),\nonumber\\
\eeqn
where $M_{11}=M_Z^2c_{\beta}^2+M_A^2s_{\beta}^2$, $M_{12}=(M_Z^2+M_A^2)s_{\beta}c_{\beta}$, $M_{22}=M_Z^2s_{\beta}^2+M_A^2c_{\beta}^2$, $c_{\beta},s_{\beta}=\cos\beta,\sin\beta$, and $M_{33}= M_A^2+ \Delta_{33}$. and
where $(c_{\beta}, s_{\beta})=(\cos\beta, \sin\beta)$.
Here
 the explicit Q dependence has been absorbed in $m_A^2$
which is given by

\beqn
m_A^2=(\sin\beta\cos\beta)^{-1}(-m_3^2\cos\theta +\nonumber\\
\frac{1}{2}\beta_{h_t}
|A_t||\mu|\cos\gamma_t f_1(m_{\tilde t_1}^2,m_{\tilde t_2}^2)\nonumber\\
+\frac{1}{2}\beta_{h_b} |A_b||\mu| \cos\gamma_b f_1
(m_{\tilde b_1}^2,m_{\tilde b_2}^2)\nonumber\\
+\frac{1}{2}\beta_{h_{\tau}} |A_{\tau}||\mu| \cos\gamma_{\tau} f_1
(m_{\tilde {\tau}_1}^2,m_{\tilde {\tau}_2}^2)\nonumber\\
 +\frac{g_2^2}{16\pi^2}|\tilde m_2|
|\mu| \cos\gamma_2 f_1(m_{\chi_1^+}^2, m_{\chi_2^+}^2))   + \Delta_{\chi}.\nonumber\\
\label{Amass}
\eeqn
Here $\Delta_{\chi}$ is the contribution arising from the neutralino exchange and
\beqn
\Delta_{\chi}=
-\frac{1}{16\pi^2}\sum_{j=1}^{4}\frac{M_{\chi_j}^2}{D_j}
(log(\frac{M_{\chi_j}^2}{Q^2}) -1)\nonumber\\
( M_{\chi_j}^4(-g_2^2|\mu||\tilde m_2|\cos\gamma_2 -g_1^2|\mu||\tilde m_1|
\cos\gamma_1)\nonumber\\
+M_{\chi_j}^2(g_2^2(|\tilde m_1|^2+|\mu|^2)|\mu| |\tilde m_2|\cos\gamma_2\nonumber\\
+g_1^2(|\tilde m_2|^2+|\mu|^2)   |\mu| |\tilde m_1|\cos\gamma_1)\nonumber\\
-g_2^2 |\tilde m_1|^2 |\mu|^3 |\tilde m_2| \cos\gamma_2
-g_1^2 |\tilde m_2|^2 |\mu|^3 |\tilde m_1| \cos\gamma_2)
\label{Amass1}
\eeqn
where $\Delta\xi =\xi_1-\xi_2$.
The first term in the second
 brace on the right hand side of Eqs.(\ref{Amass})
 is the tree term,
while the second, the
third  and the fourth terms come from the stop, sbottom, stau and chargino
exchange contributions. The remaining contributions in Eq.(\ref{Amass}) arise
from the neutralino sector.
For $\Delta_{ij}$ one has
\beq
\Delta_{ij}=\Delta_{ij\tilde t}+\Delta_{ij\tilde b}+
\Delta_{ij\tilde{\tau}}+
\Delta_{ij\chi^+} + \Delta_{ij\chi^0}
\eeq
where $\Delta_{ij\tilde t}$ is the contribution from the
stop exchange in the loops, $\Delta_{ij\tilde b}$ is the
contribution from the sbottom exchange in the loops,
$\Delta_{ij\tilde{\tau}}$ is the contribution from the stau loop,
$\Delta_{ij\chi^+}$ is the contribution from the chargino
 sector and  $\Delta_{ij\chi^0}$ is the contribution from the neutralino
 sector.  For illustration
 $\Delta_{ij\tilde t}$ are listed in Sec.(\ref{qG}). \\

 \begin{figure}
 \hspace{-.5cm}
\includegraphics*[angle=270, scale=0.3]{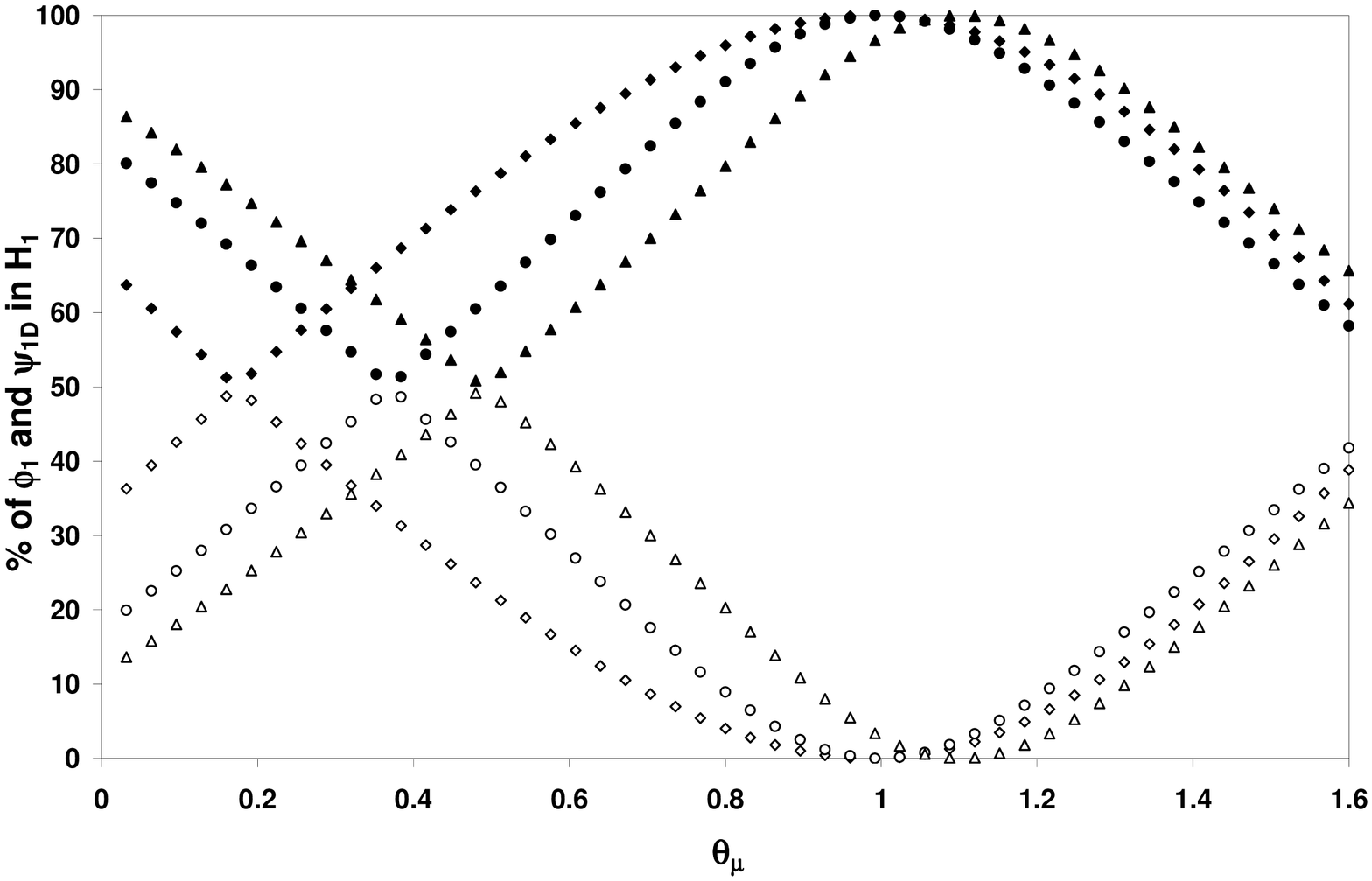}
\vspace{0.0cm}
\caption{ An exhibition of the phenomenon of CP even -CP odd Higgs mixing via the SUSY CP phases.
The CP even component $\phi_1$ of $H_1$ (upper
curves)  and the CP odd component $\psi_{1D}$ of $H_1$ (lower curves)
including the stop, sbottom, stau, chargino and neutralino sector contributions
as a function of $\theta_{\mu}$.
The common parameters are  $m_A=300$, $Q=320$,
$m_0=100$, $m_{\frac{1}{2}}=500$, $\xi_2=.5$, $\alpha_0=.3$, $|A_0|=1$.
For curves with diamonds $\tan\beta=15$, $\xi_1=1.5$,
for circles $\tan\beta=8$, $\xi_1=1.5$, and
for triangles $\tan\beta=8$, $\xi_1=0.5$  where all masses are in GeV
and all angles are in radians.
}
\label{cpevenodd}
\end{figure}

We note that the phases come to play a role here through  the squark, slepton,  chargino and neutralino eigen values of their mass matrices.
We note that in the supersymmetric limit $M_{\chi_i^0}=(0,0,M_Z,M_Z)$, $(M_{h^0},M_{H^0})=(M_Z,0)$, $M_{\chi^+_i}=M_{H^+}=M_W$ and $M_{\tilde{q_i}}=m_q$.
With this in mind one can see that all the radiative corrections to the potential vanish in the supersymmetric limit.
By introducing  a new basis $\phi_1,\phi_2,\psi_{1D},\psi_{2D}$ where
\beqn
\psi_{1D}=\sin\beta \psi_1+\cos\beta \psi_2\nonumber\\
\psi_{2D}=-\cos\beta \psi_1+\sin\beta \psi_2,
\eeqn
  one finds  that the field $\psi_{2D}$ decouples from the other three fields and is a massless state (a Goldstone field). The Higgs mass$^2$ matrix $M^2_{Higgs}$ of the remaining three fields is given by
\beqn
\left(\begin{array}{ccc} M_{11}+\Delta_{11} &
-M_{12}+\Delta_{12} &\Delta_{13}\\
-M_{12}+\Delta_{12} &
M_{22}+\Delta_{22} & \Delta_{23} \\
\Delta_{13} & \Delta_{23} &M_A^2+\Delta_{33} \end{array} \right)
\eeqn
We note that the basis fields $\{\phi_1,\phi_2,\psi_{1D}\}$ of the above matrix are the real parts of the neutral Higgs fields and a linear combination of their imaginary parts $\psi_i$. Thus these states are pure CP states where $\phi_{1,2}$ are CP even (scalars) and $\psi_{1D}$ is CP odd (a pseudoscalar).
What we are interested here is the mixing between the CP even and the CP odd Higgs states in the eigen vectors of the above matrix and this mixing is governed by the off-diagonal elements $\Delta_{12}$ and $\Delta_{23}$. These are found to be linear combination of $\sin\gamma_t$, $\sin\gamma_b$, $\sin\gamma_1$, $\sin\gamma_2$, and $\sin\gamma_{\tau}$ where these
phases are defined as in Eq.(\ref{phasecom1}).
In the limit of vanishing CP phases the matrix elements $\Delta_{12}$ and $\Delta_{23}$ vanish and thus the Higgs mass$^2$ matrix factors into a $2\times 2$ CP even Higgs matrix times a CP odd element.
The effect of phases on CP even-CP odd Higgs boson mixings have been studied by
\cite{Pilaftsis:1999qt,Choi:2000wz,Demir:1999qm,Ibrahim:2000qj}
  and   found to be  significant. It is shown that if a mixing effect among the CP even and the CP odd Higgs bosons is observed experimentally, then it is only the cancelation mechanism of EDMs that can survive\cite{Ibrahim:2001xg}.
   A more accurate determination of the  VEV of the Higgs fields would require use of two loop effective potential.
  An  improved accuracy and scale dependence should be obtained with the full two-loop effective
   potential\cite{Martin:2002wn}.

\subsection{Effect of SUSY CP phases on the  b quark mass}\label{kC}

The running b quark mass is another object in MSSM where CP phases
could have an impact. $m_b$ can be written in the form \beqn
m_b(M_Z)=h_b(M_Z) \frac{v}{\sqrt{2}}\cos\beta (1+\Delta_b) \eeqn
where $h_b(M_Z)$ is the Yukawa coupling for the b quark at the scale
$M_Z$ and $\Delta_b$ is the loop correction to $m_b$. The SUSY QCD
and electroweak corrections are large in the large $\tan\beta$
region\cite{Hall:1993gn,Carena:1994bv,Pierce:1996zz}. At the tree
level the b quark couples to the neutral component of $H_1$ Higgs
boson while the coupling to the $H_2$ Higgs boson is absent. Loop
corrections produce a shift in the $ H_1^0$ couplings and generate a
non-vanishing effective coupling with $H_2^0$. Thus the effective
Lagrangian would be written as \cite{Babu:1998bf,Carena:2002es}
\beqn -{\cal L}_{eff}=(h_b+\delta h_b)\bar{b}_R b_L H_1^0 +\Delta
h_b \bar{b}_R b_L H_2^{0*}+H.c. \eeqn where the star on $H_2^0$ is
necessary in order to have a gauge invariant Lagrangian. The
quantities $\delta h_b$ and $\Delta h_b$ receive SUSY QCD and SUSY
electroweak contributions. The QCD contribution arises from the
corrections where  gluinos and sbottoms are running in the loops. In
the electrowak contributions, the sbottoms (stops) and the
neutralinos (charginos) are running in the loops. The basic integral
that appears in the expressions of $\delta h_b$ and $\Delta h_b$
involving heavy scalars $\tilde{S}_1$, $\tilde{S}_2$ and a heavy
fermion ${\tilde f}$ is 
\beqn I=\int \frac{d^4k}{(2\pi)^4}
\frac{m_{\tilde f}+\gamma_{\mu}
k^{\mu}}{(k^2-m^2_{\tilde{f}})(k^2-m^2_{\tilde{S}_1})(k^2-m^2_{\tilde{S}_2})}
\eeqn 
In the approximation of the zero external momentum this
integral could be written in the closed form \beqn
I=\frac{m_{\tilde{f}}}{(4\pi)^2} f(m^2_{\tilde{f}},
m^2_{\tilde{S}_1}, m^2_{\tilde{S}_2}) \eeqn where the function
$f(m^2, m^2_i, m^2_j)$ is given by \beqn
f(m^2, m^2_i, m^2_j)=\nonumber\\
\left[
\left(m^2-m^2_i)(m^2-m^2_j)(m^2_j-m^2_i \right)\right]^{-1}  \nonumber\\
\times \left(m^2_jm^2\ln \frac{m^2_j}{m^2}+m^2m^2_i\ln
\frac{m^2}{m^2_i}+m^2_im^2_i\ln \frac{m^2_i}{m^2_j}\right) \eeqn for
the case $i\neq j$ and \beqn f(m^2, m^2_i, m^2_j)=
\frac{1}{(m^2_i-m^2)^2} (m^2\ln \frac{m^2_i}{m^2}+(m^2-m^2_i))
\nonumber \eeqn for the case $i=j$. In the SUSY QCD the heavy
fermion is the gluino and the heavy scalars are the sbottoms. In the
chargino contribution, the chargino is the heavy fermion and the
heavy scalars are the stops. In the neutralino part, the neutralino
is the heavy fermion and the heavy scalars are the sbottoms.

The couplings $\delta h_b$ and $\Delta h_b$ are generally complex due to CP phases in the soft SUSY breaking terms.
Electroweak symmetry is broken spontaneously by giving expectation values to $H_1^0$ and $H_2^0$. Thus one finds for the mass term
\beqn
-{\cal L}_{m}=M_b \bar{b}_R b_L +H.c.,
\eeqn
where
\beqn
M_b=\frac{h_b v \cos\beta}{\sqrt{2}}(1+\frac{\delta h_b}{h_b}+\frac{\Delta h_b}{h_b}\tan\beta)
\eeqn
Here $M_b$ is complex. By rotating the b quark field
\beqn
b=e^{i/2\gamma_5 \chi_b}b',
~~\tan\chi_b=\frac{Im M_b}{Re M_b}
\eeqn
one gets
\beqn
-{\cal L}_{m}=m_b \bar{b'}_R b'_L +H.c.,
\eeqn
where $m_b$ is real and positive and $b'$ is the physical field.
\beqn
m_b=\frac{h_b v \cos\beta}{\sqrt{2}}
\left((1+\delta_R)^2 + \delta_I^2\right)^{\frac{1}{2}}\nonumber\\
\delta_R =
Re(\frac{\delta h_b}{h_b})+Re(\frac{\Delta h_b}{h_b})\tan\beta\nonumber\\
\delta_I=Im(\frac{\delta h_b}{h_b})+Im(\frac{\Delta h_b}{h_b})\tan\beta
\eeqn
Thus one finds for the mass correction
\beqn
\Delta_b\approx Re\frac{\Delta h_b}{h_b}\tan\beta+Re\frac{\delta h_b}{h_b}
\eeqn
The SUSY CP violating phases in the SUSY QCD corrections are $\xi_3$, $\alpha_{A_b}$ and $\theta_{\mu}$. These come from the vertices of $b\tilde{b}\tilde{g}$ and $\tilde{b}\tilde{b}H$. In the chargino part one finds the phases $\xi_2$, $\alpha_{A_t}$ and $\theta_{\mu}$. In the case of neutralino we have
$\xi_2$,
 $\xi_1$, $\alpha_{A_b}$ and $\theta_{\mu}$. The corrections of the b quark mass are found to be very sensitively dependent on $\theta_{\mu}$, $\xi_3$ and $\alpha_{A_0}$ as the values of these phases affect both the sign and the magnitude of the correction. Thus the correction can vary from zero to as much as 30$\%$ in some regions of the parameter space and can also change its sign depending on the value of these phases. The effect of $\xi_2$ is less important and $\xi_1$ is found to be the least important
 phase\cite{Ibrahim:2003ca}.
  Similar results hold for the $\tau$ lepton mass and  for the top quark mass. For the $\tau$ lepton the numerical size of the correction is as much as 5$\%$ and for the top quark is typically less than a percent.

\subsection{SUSY CP phases and the decays $h\to b\bar b, ~h\to \tau\bar \tau$}
\label{kD}

As was mentioned above, the spectrum of the neutral Higgs sector and its CP properties are sensitive to the CP violating phases through radiative corrections. The couplings of the quarks with the Higgs are also found to be dependent of these phases. Thus one can deduce the corrected effective interaction of the b quark with the lightest Higgs boson $H_2$ as
\beqn
-{\cal L}_{int}=\bar{b}(C^S_b+i\gamma_5C^P_b)b H_2
\eeqn
where
\beqn
\left(\begin{array}{c} C^S_b\\
C^P_b\end{array} \right)=
\left(\begin{array}{cc} \cos\chi_b & -\sin\chi_b\\
\sin\chi_b & \cos\chi_b\end{array} \right)
\left(\begin{array}{c} C^1_b\\
C^2_b\end{array} \right)
\eeqn
where
\beqn
C_b^1=\frac{1}{\sqrt{2}}(Re(h_b+\delta h_b)R_{21} +
\{-Im(h_b+\delta h_b)\sin\beta\nonumber\\
+Im(\Delta h_b)\cos\beta\}R_{23}
+Re(\Delta h_b)R_{22})\nonumber\\
C_b^2=
-\frac{1}{\sqrt{2}}(-Im(h_b+\delta h_b)R_{21}
+\{-Re(h_b+\delta h_b)\sin\beta\nonumber\\
+Re(\Delta h_b)\cos\beta\}R_{23}
-Im(\Delta h_b)R_{22})\nonumber
\eeqn
The matrix $R$ is the diagonalizing matrix of the Higgs mass$^2$ matrix
\beqn
R M^2_{Higgs}R^T=diag(m^2_{H_1}, m^2_{H_2}, m^2_{H_3})
\eeqn
where we use the convention that in the limit of vanishing CP phases, one has $H_1\rightarrow H$, $H_2\rightarrow h$ and $H_3\rightarrow A$. These elements $R_{ij}$ and the corrections $\delta h_b$ and $\Delta h_b$ are found to be sensitive functions of the CP violating phases and their values are all determined by SUSY radiative corrections of MSSM potential.
The quantity $R_{b/\tau}$ defined as
\beqn
R_{b/\tau}=\frac{BR(h\rightarrow \bar{b}b)}{BR(h\rightarrow \bar{\tau}\tau)}
\eeqn
is found to be an important tool to discover supersymmetry.
In Standard Model, it is given by
\beqn
R^{SM}_{b/\tau}=3(\frac{m^2_b}{m^2_{\tau}})[\frac{m^2_h-4m^2_b}{m^2_h-4m^2_{\tau}}]^{3/2}(1+w)
\eeqn
where $(1+w)$ is the QCD enhancement factor\cite{Gorishnii:1990zu}.

\beqn
1+w=1+5.67 \frac{\alpha_s}{\pi}+29.14\frac{\alpha^2_s}{\pi^2}
\eeqn
By identifying $m_h$ with $m_{H_2}$, the lightest Higgs boson in MSSM, we find a shift in $R_{b/\tau}$ value due to supersymmetric effect including the effects due to CP phases as follows
\beqn
\Delta R_{b/\tau}=\frac{R_{b/\tau}-R^{SM}_{b/\tau}}{R_{b/\tau}}
\eeqn
The quantity $R_{b/\tau}$ in MSSM depends on  the CP phase via
$C^S_b$ and $C^P_b$.
Thus if a neutral Higgs is discovered and $R_{b/\tau}$  measured and  found to be
different from what one expects in the Standard Model, then it would point to a non-standard Higgs  boson
such as from MSSM \cite{Babu:1998er}.
 The analysis of \cite{Ibrahim:2003jm}
that the supersymmetric effects with CP phases can change the branching ratios by as much as 100 $\%$ for the lightest Higgs boson decay into $\bar{b}b$ and $\bar{\tau}\tau$. Similar results are reported for the other heavier Higgs bosons.
Thus the deviation from the Standard Model result for $R_{b/\tau}$  depends on the CP phases and it can be used as a possible signature for supersymmetry and CP effects. Similar analyses can also be given for the decay of the heavy
Higgs, e.g., for $H^0\to t\bar t, b\bar b$ and to $ \chi^+\chi^-$  \cite{Ibrahim:2007ne,Eberl:2004ic}
 if allowed kinematically.

\subsection{SUSY CP phases and charged  Higgs  decays  $H^-\to \bar t b$, $H^-\to \bar\nu_{\tau} \tau$}
\label{kE}

In the neutral Higgs sector, the ratio
$R^{h_0}=BR(h^0\rightarrow b\bar{b})/BR(h^0\rightarrow \tau\bar{\tau})$ is found to be sensitive to the supersymmetric loop corrections and to the CP phases.  In an analogous fashion we may define the ratio $R^{H^-}=BR(H^-\rightarrow b\bar{t})/BR(H^-\rightarrow \tau\bar{\nu_{\tau}})$ and it  is also  affected by SUSY loop corrections,
and is sensitive to  CP phases.
Thus the tree level couplings of the third generation quarks to the Higgs  bosons
\beqn
-{\cal L}=\epsilon_{ij} h_b \bar{b}_R H^i_1 Q^j_L-\epsilon_{ij} h_t \bar{t}_R H^i_2 Q^j_L+H.c.
\eeqn
receive SUSY QCD and the SUSY electroweak loop corrections which
produce shifts in  couplings similar to the case for the neutral Higgs bosons.
 Thus the general effective interaction may be written as
\beqn
-{\cal L}_{eff}=\epsilon_{ij} (h_b+\delta h^i_b) \bar{b}_R H^i_1 Q^j_L+\Delta h^i_b \bar{b}_R H^{i*}_2 Q^i_L\nonumber\\
-\epsilon_{ij} (h_t+\delta h^i_t) \bar{t}_R H^i_2 Q^j_L
+\Delta h^i_t \bar{t}_R H^{i*}_1 Q^i_L
+H.c.
\eeqn
We note that in the approximation
\beqn
\delta h^1_f=\delta h^2_f,
~\Delta h^1_f=\Delta h^2_f
\eeqn
one finds that the above Lagrangian preserves weak isospin. This is the approximation that is often used in the literature\cite{Carena:2002es}.
However, in general, the above approximation will not hold and there will be violations of weak isospin.
In the neutral Higgs interaction with the quarks and leptons of third generation, we examined $\delta h^1_{b,\tau}$, $\Delta h^2_{b,\tau}$, $\delta h^2_{t}$ and $\Delta h^1_{t}$.
In the charged Higgs interaction with these particles we should similarly examine $\delta h^2_{b,\tau}$, $\Delta h^1_{b,\tau}$, $\delta h^1_{t}$ and $\Delta h^2_{t}$.
The latter corrections have SUSY QCD contributions when gluinos, stops and sbottoms are running in the loops and SUSY electroweak contributions when neutralinos and/or charginos, stops and/or sbottoms are running in the loops.
The CP violating phases that enter $\delta h^{2g}_{b}$ and $\Delta h^{1g}_{b}$ are $\xi_3$, $\alpha_{A_t}$, $\alpha_{A_b}$ and $\theta_{\mu}$. The phases that appear in
 $\delta h^{2E}_{b,\tau}$ and $\Delta h^{1E}_{b,\tau}$ are
$\xi_1$, $\xi_2$, $\alpha_{A_t}$, $\alpha_{A_b}$, $\alpha_{A_{\tau}}$ and $\theta_{\mu}$. The phases that enter the corrections
$\delta h^1_{t}$ and $\Delta h^2_{t}$ are the same as in
$\delta h^2_{b}$ and $\Delta h^1_{b}$.
One can measure the size of the violation of weak isospin by defining $r_b$
\beqn
r_b =   (|\Delta h^1_b|^2+|\delta h^2_b|^2)^{\frac{1}{2}}
(|\Delta h^2_b|^2+|\delta h^1_b|^2)^{-\frac{1}{2}}
\eeqn
Similar ratios could be defined for the top and tau, $r_t$ and $r_{\tau}$. The deviation of these quantities from unity is an indication of the violation of weak isospin in the Higgs couplings. It is found that such deviations from unity can be as much as $50\%$ or more depending on the region of the parameter space one is in. It is also seen
 that these measures are sensitive functions of CP violating phases \cite{Ibrahim:2003tq}.
The interactions of the charged Higgs  are thus governed by the Lagrangian
\beqn
-{\cal L}=\bar{b}(B^s_{bt}+B^p_{bt}\gamma_5)tH^{-}
+\bar{\tau}(B^s_{\nu\tau}+B^p_{\nu\tau}\gamma_5)\nu H^{-}+H.c.\nonumber\\
\eeqn
where
\beqn
B^s_{bt}=-\frac{1}{2}(h_b+\delta h_b^2)e^{-i\theta_{bt}}\sin\beta+\frac{1}{2}\Delta h_b^1 e^{-i\theta_{bt}}\cos\beta\nonumber\\
-\frac{1}{2}(h_t+\delta h_t^{1*})e^{i\theta_{bt}}\cos\beta+\frac{1}{2}\Delta h_t^{2*} e^{i\theta_{bt}}\sin\beta\nonumber\\
B^p_{bt}=-\frac{1}{2}(h_t+\delta h_t^{1*})e^{i\theta_{bt}}\cos\beta+\frac{1}{2}\Delta h_t^{2*} e^{i\theta_{bt}}\sin\beta\nonumber\\
+\frac{1}{2}(h_b+\delta h_b^{2})e^{-i\theta_{bt}}\sin\beta-\frac{1}{2}\Delta h_b^{1} e^{-i\theta_{bt}}\cos\beta\nonumber\\
B^s_{\nu\tau}
=-\frac{1}{2}(h_{\tau}+\delta h_{\tau}^2)e^{-i\chi_{\tau}/2}\sin\beta
+\frac{1}{2}\Delta h_{\tau}^1 e^{-i\chi_{\tau}/2}\cos\beta\nonumber\\
\eeqn
and where
$B^p_{\nu\tau}=-B^s_{\nu\tau}$, and
 $\theta_{bt}=(\chi_b+\chi_t)/2$. The same holds for $\chi_{\tau}$ with $b$ replaced by $\tau$. For $\tan\chi_t$ a similar expression holds with $b$ replaced by $t$.
The loop corrections to the charged Higgs couplings can be quite significant.  Also
the loop corrections can generate  significant violations  of the weak isospin  in this sector.

  \begin{figure}
% \vspace{-2cm}
 \hspace{-.5cm}
\includegraphics*[width=8cm, height=4cm]{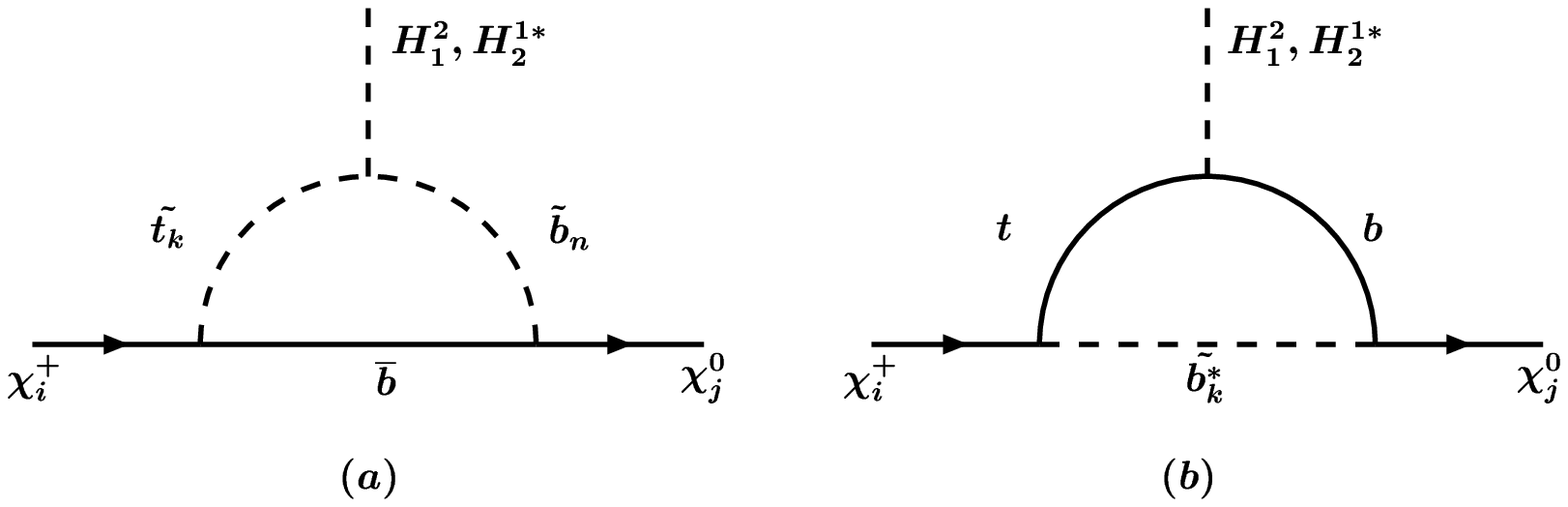}
\includegraphics*[width=8cm, height=4cm]{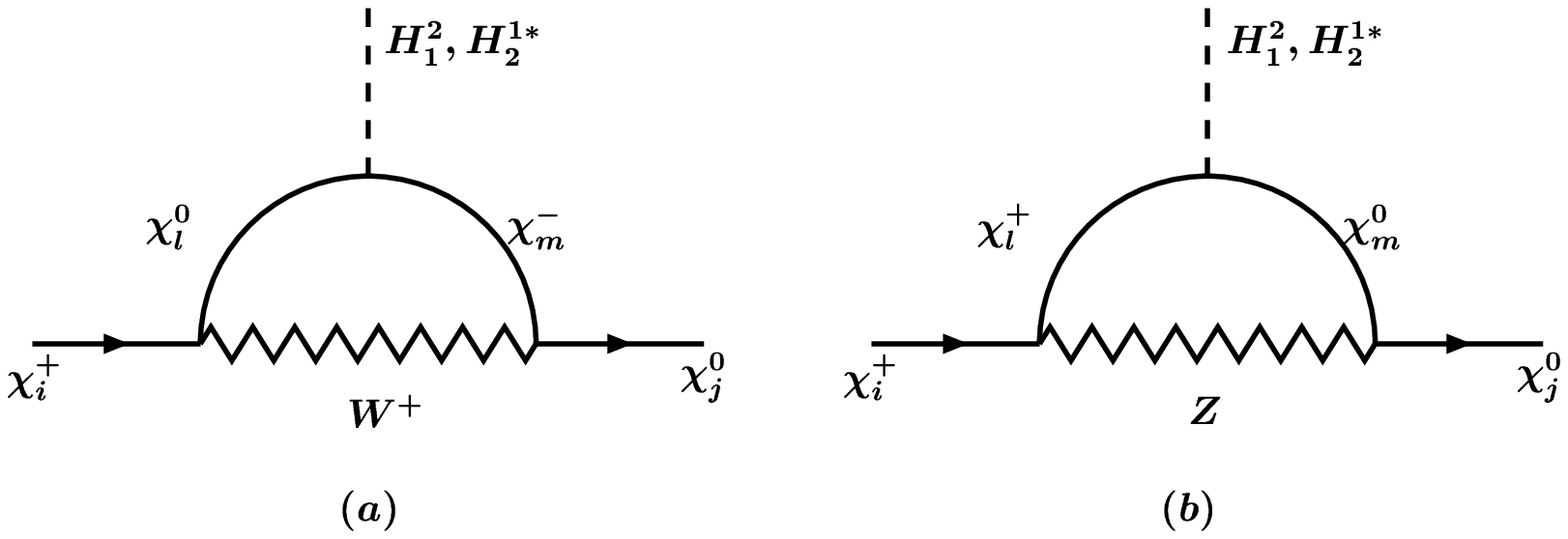}
\vspace{-.5cm}
\caption{Loop diagrams with CP dependent vertices that contribute to charged Higgs decays
into charginos and neutralinos.}
\label{stopcp}
\end{figure}

\subsection{SUSY CP phases and charged  Higgs  decays  $H^{\pm}\to \chi^{\pm} \chi^0$}
\label{kF}

The decay $H^{\pm}\to \chi^{\pm} \chi^0$ is sensitive to CP violation phases even at the tree level.
Inclusion of  the loop corrections  further enhance the effects of the CP phases.
The tree level lagrangian for $H^{\pm}\chi^{\mp}\chi^0$ is
\beqn
{\cal L}=\xi_{ji}H_2^{1*}\bar{\chi^0}_jP_L\chi^+_i+\xi'_{ji}H_1^{2}\bar{\chi^0}_jP_R\chi^+_i+H.c.,
\eeqn
where $\xi_{ij}$ and $\xi'_{ij}$ are given by
\beqn
\xi_{ji}=-gX_{4j}V^*_{i1}-\frac{g}{\sqrt{2}}X_{2j}V^*_{i2}-\frac{g}{\sqrt{2}}\tan\theta_WX_{1j}V^*_{i2}\nonumber\\
\xi'_{ji}=-gX^*_{3j}U_{i1}+\frac{g}{\sqrt{2}}X^*_{2j}U_{i2}+\frac{g}{\sqrt{2}}\tan\theta_WX^*_{1j}U_{i2}\nonumber\\
\eeqn
The phases that enter the couplings $\xi_{ji}$ and $\xi'_{ji}$ are $\xi_1$, $\xi_2$ and $\theta_{\mu}$.
The loop corrections produce shifts in the couplings and the effective Lagrangian with loop corrected couplings is given by
\beqn
{\cal L}_{eff}=(\xi_{ji}+\delta \xi_{ji})H_2^{1*}\bar{\chi^0}_jP_L\chi^+_i+\Delta \xi_{ji}H^2_1 \bar{\chi^0}_jP_L\chi^+_i\nonumber\\
+(\xi'_{ji}+\delta \xi'_{ji})H_1^{2}\bar{\chi^0}_jP_R\chi^+_i+
+\Delta \xi'_{ji}H^{1*}_2\bar{\chi^0}_jP_R\chi^+_i+H.c.\nonumber\\
\eeqn
The phases that enter the corrections $\Delta \xi_{ij}$, $\delta \xi_{ij}$ are $\xi_1$, $\xi_2$, $\alpha_{A_t}$, $\alpha_{A_b}$ and $\theta_{\mu}$. This dependence arises from the shifts in the vertices  of  the charginos with top and sbottoms, charginos with bottoms and stops, neutralino with bottom and sbottoms, neutralino with tops and stops, W bosons with charginos and neutralinos, Z bosons with charginos and neutralinos, charged Higgs with neutralinos and charginos and charged Higgs with stops and sbottoms. All these vertices enter in the loop corrections.
Thus ${\cal L}_{eff}$ may be written in terms of the mass eigenstates as follows
\beqn
{\cal L}_{eff}=H^-\bar{\chi^0}_j(\alpha^S_{ji}+\gamma_5\alpha^P_{ji})  \chi_j^++H.c.,
\eeqn
where
\beqn
\alpha^S_{ji}=\frac{1}{2}(\xi'_{ji}+\delta \xi'_{ji})\sin\beta+\frac{1}{2}\Delta \xi'_{ji}\cos\beta\nonumber\\
+\frac{1}{2}(\xi_{ji}+\delta \xi_{ji})\cos\beta+\frac{1}{2}\Delta \xi_{ji}\sin\beta,\nonumber\\
\alpha^P_{ji}=\frac{1}{2}(\xi'_{ji}+\delta \xi'_{ji})\sin\beta+\frac{1}{2}\Delta \xi'_{ji}\cos\beta\nonumber\\
-\frac{1}{2}(\xi_{ji}+\delta \xi_{ji})\cos\beta-\frac{1}{2}\Delta \xi_{ji}\sin\beta.
\eeqn
From the above Lagrangian one can write down the decay rate of the charged Higgs into charginos and neutralinos.
\beqn
\Gamma_{ji} (H^-\rightarrow \chi^0_j\chi^-_i)\nonumber\\
=\frac{1}{4\pi M^3_{H^-}}
\left(
(m^2_{\chi^0_j}+m^2_{\chi^+_i}-M^2_{H^-})^2-4m^2_{\chi^+_i}m^2_{\chi^0_j} \right)^{\frac{1}{2}}\nonumber\\
\times (0.5(|\alpha^S_{ji}|^2+|\alpha^P_{ji}|^2)(M^2_{H^-}-m^2_{\chi^0_j}-m^2_{\chi^+_i})\nonumber\\
-0.5  (|\alpha^S_{ji}|^2-|\alpha^P_{ji}|^2)(2m_{\chi^+_i}m_{\chi^0_j}))~~~
\eeqn
The charged Higgs  decays are  found to be more sensitive to the
phases that enter  both at the tree level as well as at  the   loop level such as $\theta_{\mu}$
\cite{Ibrahim:2004cf}
relative to the phases  such as $\alpha_A$ which enter only at the loop level.

\subsection{Effect of CP phases on neutralino dark matter}
\label{kG}

If the lightest neutralino is the LSP then with R parity invariance it is a possible candidate
for cold  dark matter,  and in this case the relic density
\cite{Falk:1995fk,Falk:1998pu,Chattopadhyay:1998wb,Gomez:2005nr}
as well as  the rates in experiments
to detect neutralinos will be affected by the presence of CP phases
\cite{Chattopadhyay:1998wb,Falk:1998xj,Falk:1999mq}.
 We give a brief
discussion of neutralino dark matter and highlight the effects of
CP on neutralino dark matter analyses.  A quantity of interest in
experimental measurements is $\Omega_{dm} h_0^2$ where
$\Omega_{dm}=\rho_{dm}/\rho_c$, where $\rho_{dm}$ is the dark
matter density, and $\rho_c$ is the critical matter density needed
to close the universe where \beqn \rho_c= 3H_0^2/8\pi G_N \sim 1.88
\times h_0^2 \times 10^{-29}  gm/cm^3. \eeqn Here $H_0$ is the
Hubble constant, and $h_0$ is its value in units of 100km/sec.Mpc,
and $G_N$ is the Newtonian constant.  The current limit  from WMAP3
on cold dark matter is \cite{Spergel:2006hy} \beqn
\Omega_{cdm}h^2=0.1045^{0.0072}_{-0.0095}. \eeqn

\begin{figure}
\includegraphics*[width=6cm, height=6cm]{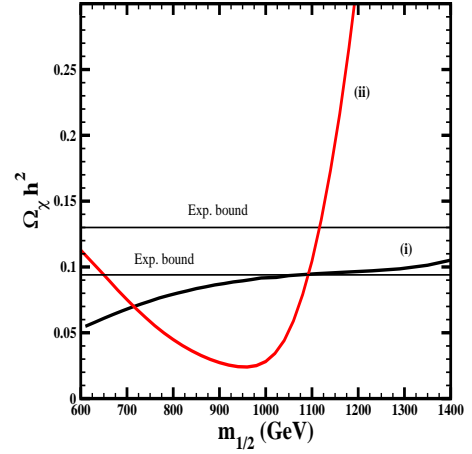}
\caption{An exhibition of the satisfaction of the relic density
constrains with large phases from the analysis of
\cite{Gomez:2005nr}. The curve labeled (i) is for the case
$m_0=1040$, $|A_0|=0$, $\tan\beta=40$ , $\theta_{mu}=2.9$,
$\alpha_A=0$,$\xi_1=1.0$, $\xi_2=0.15$, $\xi_3=0.5$, while the curve
labeled (ii) corresponds to $m_0=1080$, $|A_0|=0$, $\tan\beta=40$,
$\theta_{\mu}=0.6$, $\alpha_A=0$,$\xi_1=0.5$, $\xi_2=-0.6$,
$\xi_3=1.6$.  For case (i) EDM constraints are satisfied when
$m_{1/2}=1250$ and for case (ii) they are satisfied when
$m_{1/2}=1100$. All masses are in units of GeV and all angles in
radians.} \label{cprelic1}
\end{figure}

In the Big Bang scenario the neutralinos will be produced at the
time of the Big Bang and will be in thermal equilibrium with the
background till the  time of freeze out when they will  go out of
equilibrium. The procedure for computation of the  density of the
relic neutralinos is well known using the Boltzman equations.  In
general the analysis  will  involve co-annihilations and one will
have  processes of the type
 \beqn
\chi_i^0+\chi_j^0~~\to~~ f\bar f,\, WW,\, ZZ,\, WH,\, \cdots~ \ ,
\label{coann} \eeqn Additionally co-annihilations with staus,
charginos, and other sparticle species can also contribute.   Thus
the relic  density of neutralinos $n=\sum_i n_i$     is governed by
the Boltzman equation\cite{Griest:1990kh,Gondolo:1990dk,Lee:1977ua}
\beqn \frac{dn}{dt} = -3Hn -\sum_{ij} \langle \sigma_{ij}v\rangle
(n_in_j-n_i^{\rm eq}n_j^{\rm eq}) \ , \eeqn Here
 $\sigma_{ij}$ is the cross-section for annihilation of
particle types  $i,j$, and $n_i^{\rm eq}$ the number density of
$\chi^0_i$ in thermal equilibrium.  Under the  approximation $n_i/n=n_i^{\rm
eq}/n^{\rm eq}$ one has the well known result
 \beqn
\frac{dn}{dt} =-3nH -\langle \sigma_{\rm eff}\rangle(n^2-(n^{\rm
eq})^2)\ , \eeqn
where $ \sigma_{\rm eff}=\sum_{i,j}
\sigma_{ij}\gamma_i\gamma_j$, and  $\gamma_i$ are the Boltzman
suppression factors $ \gamma_i=\frac{n_i^{\rm eq}}{n^{\rm eq}}$.
Explicitly one finds that the freeze-out temperature is given
by
 \beqn x_f= \ln\left[ x_f^{-\frac{1}{2}} \langle\sigma_{\rm
eff} v\rangle_{x_f} m_{1} \sqrt{\frac{45}{8\pi^6N_f
G_N}}\hspace{.2cm}\right] \ , \eeqn where $N_f$ is the number of
degrees of freedom at freeze-out and $G_N$ is Newton's constant. The
relic abundance  of neutralinos  at current temperatures is then
given by \beqn \Omega_{\chi^0} h_0^2 = \frac {1.07\times 10^9
\rm{GeV}^{-1}}{N_f^{\frac{1}{2}} M_{\rm Pl} }
\left[\int_{x_f}^{\infty} \langle \sigma_{\rm eff} v\rangle
\frac{dx}{x^2}\right]^{-1}\ . \eeqn Here $x_f={m_1}/{T_f}$, $T_f$ is
the freeze-out temperature, $M_{\rm Pl} =1.2\times 10^{19}$ GeV, and
$<\sigma v>$ is the thermal average of $\sigma v$ so that \beqn
<\sigma v>=\int_{0}^{\infty} dv v^2 (\sigma v)
e^{v^2/4x}/\int_{0}^{\infty} dv v^2 e^{v^2/4x}. \eeqn 
The diagrams
that contribute to $<\sigma v>$ include the s channel $Z$, $h$,
and $A^0$ poles, and  the $t$ and $u$ channel  squark and slepton
exchanges. The Higgs boson, and sparticle masses are affected by the
CP phases of the soft parameters. Further, the vertices are also
affected.  Inclusion of the loop corrections to the vertices further
enhances the dependence on phases \cite{Gomez:2004ek}. Specifically
the Yukawa couplings of bottom quark and neutral Higgs bosons  are
found to be sensitive to $\xi_3$  if one includes SUSY QCD
corrections in the analysis.
 A detailed analysis to study the sensitivity of dark matter to the $b$ quark mass and to the neutral
 Higgs boson mixings is given  in \cite{Gomez:2004ek}.
It is found that the relic density is very sensitive to the mass of the $b$ quark for large $\tan\beta$
and consequently also to the CP phases since the b quark mass  is sensitive to the phases.
In Fig.(\ref{cprelic1}) we give an exhibition of the relic density and its sensitivity to phases.
In the analysis presented in Fig.(\ref{cprelic1}), the relic density was satisfied due to the
annihilation through resonant Higgs poles, and one observes the  sensitivity of the relic density to CP violating phases. The analysis of the relic density with inclusion of Yukawa unification constraint with inclusion
of CP phases is given in \cite{Gomez:2005nr}. An analysis of relic density in the presence of
CP phases is also given in  \cite{Argyrou:2004cs,Falk:1998pu,Belanger:2006qa,Nihei:2004bc}.

Typical dark matter  experiments
involve scattering of neutralinos of the Milky Way
that reside  in our  vicinity  with target nuclei.
The basic lagrangian that governs such scattering is the neutralino-quark scattering with
neutralino and quarks in the initial and final states.
The relative velocity of the LSP hitting the target is small, and so, one can approximate the effective interaction governing the neutralino-quark scattering by an effective four-fermi interaction
\beqn
{\cal L}_{eff}=\bar{\chi}\gamma_{\mu}\gamma_5\chi\bar{q}\gamma^{\mu}(AP_L+BP_R)q
+C\bar{\chi}\chi m_q\bar{q}q\nonumber\\
+D\bar{\chi}\gamma_5\chi m_q\bar{q}\gamma_5q
+E\bar{\chi}i\gamma_5\chi m_q\bar{q}q +F\bar{\chi}\chi m_q
\bar{q}i\gamma_5 q~ \label{4fermi}
\label{darkz}
\eeqn
The deduction of Eq.(\ref{darkz}) requires Fierz rearrangement which is discussed in
Sec.(\ref{qH}) and further  details are  given in Sec.(\ref{qI}).
 The first two terms $A, B$ in Eq.(\ref{darkz}) 
are spin-dependent interaction and arise from the $Z$ boson and the
sfermion exchanges. The effect of CP violating phases enter via the
neutralino eigen vector components and the matrix $D_{\tilde{q}}$
that diagonalizes the squark mass matrix. Then the phases that play
a role here are $\theta_{\mu}$, $\xi_1$, $\xi_2$ and $\alpha_{A_q}$.
The $C$ term represents the scalar interaction which gives rise to
coherent scattering. It receives contributions from the sfermion
exchange, and from the exchange of the neutral Higgs $H_i$ mass
eigenstates. The term $D$ is non vanishing in the limit when CP
phases vanish. However,  this term is mostly ignored in the
literature as its contribution is suppressed because of the small
velocity of the relic neutralinos. In fact the contributions of $D,
E$ and $F$ are expected to be relatively small and could be ignored.
A significant body of work exists on the analysis of detection rates
in the absence  of CP
phases \cite{Nath:1994ci,Arnowitt:1995vg,Nath:1997qm}, but much less
so with inclusion of CP phases.  Inclusion of the CP phases shows a
very significant effect of CP phases on the detection rates
\cite{Chattopadhyay:1998wb,Falk:1999mq}.
The CP effects can be significant even with inclusion of the EDM
constraints\cite{Gomez:2004yv,Nihei:2004bc}.

\subsection{Effect of CP phases  on proton stability}
\label{kH}

CP violating phases can affect the nucleon stability in supersymmetric grand unified models with baryon and lepton number
 violating dimension five operators (\cite{Weinberg:1981wj,Sakai:1981pk}. For  a recent review see \cite{Nath:2006ut}).
 Thus in a wide class of unified models including grand unified models, string and  brane models baryon and
lepton number violation arises  via dimension  LLLL and RRRR chiral operators of the form
 \beqn
{\cal L}_{5L}= \frac{1}{M} \epsilon_{abc}(Pf_1^uV)_{ij}(f_2^d)_{kl}
( \tilde u_{Lbi}\tilde d_{Lcj})\nonumber\\
(\bar e^c_{Lk}(Vu_L)_{al}
-\bar \nu^c_kd_{Lal})+..
+H.c~~.~~
\label{llll}
\eeqn
\beqn
 {\cal L}_{5R}= -\frac{1}{M} \epsilon_{abc}(V^{\dagger} f^u)_{ij}(PVf^d)_{kl}
(\bar e^c_{Ri}u_{Raj}\tilde u_{Rck}\tilde d_{Rbl}+\nonumber\\..
+H.c.~~~~~
\label{rrrr}
\eeqn
Here ${\cal L}_{5L}$ in the LLLL and ${\cal L}_{5R}$ is the RRRR
lepton and baryon number violating dimension 5 operators,
 V is the CKM matrix and $f_i$ are  related to quark masses,
and $P_i$ appearing in Eqs(\ref{llll}) and (\ref{rrrr}) are the generational phases
 given by $P_i=(e^{i\gamma_i})$ with the constraint  $\sum_i \gamma_i=0$  (i=1,2,3).

Using the above one  generates  the baryon and the lepton number
violating dimension six operators by dressing the dimension five
operators by the chargino, the gluino and the neutralino exchanges.
 The  dressing loops contain
the CP phases  both via the sparticle  spectrum as  well as via the vertices.
This can be explicitly seen by elimination the sfermion fields above via the relations

\beqn
\tilde u_{iL}=2\int [\Delta_{ui}^L L_{ui}+\Delta_i^{LR} R_{ui}]\nonumber\\
\tilde u_{iR}=2\int [\Delta_{ui}^R R_{ui}+\Delta_i^{RL} L_{ui}]
\eeqn where $L_{ui}=\delta \it L_I/\delta \tilde u_{iL}^{\dagger}$,
$R_{ui}= \delta \it L_I/\delta \tilde u_{iR}^{\dagger}$. Here $L_I$
is the sum of fermion-sfermion-gluino, fermion-sfermion-chargino and
fermion-sfermion-neutralino and  $\Delta's$ are the propagators.   A
detailed analysis of the specific mode $p\to \bar \nu K^+$ which is
typically the  dominant mode in supersymmetric decay modes of the
proton is then given by the following with the inclusion of CP
phases

\beqn
\Gamma(p\rightarrow\bar\nu_iK^+)=\frac{\beta_p^2m_N}{M_{H_3}^232\pi f_{\pi}^2}
(1-\frac{m_K^2}{m_N^2})^2\nonumber\\
|{\cal A}_{\nu_iK}|^2 A_L^2(A_S^L)^2\nonumber\\
|(1+\frac{m_N(D+3F)}{3m_B})(1+{\cal Y}_i^{tk}+(e^{-i\xi_3}{\cal Y}_{\tilde g}+
{\cal Y}_{\tilde Z})\delta_{i2}+\nonumber\\
\frac{A_S^R}{A_S^L}{\cal Y}_1^R\delta_{i3})
+\frac{2}{3}\frac{m_N}{m_B}D(1+{\cal Y}_3^{tk}-
(e^{-i\xi_3}{\cal Y}_{\tilde g}-{\cal Y}_{\tilde Z})\delta_{i2}\nonumber\\
+\frac{A_S^R}{A_S^L}{\cal Y}_2^R \delta_{i3})|^2~~~
\eeqn
where
\beqn
{\cal A}_{\nu_iK}=(\sin 2\beta M_W^2)^{-1}\alpha_2^2P_2m_cm_i^dV_{i1}^{\dagger}
V_{21}V_{22}\nonumber\\
({\cal F}(\tilde c; \tilde d_i; \tilde W)+
 {\cal F}(\tilde c; \tilde e_i ; \tilde W)\nonumber\\
\eeqn
In the above $A_L(A_S)$ are the long (short) suppression factors,
D,F, $f_{\pi}$ are the effective Lagrangian parameters, and
$\beta_p$ is defined by
$\beta_p U_L^{\gamma}=\epsilon_{abc}\epsilon_{\alpha \beta} <0|d_{aL}^{\alpha}
u_{bL}^{\beta}u_{cL}^{\gamma}|p>$ where $U_L^{\gamma}$ is the
proton wavefunction. Theoretical determinations of $\beta_p$ lie in the
range $0.003-0.03~GeV^3$.
Perhaps the more reliable estimate is from lattice gauge calculations
which gives\cite{Tsutsui:2004qc}
$|\beta_p|=0.0096(09)(^{+6}_{-20})$ GeV$^3$.\\

  CP violating phases of the soft SUSY breaking
 sector enter in the proton decay amplitude. The CP phases enter the
dressings in two ways,  via the mass matrices of the charginos, the
neutralinos and the sfermions, and via the interaction vertices.
Taking account of this additional complexity, the analysis for
computing the proton decay amplitudes follows the usual procedure.
This effect is exhibited by considering $R_{\tau}$ \beqn
R_{\tau}=\frac{\tau(p\rightarrow \bar{\nu}+K^+)}{\tau_0(p\rightarrow
\bar{\nu}+K^+)} \eeqn where $\tau(p\rightarrow \bar{\nu}+K^+)$ is
the proton lifetime with CP violating phases and
$\tau_0(p\rightarrow \bar{\nu}+K^+)$ is the lifetime without CP
phases. This ratio is largely model independent. All the model
dependent features would be contained mostly in the front factors
which cancel out in the ratio. Since the dressing loop integrals
enter in the proton decay lifetime in GUTs which contain the baryon
and the lepton number violating dimension five operators, the
phenomena of CP violating effects on the proton life time should
hold for a wide range of models of GUTs. The baryon and lepton
number violating operators must be dressed by the chargino, the
gluino and the neutralino exchanges to generate effective baryon and
lepton number violating dimension six operators at low energy. These
dressing loops have vertices of quark-squark-chargino,
quark-squark-neutralino and quark-squark-gluino. From this structure
one can read the phases that might enter the analysis. The chargino
one has the phases of $\theta_{\mu}$, $\alpha_{A_q}$ and $\xi_2$.
The neutralino vertex has beside the above set, an extra phase
$\xi_1$. The gluino vertex has the set of $\theta_{\mu}$,
$\alpha_{A_q}$, $\xi_3$. Following the standard procedure
\cite{Nath:1985ub,Weinberg:1981wj} one can obtain the effective
dimension six operators for the baryon and lepton violating
interaction arising from dressing of the dimension five operators.
By doing so and estimating $R_{\tau}$, one  finds that this ratio is
a sensitive function of CP phases \cite{Ibrahim:2000tx}.
Modifications of the proton lifetime by as much as a factor of 2 due
to the effects of the CP violating phases can occur. It is found
also that the CP phase effects could increase or decrease the proton
decay rates and that the size of their effect depend highly on the
region of the parameter space one is in.
 \begin{figure}
 \hspace{-1.0cm}
\includegraphics*[width=8cm, height=5cm]{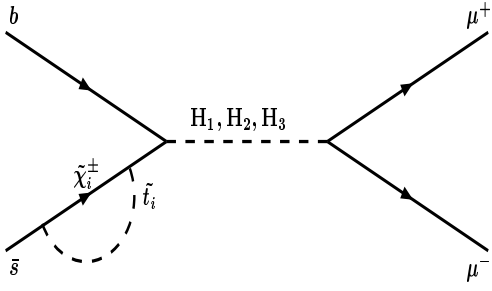}
\caption{
The counter term diagram which produces the leading term in amplitude proportional to
$\tan^3\beta$  in the branching ratio $B^0_{s}\to l^+l^-$}
\label{counterterm}
\end{figure}

\subsection{SUSY CP phases and  the decay $B^0_s\rightarrow \mu^+\mu^-$.}
\label{kI}

The branching ratio of the rare process $B^0_s\rightarrow \mu^+\mu^-$ is another area where CP violating phase effects arise. It is known that the standard model value is rather small while in supersymmetric models it can get three orders of magnitude larger for
large
 $\tan \beta$  \cite{Choudhury:1998ze,Babu:1999hn,Bobeth:2001sq,Chankowski:2001uz,Isidori:2001fv,Huang:2000sm,Buras:2002wq,Xiong:2002up,Dedes:2002zx,Arnowitt:2002cq,Baek:2002wm,Mizukoshi:2002gs}.\\

\begin{figure}
 \hspace{-0.5cm}
\includegraphics*[width=8cm, height=8cm]{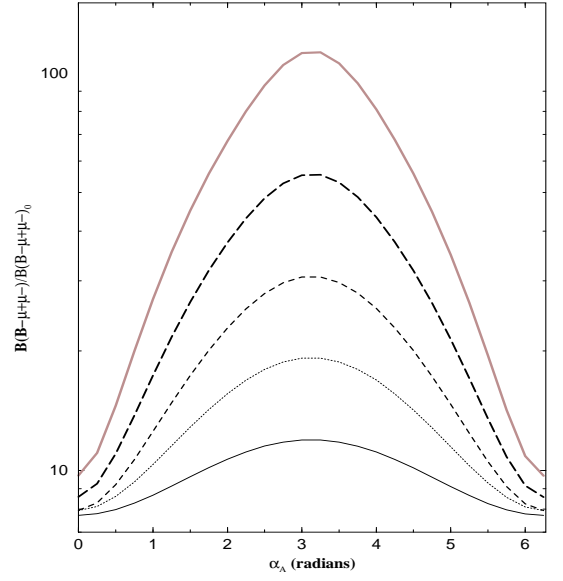}
\caption{An exhibition of  the strong dependence on $\alpha_A$ of  the ratio of the branching
ratios $B(B^0_s\rightarrow \mu^+\mu^-)/B(B^0_s\rightarrow \mu^+\mu^-)_0$,
where $B(B^0_s\rightarrow \mu^+\mu^-)_0$ is the branching ratio when all phases
are set to zero \cite{Ibrahim:2002fx}. The curves in ascending order are  for values
of $|A_0|$ of 1,2,3,4,5.
The other parameters are $m_0=200$ GeV, $m_{1/2}=200$ GeV,
 $\tan\beta =50$, $\xi_1=\xi_2=\pi/4$, $\xi_3=0$, and $\theta_{\mu}=2$.}
\label{Bdecay}
\end{figure}

Detecting such large values of $B^0_s$ would be a positive test for
SUSY even before any sparticles are found. This decay is governed by
the effective Hamiltonian \beqn
H_{eff}=-\frac{G_Fe^2}{4\sqrt{2}\pi^2}V_{tb}V^*_{td'}\nonumber\\
\times (C_S O_S + C_P O_P+C'_S O'_S+C'_P O'_P+C_{10} O_{10})_Q~
\eeqn where $C's$ are the coefficients of the Wilson operators $O's$
defined by \beqn O_S=m_b(\bar{d'}_{\alpha}P_Rb_{\alpha})\bar{l}l,
~O_P=m_b(\bar{d'}_{\alpha}P_Rb_{\alpha})\bar{l}\gamma_5 l,\nonumber\\
O'_S=m_{d'}(\bar{d'}_{\alpha}P_Lb_{\alpha})\bar{l}l,
~O'_P=m_{d'}(\bar{d'}_{\alpha}P_Lb_{\alpha})\bar{l}\gamma_5 l,\nonumber\\
O_{10}=(\bar{d'}_{\alpha}\gamma^{\mu}P_Lb_{\alpha})\bar{l}\gamma_{\mu}\gamma_5l,~~
\eeqn and $Q$ is the scale where the coefficients are evaluated. The
branching ratio is a function of the coefficients $C_{S,P}$ and
$C'_{S,P}$. In the counter term diagram (see Fig.(\ref{counterterm})
which contributes to this ratio one can find vertices of  $\bar{b}
bH_i$, $\bar{s}\chi^-\tilde{t}$ and $\bar{\mu}\mu H_i$. The first
two vertices are sensitive functions of the CP violating phases as
was explained in the different applications above. The phases that
play a major role here are $\theta_{\mu}$, $\xi_2$ and
$\alpha_{A_q}$. Gluino and neutralino exchange diagrams  also
contribute  which brings a dependence on additional phases $\xi_1$
and $\xi_3$.  Inclusion of these \cite{Ibrahim:2002fx} shows  that
the branching ratio can vary in some parts of the parameter space by
up to 1-2 orders of magnitude due to the effect of  CP phases. A
demonstration of the strong effect of the phases on B decay
branching ratio is given in Fig.(\ref{Bdecay}).
An analysis of this process using the so called resummed effective lagrangian approach for
Higgs mediated interactions in the CP violating MSSM is given in \cite{Dedes:2002er}.

\subsection{CP effects on squark decays}
\label{kJ}

The interactions of $\bar{q}\tilde{q'_i}\chi^+_j $ and
$\bar{q}\tilde{q_i}\chi^0_j$ do have CP violating phases at the tree
level. These interactions are important for squark decays into
fermions and such decays are expected to show up in the Large Hadron
Collider when squarks become visible. The Lagrangian that governs
the squark decays is given by \beqn {\cal
L}=g\bar{t}(R_{bij}P_R+L_{bij}P_L)\tilde{\chi^+}_j\tilde{b}_i
+g\bar{b}(R_{tij}P_R+\nonumber\\
L_{tij}P_L)\tilde{\chi^c}_j\tilde{t}_i
+g\bar{t}(K_{tij}P_R+M_{tij}P_L)\tilde{\chi^0}_j\tilde{t}_i\nonumber\\
+g\bar{b}(K_{bij}P_R+M_{bij}P_L)\tilde{\chi^0}_j\tilde{b}_i+H.c.
\eeqn
where
\beqn
\kappa_{t(b)}=\frac{m_{t(b)}}{\sqrt{2}m_W\sin\beta(\cos\beta)}
\eeqn
and where
\beqn
L_{bij}=\kappa_tV^*_{j2}D_{b1i}\nonumber\\
R_{bij}=-(U_{j1}D_{b1i}-\kappa_bU_{j2}D_{b2i})\nonumber\\
K_{bij}=-\sqrt{2}[\beta_{bj}D_{b1i}+\alpha^*_{bj}D_{b2i}]\nonumber\\
M_{bij}=-\sqrt{2}[\alpha_{bj}D_{b1i}-\gamma_{bj}D_{b2i}]
\eeqn
The corresponding quantities with subscript $t$ can be  obtained by the
substitution $b\to t$, $U \leftarrow\rightarrow  V$.
 \begin{figure}
\hspace{-2.0cm}
\includegraphics*[width=11cm, height=12cm]{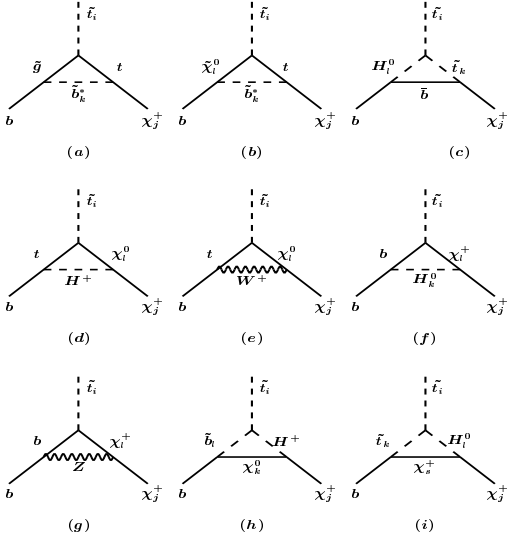}
\vspace{-3.0cm}
\caption{A sample of one loop diagrams with CP phase dependent vertices
that contribute to the decay of the stops. }
\label{stopdiagrams}
\end{figure}
The couplings $R$ and $L$ are functions of the phases
$\theta_{\mu}$, $\xi_2$ and $\alpha_{A_q}$. The set of phases that
enter the couplings $K$ and $M$ is the same above set with an extra
phase $\xi_1$. The loop corrections produce shifts in the couplings
as follows

\beqn
{\cal L}_{eff}=g\bar{t}(\tilde R_{bij}P_R+ \tilde  L_{bij}P_L)\tilde{\chi^+}_j\tilde{b}_i\nonumber\\
+g\bar{b}(\tilde R_{tij}P_R+\tilde  L_{tij} P_L)
\tilde{\chi^c}_j\tilde{t}_i
+g\bar{t}(\tilde K_{tij}P_R+ \tilde M_{tij}P_L)\nonumber\\
\tilde{\chi^0}_j\tilde{t}_i
+g\bar{b}( \tilde K_{bij} P_R+\tilde M_{bij} P_L)
\tilde{\chi^0}_j\tilde{b}_i
+H.c.~~~~
\eeqn
where $\tilde R_{bij} =R_{bij}+\Delta R_{bij}$ where $\Delta R_{bij}$ is the loop correction
and other tilde are similarly defined.
The loops that enter the analysis of $\Delta$'s have gluinos, charginos, neutralinos, neutral Higgs, charged Higgs,
squarks, $W$ and $Z$ boson exchanges.   The  masses of sparticles as well as the vertices where they enter are sensitive
to the CP phases.
The analysis using the loop corrected lagrangian enhances the CP dependence of the  masses and the vertices that already
appear at the tree level.  Recent analyses of stop and sbottom decays can be found in
 \cite{Bartl:2003pd,Bartl:2003he,Ibrahim:2004gb}

\subsection{$B\to \phi K$ and CP asymmetries}
\label{kK}

Like $B\to X_s +\gamma$, the decay $B\to \phi K_S$ has no tree level contribution and proceeds
only via loop corrections. Thus the process presents a  good  testing ground for  new physics since new physics
also enters at the loop level.  An interesting phenomenon concerns the fact that in the SM, the
CP asymmetry predicted for $B\phi K_S$ is the same as in $B\to J/\Psi K_s$ to  $O(\lambda^2)$\cite{Grossman:1996ke}.\\

The current value of  the $B\to J/\Psi K_S$ experimentally is
\beqn
S_{J/\Psi K_s} = 0.734\pm 0.055
\eeqn
which is in excellent agreement with SM prediction of $\sin 2\beta =0.715^{+0.055}_{-0.045}$.
Although currently the experimental value for $S_{\phi K_S}$ \cite{Aubert:2004dy}
\beqn
S_{\phi K}= 0.50\pm 0.25 (stat) ^{+0.07}_{-0.04} (syst)
\eeqn
is consistent within $1\sigma$ of the SM prediction, its value has significantly in the past  showing
a $2.7\sigma$ deviation from the SM prediction,
which triggered much theoretical activity to explain the large
deviation \cite{Arnowitt:2003ev,Cheng:2003im,Chakraverty:2003uv,Kane:2003zi,Agashe:2003rj,Kundu:2003tz,Chiang:2003jn,Baek:2003kb,Khalil:2002fm,Dutta:2002ah,Ciuchini:2002pd,Datta:2002nr,Hiller:2002ci}.\\

Although the discrepancy has
largely disappeared it  is still instructive to review briefly the possible processes that could make a
large contribution to the $B\to \phi K_S$ process.  It should be noted  that the branching ratio
$BR(B^0\to \phi K_S)= (8.0\pm 1.3)\times 10^{-6}$ is quite consistent with the SM result.

The time dependent asymmetries in $B\to \phi K_S$ are defined so that
\beqn
{\cal A}_{\phi K}(t) = \frac{\Gamma(\bar B (t) \to \phi K_S)-\Gamma(B(t)\to \phi K_S}
{\Gamma(\bar B (t) \to \phi K_S) + \Gamma(B(t)\to \phi K_S}\nonumber\\
= -C_{\phi K} \cos(\Delta m_B t) + S_{\phi K} \sin (\Delta m_B t)
\eeqn
where $S_{\phi K_S}$ and $C_{\phi K_S}$ are given by
\beqn
C_{\phi K_S} = \frac{1+ |\lambda_{\phi K_S}|^2}
{1+ |\lambda_{\phi K_S}|^2}, ~
 S_{\phi K_S} = \frac{2 Im \lambda_{\phi K_S}}{1+ |\lambda_{\phi K_S}|^2}
\eeqn
where $\lambda_{\phi K_S}$ is defined by

\beqn
\lambda_{\phi K_S} = -e^{-2i(\beta +\delta \beta)} \frac{ \bar A( \bar B^0 \to \phi K_S)} {A(B^0 \to \phi K_S)}
\eeqn
where $\beta$ is as defined in the SM, and $\delta  \beta$ is any possible new physics contribution.
Much of the work in trying to produce large  effects within supersymmetric models has focussed  on
generating corrections from flavor mixing in the quark sector using the  mass insertion method\cite{Hall:1985dx,Gabbiani:1996hi}.
\\

 Thus, for example,
an LL type mass insertion in the down quark sector will have the
form \beqn (\delta_{LL}^d)_{ij} = \left(V_L^{d\dagger} (M_d)^2_{LL}
V_L^d\right)_{ij} \eeqn Here $(M_d^2)_{LL}$ is the LL down squark
mass  matrix, $V_L^d$ is the rotation matrix that diagonalizes the
down squark mass matrix, and $\tilde m$ is the average  squark mass.
Similarly one defines the mass insertions $(\delta_{RR}^d)_{ij}$,
$(\delta_{LR}^d)_{ij}$ and  $(\delta_{RL}^d)_{ij}$.  Among the
supersymmetric contributions considered are the gluino-mediated
$b\to sq\bar q$ with $q=u,d,s,c,b$ and Higgs  mediated $b\to ss\bar
s$. Typically it is found that the  LL and RR insertions give too
small an effect but chirality flipping $LR$ and $RL$ insertions can
generate sizable corrections to $B\phi K_S$. Thus, for example,
$|(\delta_{LR}^d)_{23}|\leq 10^{-2}$ can significantly affect   $B\to
\phi K_S$ while the constraints on $B\to X_s \gamma$ and $\Delta
M_s$ are obeyed. The analysis in $B\to \phi K_S$ in supergravity
grand unification with inclusion of CP phases was  carried out  by
\cite{Arnowitt:2003ev} and it was concluded that significant
corrections to the asymmetries can arise with inclusion in the
trilinear soft parameter $A$ with mixings in the second and third
generations either in the up sector or in the down sector.  A
similar analysis of  asymmetries in  $B\to \eta' K$ have also been
carried out  by \cite{Gabrielli:2004yi}.

\subsection{T and CP odd operators and their observability at colliders}
\label{kL}

In the previous sections we have discussed the effects of CP violation on several phenomena.
The list of CP odd or T odd (assuming CPT invariance) is rather large (For a sample, see,
  \cite{Kane:1991bg,Valencia:1994zi,Bernreuther:1991gh,DeRujula:1990db}.).
We discuss briefly now the possibilities for the observation of CP
in collider experiments. First we note that CP phases affect decays
and scattering cross sections in two different ways. Thus in
addition to generating a CP violating contribution to the
amplitudes, they also affect the CP even part of the amplitudes
which can affect the over all magnitude of decay widths and
scattering cross sections.  However, definite tests of CP violation
can arise only via the observation of T odd or CP odd  parts.  As an
example of the size of the effects induced  by CP odd operators in
supersymmetry on cross sections consider the process $e^+e^-\to
t\bar t$. Here an analysis in MSSM including loop effects with CP
phases gives \cite{Christova:1992ee} \beqn \frac{d\sigma}{d\Omega} =
\frac{d\sigma_0^{t\bar t}}{d\Omega} \left( 1+ c \frac{\alpha_s}{\pi}
\sin(\alpha_{A_t}-\phi_{\tilde g}) \frac{(\vec J.\vec p\times \vec
k)}{|\vec p\times \vec k|}\right) \eeqn where  $\vec k$ ($\vec p$)
are the center of mass momentum of one of the initial  (final)
particles and $\vec J$ is the unit  polarization vector of one the
produced t quarks perpendicular to the production plane. c depends
on the details of the sparticle spectrum and can vary significantly
depending on the sparticle spectrum. The choice $c\sim .1$ give the
correction of the  T-odd
 observable to be  of size   $(10^{-1} \frac{\alpha_s}{\pi}) $ which is typically of the same size
 as the radiative corrections from the Standard Model.  More generally in $e^+e^-$ colliders
 in the process $e^+e^-\to X$ with momenta $\vec p_1, \vec p_2, \vec p$
 a product of the type $(\vec \xi_i\times \vec \xi_j).\vec \xi_k$
 where $\xi_i$ is either a momentum or a polarization will give a T-odd observable. For
 example one will have   T-odd operators of the type \cite{Gavela:1988jx}
  \beqn
 {\cal T}_1= (\vec p_1\times \vec p_2). \vec S_{e^-},\nonumber\\
  {\cal T}_2= \vec p.(\vec S_{e^-}\times \vec S_{e^+})
 \eeqn
 More  generally with several particles  $(i=1,..,n, n>4)$ one can form T odd operator
 such as
 \beqn
 \epsilon_{\alpha\beta\gamma\delta}p^{\alpha}_ip^{\beta}_jp^{\gamma}_k p^{\delta}_l.
 \eeqn
 An example of such an operator is the squark decay  $\tilde t \to t+l^+l^-+\chi_1^0$
 which can also lead to an observable signal at the LHC
  \cite{Langacker:2007ur}.  A study of the  effects of  CP-violating phases of
                  the MSSM on  leptonic high-energy observables is given in  \cite{Choi:2004rf}.
 An efficient way to observe CP violation is via use of polarized beams in $e^+e^-$
 colliders which is of interest in view of the proposed International Linear collider.
 A discussion on tests  of  supersymmetry at linear colliders can be found in \cite{Baer:2003ru}
  and a detailed discussion of tests  of CP asymmetries is   given in \cite{Moortgat-Pick:2005cw}.
  A number of  works related to the effects of CP on the  Higgs and sparticle phenomena discussed in this  section are
  \cite{Ghosh:2004cc,Akeroyd:2001kt,Alan:2007rp,Bartl:2006hh,Boz:2000wr,Cheung:2005pv,Hollik:1998wk,Hollik:1998vz,Accomando:2006ga,Heinemeyer:2004sv,Bartl:2003tr,Choi:2003pq,Bartl:2004vi}.  
    An interesting issue concerns the possibility of expressing CP odd quantities
  in terms of basis independent quantities for the supersymmetric case  similar to the Jarlskog invariant for the
 case of the Standard Model. Recent works in this direction can be found in \cite{Dreiner:2007yz,Lebedev:2002wq}.\\

Finally we note that the computation of SUSY phenomena with CP phases is more
difficult than computations without CP phases. In Sec.(\ref{qJ}) we give  a brief discussion
of the tools necessary for the computation of SUSY phenomena with CP phases.

\section{Flavor and CP phases}\label{m}
CP violation can influence flavor physics (for  recent reviews see
\cite{Bigi:2007sq,Fleischer:2006fx,Schopper:2006he}) and thus such
effects could be  used as probes of the SUSY CP violation effects.
This can happen in several ways.  This could happen in CP violation
effects in K and B physics,  or if EDMs of leptons are measured and
turn out to be in violation of scaling, and in possible  future
sparticle decays which may contain flavor dependent CP violating
effects.   Let us consider first CP violation in K and B physics.
Essentially all of the phenomena seen here can be explained in terms
of the CP violation with a Standard Model origin, i.e., arising from
the phase $\delta_{CKM}$. This means that unless  some deviations
from the Standard Model predictions are seen, the supersymmetric CP
violation must be small. On the other hand if significant deviations
occur from the Standard Model predictions then one would need in
addition to the large CP phases a new flavor structure.  An example
of this is
 flavor changing terms arising from the off diagonal component in the LR mass matrix
$(\delta_{ij})_{LR}(d)= (m^2_{LR}(d))_{ij}/\tilde m_q^2 $
\cite{Dine:1993np,Dine:2001ne,Masiero:1999ub,Khalil:1999ym,Demir:1999ky,Demir:1999qm}.\\

Further, if one adopts the view point that the entire CP phenomena
in the K and B system arise from the supersymmetric CP phases
\cite{Frere:1983aq,Brhlik:1999hs} then  one will need a new flavor
structure.  But such an assumption appears to be drastic since Yukawa
couplings arising from string  compactification will typically be
complex. However, there are other ways  in which CP violation can
act as strong probes of flavor physics and vice versa.  For
instance,  SUSY CP effects would be relevant in flavor changing
neutral current processes such as $b\to s+\gamma$ and in $\mu\to
e+\gamma$. Also if the EDM of the electron and the muon are
eventually determined and a scaling violation is found, then such
effects give us a connection between CP violation and flavor.
Similarly the connection between and CP and flavor can be obtained
from collider data in the decays of sparticles. In the following we
discuss two specific phenomena where CP and flavor affects can be
significant. Issue of flavor and CP violation is discussed in many
papers \cite{Demir:1999qm,Masiero:1999ub,Chang:2002mq,Ayazi:2007kd,Demir:2005ya}.
Additional
papers that discuss these issues
are\cite{Farzan:2007us,Ellis:2006mg,Pospelov:2006jm,Gronau:2006gn,Pospelov:2005ks}.
 CP and flavor violation in SO(10) is discussed in \cite{Chen:2004ww,Dutta:2005ue,Harvey:1980je,Babu:2004dp,Nath:2001uw}.

\subsection{$d_{\mu}$ vs $d_e$ and possible scaling violations}
\label{mA}

The EDM of the muon and the electron are essentially scaled by their masses, so that
\beqn
d_{\mu}/d_e \simeq m_{\mu}/m_e
\label{muEDM1}
\eeqn
The current experimental limits on the muon EDM are much less stringent than on the
electron EDM, and thus it is reasonable to ask if the EDM of the muon could be much
larger than the EDM of the electron.  If so the improved experiments on the muon EDM
may be able to detect it.  Thus we explore the  conditions under which significant
  violations of scaling may occur.  Now we recall from our discussion of the EDM of the
  electron, that large EDM for the electron generated by the chargino exchange
  may be canceled by the contributions from the neutralino exchange. Thus one
  possibility in generating  a large muon EDM is to upset this cancelation for the muon
  case. This appears possible by inclusion of flavor dependent nonuniversalities in
  the soft parameters. To make this idea concrete we consider the
  chargino and neutralino exchange contributions to a lepton EDM are

   \beqn
d_{\it l}=\frac{e\alpha_{EM}}{4\pi \sin^2\theta_W}
\frac{\kappa_{\it l}}{m_{\tilde \nu_{\it l}}^2}
\sum_{1=1}^{2}\tilde m_{\chi_i^+}Im(U^*_{i2}V^*_{i1})
A(\frac{\tilde m_{\chi_i^+}^2}{m_{\tilde \nu_{\it l}}^2})\nonumber\\
+\frac{e\alpha_{EM}}{4\pi \sin^2\theta_W}
\sum_{k=1}^{2}\sum_{i=1}^{4} Im (\eta_{ik}^{\it l})
\frac{\tilde m_{\chi_i^0}}{M_{\tilde {\it l}_k}^2}
Q_{\tilde {\it l}}B(\frac{\tilde m_{\chi_i^0}^2}{M_{\tilde {\it l}_k}^2})
\label{muEDM2}
\eeqn
  where A, B and $\kappa_{\it l}$ are defined earlier,
  and where $\eta_{ik}^{\it l}$ is given by
\beqn
\eta_{ik}^{\it l}=[-\sqrt 2\{\tan\theta_W(Q_{\it l}-T_{3\it l})X_{1i}
+T_{3\it l}X_{2i}\}D_{\it l1k}^*-\nonumber\\
-\kappa_{\it l} X_{3i}D_{\it l2k}^*]
(\sqrt 2 \tan\theta_W Q_{{\it l}} X_{1i}D_{{\it l} 2k}-\kappa_{{\it l}}
X_{3i}D_{{\it l}1k})~
\label{muEDM3}
\eeqn
Here
 $X$ diagonalizes the  neutralino matrix $M_{\chi^0}$, and
 $D_{\it l}$ diagonalizes the slepton (mass)$^2$ matrix.
The chargino exchange contribution depends on the single phase
combination $\xi_2+\theta_{\mu}$, while the neutralino exchange
contribution depends additionally on the phase combinations
$\theta_{\mu}+\xi_1$, and $\theta_{\mu}+\alpha_{A_{\tilde {\it
l}}}$. Non-universalities can be  introduced in two ways: via
sneutrino masses which enter in the chargino exchange, and via
slepton masses that enter in the neutralino exchange diagram. One
efficient way to introduce nonuniversalities in the slepton sector
is  via the trilinear coupling parameter $A_{\it l}$ which can be
 chosen to be flavor dependent at the GUT scale. In this case the cancelation
 in the electron EDM sector would not imply the same exact cancelation in the
 muon sector and significant violations of the scaling relation can be obtained.

\begin{figure}
\includegraphics*[angle=270, scale=0.35]{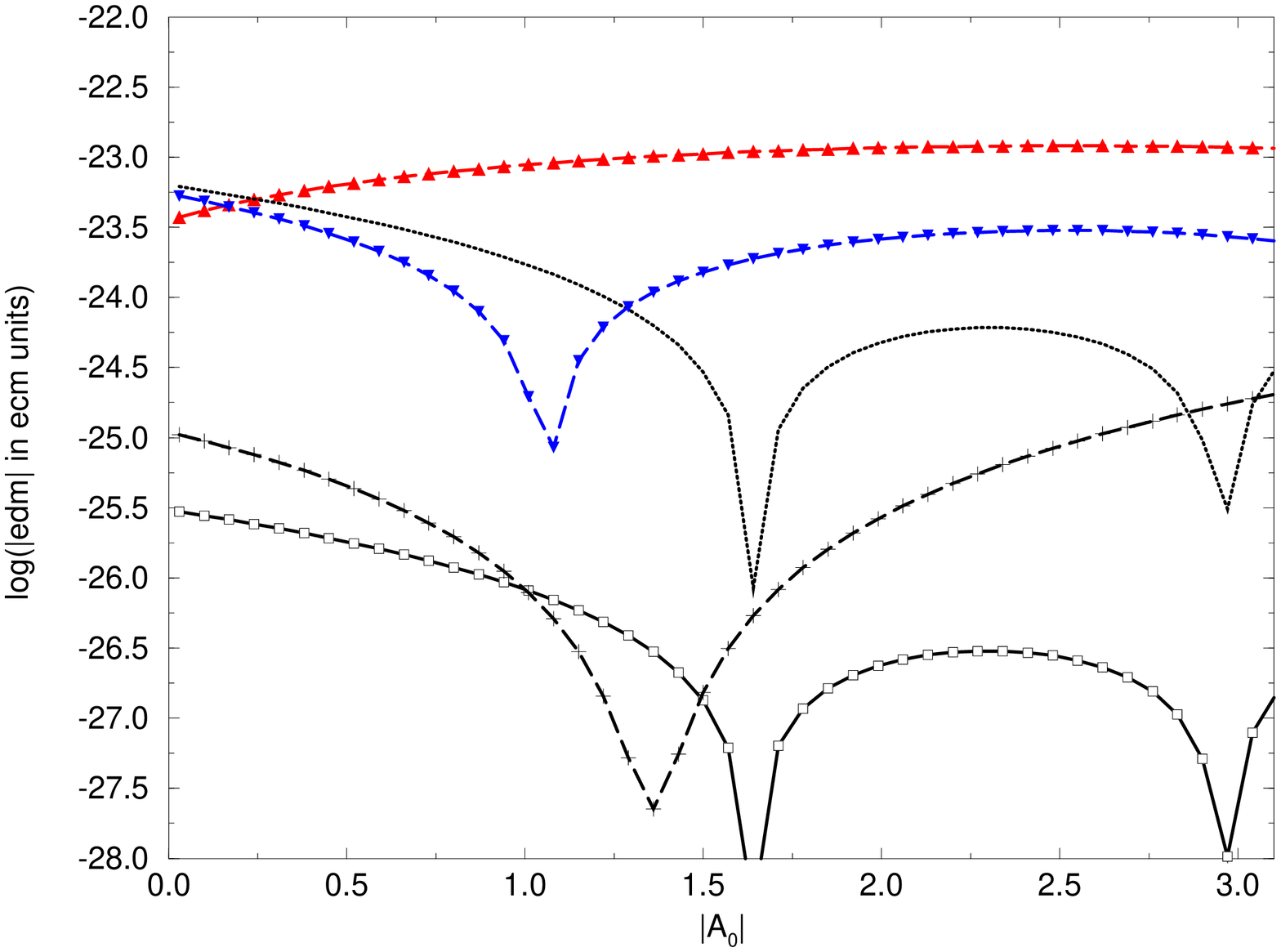}
\vspace{.5cm} \caption[]{An exhibition of the strong flavor
dependence via $A$  nonuniversalities in enhancing the muon EDM
relative to the electron EDM in the cancelation region from the
analysis of \cite{Ibrahim:2001jz}.
 Plotted are the electron EDM $d_e$
(solid line with squares),
 of the neutron EDM $d_n$ (dashed line with plus signs), and of the
 muon EDM $d_{\mu}$
as a function of
 $|A_0|$ for the case when $\tan\beta =20$,
$m_0=200$, $m_{\frac{1}{2}}=246$, $\xi_1=.28$,
$\xi_2=-.51$, $\xi_3=-.11$, $\theta_{\mu}=.4$ and $\alpha_{A_e}=1.02$
where all masses are in GeV.
The curve with dashed line with triangles pointed down is a  plot of the
 muon EDM $d_{\mu}$ which have all the same parameters as for
 $d_e$ and $d_n$ except that $\alpha_{A_{\mu}}=0.0$  and the
 curve with dashed line with triangles pointed up is a  plot of the
 muon EDM $d_{\mu}$ which have all the same parameters as for
 $d_e$ and $d_n$ except that $|A_{\mu}|= 6.0$, and
 $\alpha_{A_{\mu}}=-2.0$.}
\label{muonEDM}
\end{figure}

Since the violations of scaling arise from the neutralino sector  we discuss this in further
detail.
Here  the leading dependence of
the lepton mass arises from $n_{ik}^l$ while subleading
dependence arises from the outside smuon mass factors in Eq.(\ref{muEDM2}).
Thus to understand the scaling phenomenon and its breakdown we
focus on $n_{ik}^l$ which can be expanded as follows using Eq.(\ref{muEDM3}).
\beqn
\eta_{ik}^{\it l}=a_0c_0X^2_{1i}D^*_{\it l 1k}D_{\it l 2k}+b_0c_0 X_{1i}X_{2i}D^*_{\it l 1k}D_{\it l 2k}\nonumber\\
-\kappa_{{\it l}}a_0X_{1i}X_{3i}|D_{\it l 1k}|^2
-\kappa_{{\it l}}b_0X_{2i}X_{3i}|D_{\it l 1k}|^2-\nonumber\\
\kappa_{{\it l}}c_0X_{1i}X_{3i}|D_{\it l 2k}|^2
+\kappa^2_{\it l}X^2_{3i}D_{\it l 1k}D^*_{\it l 2k}
\label{muEDM4}
\eeqn
where $a_0$, $b_0$ and $c_0$ are independent of the lepton mass.
 The first  two terms on the right hand side of Eq.(\ref{muEDM4})
 are linear in lepton
mass through the relation
\beqn
Im(D^*_{\it l11}D_{\it l21})=-Im(D^*_{\it l12}D_{\it l22})=\nonumber\\
\frac{m_{\it l}}{M^2_{\tilde{\it l1} }-M^2_{\tilde{\it
l2}}}(m_0|A_{\it l}|\sin\alpha_f +|\mu|\sin\theta_{\mu}\tan\beta).
\label{muEDM5}) \eeqn The third, fourth and fifth terms on the right
hand side of Eq.(\ref{muEDM4}) have a leading linear dependence on
the lepton mass through the parameter $\kappa_{\it l}$ and have
additional weaker dependence on the lepton mass through the
diagonalizing matrix elements $D_{ij}$. The last term in
Eq.(\ref{muEDM4}) is cubic in the  lepton mass. However, in most of
the parameter space considered, the first term in Eq.(\ref{muEDM4})
is the dominant one and controls the scaling behavior. Thus for the
case when all the soft SUSY breaking parameters including A are
universal (i.e., $A_l=A$) in Eq. (\ref{muEDM5})), one finds  that
scaling results, i.e.,   $d_{\mu}/d_e\simeq m_{\mu}/m_e$.
 However, for the nonuniversal
case since the contribution from the A parameter is flavor dependent
we have a breakdown of scaling here.  An analysis is given in Fig.(\ref{muonEDM}).
This breakdown can be seen by comparing
$d_{\mu}$ for the
nonuniversal cases (dashed line with triangles pointed down and
dashed line with triangles pointed up) with $d_e$ (solid line with
squares) in Fig.(\ref{muonEDM}).
  With the inclusion of nonuniversalities the $d_{\mu}$ can be several orders of magnitude
 larger than $d_e$. Specifically values of $d_{\mu}$ could be as large as  ($10^{-24} -10^{-23}$)ecm
and within reach of the proposed experiments which extend the search for the muon EDM
to the range $10^{-24}$ ecm.
An enhanced  EDM for the muon relative to
the electron EDM in excess of what scaling law allows can be generated
with large neutrino mixings arising from the  See-Saw mechanism  \cite{Babu:2000dq}.
Another analysis  where lepton flavor violations are used to generate an enhancement of
the muon EDM is given by \cite{Feng:2001sq}.

\subsection{SUSY CP phases and the FCNC process $B\rightarrow X_s\gamma$}
\label{mB}

There are other effects of the CP violating phases on the
phenomenological constraint arises from the measurement of the rare
decay $B\rightarrow X_s \gamma$. This decay only occurs at the one
loop level in the standard model
\cite{Deshpande:1987nr,Altomari:1988xf,Dominguez:1988wa,Grinstein:1987vj,Casalbuoni:1993nh,Colangelo:1993ux,Falk:1993dh}.
The supersymmetric radiative corrections might be of the same order
of magnitude as the standard model contribution
\cite{Bertolini:1990if,Goto:1994ck,Hewett:1992is,Barger:1992dy,Diaz:1993km,Nath:1994tn,Barbieri:1993av,Lopez:1993kt,Garisto:1993jc,Bertolini:1994cv,Baer:1997jq}.
It is recently been recognized that supersymmetric contributions can
receive  significant contributions from the next-to-leading order
corrections (NLO) which are  enhanced by large $\tan\beta$.  These
are typically parameterized by $\epsilon$'s.  In addition
 to the $\epsilon$'s there are other two loop  (NLO) corrections which, however, are  small and  can be  absorbed  in a  redefinition of
 the SUSY parameters\cite{Degrassi:2000qf,Carena:2000uj}.  Currently the branching ratio of $B\rightarrow X_s\gamma$ is
 fairly accurately known experimently\cite{Abe:2001hk,Barate:1998vz,Chen:2001fj,Aubert:2002pb,Aubert:2002pd,Aubert:2005cb}
   and imposes very significant constraints  on model building.
 The current experimental value is
 \beqn
BR(B\rightarrow X_s\gamma) = (355\pm 24^{+9}_{-10}\pm 3) \times 10^{-6}
 \eeqn
as given by the Heavy Flavor  Averaging group\cite{Barberio:2006bi}. The Standard  Model result with QCD corrections\cite{Chetyrkin:1996vx}
 including NLO gives \cite{Gambino:2001ew}
$BR(B\rightarrow X_s\gamma) = (3.73\pm .30) \times 10^{-4}$. A
similar robust prediction for supersymmetric models is needed.
  To analyze the  NLO corrections for the supersymmetric case  (for recent analyses  see, one has to examine the effective Lagrangian describing the interaction of quarks with the charged Higgs fields $H^{\pm}$ and the charged Goldstones $G^{\pm}$
  which we display below (see, e.g.,
  \cite{Demir:2001yz,Belanger:2001fz,Gomez:2006uv,Gomez:2005nr})
\beqn
{\cal {L}}_{\rm eff} &=& \frac{g}{\sqrt 2 M_W} G^+ \{\sum_d m_t V_{td}
\frac{1+\epsilon_t(d) \cot\beta}{1+\epsilon_{tt}\cot\beta}\bar t_R d_L\nonumber\\
&-&
\sum_u m_b V_{ub} \frac{1+\epsilon_b^{\prime}(u)
\tan\beta}{1+\epsilon_{bb}^*\tan\beta}
\bar u_L b_R\} \nonumber \\
&+&\frac{g}
{\sqrt 2 M_W} H^+ \{\sum_d m_t V_{td}\frac{1+\epsilon^{\prime}_t(d)\tan\beta}
{1+\epsilon_{tt}\cot\beta} \cot\beta \bar t_R d_L\nonumber \\
&+&\sum_u m_b V_{ub} \frac{1+\epsilon_b(u)\cot\beta}
{1+\epsilon_{bb}^*\tan\beta} \tan\beta \bar u_L b_R \}
+ H.c.\nonumber\\
\eeqn
where
\beqn
\epsilon_t(b) &=&
\frac{\Delta h_t^2}{h_t}+\tan\beta \frac{\delta h_t^1}{h_t}\nonumber\\
\epsilon^{\prime}_b(t) &=& \frac{\Delta h_b^{1*}}{h^{*}_b}
+\cot\beta \frac{\delta h_b^{2*}}{h^{*}_b}\nonumber\\
\epsilon^{\prime}_t(b) &=& -\frac{\Delta h_t^2}{h_t}
+\cot\beta \frac{\delta h_t^1}{h_t}\nonumber\\
\epsilon_b(t) &=& -\frac{\Delta h_b^{1*}}{h^{*}_b}
+\tan\beta \frac{\delta h_b^{2*}}{h^{*}_b},
\eeqn
and where $\epsilon_{bb}$ and $\epsilon_{tt}$ is given by

\beqn
\epsilon_{bb}= \frac{\Delta h_b^2}{h_b} +\cot\beta  \frac{\delta h^1_b}{h_b},\nonumber\\
\epsilon_{tt}= \frac{\Delta h_t^1}{h_t} +\tan\beta \frac{\delta h^2_t}{h_t}.
\eeqn

 \begin{figure}
 \vspace{-2cm}
 \hspace{-1.3cm}
\includegraphics*[angle=0, scale=0.45]{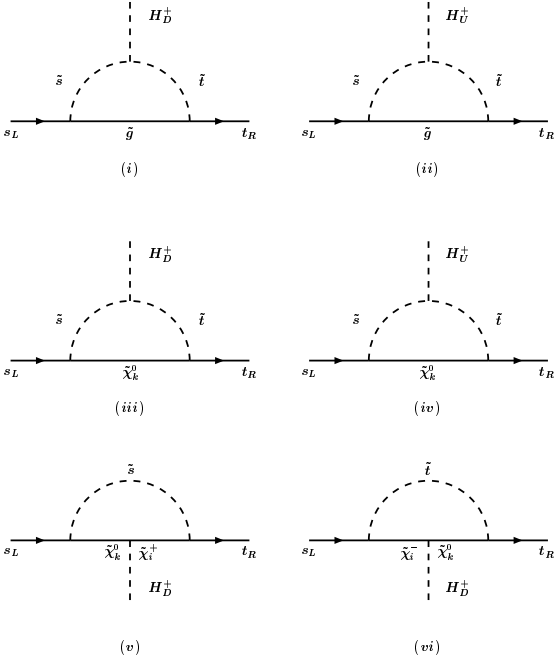}
\vspace{-1.5cm}
\caption{A sample of diagrams with CP dependent  vertices
 that contribute to the  NLO corrections to the epsilons  in $b\to s+\gamma$
decay. There are a  total of 20 such diagrams.}
\label{cpevenodd}
\end{figure}
Using the above Lagrangian along with the interaction of quarks and $W$ bosons one can write down the contributions to Wilson coefficients $C_7$ and $C_8$ in the effective Hamiltonian that governs the decay $b\rightarrow s\gamma$ (for further details see
\cite{Belanger:2001fz,Belanger:2004yn,Kagan:1998bh,Kagan:1998ym})

\beqn
H_{eff} =
-\frac{4 G_F}{\sqrt{2}} V^*_{ts} V_{tb} \sum_{i=1}^{8} C_i(Q) O_i(Q)
\eeqn
where
\beqn
O_2 &=&     (\bar{c}_L \gamma^{\mu}     b_L) (\bar{s}_L \gamma_{\mu} c_L)
\nonumber \\
O_7  &=&  \frac{e}{16 \pi^2} m_b (\bar{s}_L \sigma^{\mu \nu} b_R) F_{\mu \nu} \nonumber\\
O_8  &=&  \frac{g_s}{16 \pi^2} m_b (\bar{s}_L \sigma^{\mu \nu} T^a
b_R) G_{\mu \nu}^a
\eeqn
as
\beqn
C_{7,8}^{W}(Q_W ) &=& F_{7,8}^{(1)}(x_t) +
\frac{(\epsilon_{bb}^*-\epsilon_{b}'(t)) \tan\beta}
{1+\epsilon_{bb}^*\tan\beta} F_{7,8}^{(2)} (x_t)\nonumber\\
\eeqn

\beqn
 C_{7,8}^{H^{\pm}}(Q_W ) &=& \frac{F_{7,8}^{(1)}(y_t)}  {3\tan^2\beta}+
\frac{1+\epsilon_{t}'(s)^*\tan\beta}{1+\epsilon_{bb}^*\tan\beta}
 F_{7,8}^{(2)}(y_t)\nonumber\\
\eeqn
where $x_t$ and $y_t$ are defined  by
\beqn
x_t= \frac{m_t^2(Q_W)}{M_W^2}, ~~y_t= \frac{m_t^2(Q_W)}{M_H^2}
\eeqn
and $F_{7,8}^{(1)}$ and  $F_{7,8}^{(2)}$ are given by
\beqn
F_7^{(1)}(x) &=& \frac{x(7-5x-8x^2)}{24(x-1)^3}
+\frac{x^2(3x-2)}{4(x-1)^4}\ln x  \nonumber\\
F_7^{(2)}(x) &=& \frac{x(3-5x)}{12(x-1)^3}
+\frac{x(3x-2)}{6(x-1)^3}\ln x \nonumber\\
F_8^{(1)}(x) &=& \frac{x(2+5x-x^2)}{8(x-1)^3}
-\frac{3x^2}{4(x-1)^4}\ln x  \nonumber\\
F_8^{(2)}(x) &=& \frac{x(3-x)}{4(x-1)^3} -\frac{x}{2(x-1)^3}\ln x
\eeqn

The $C_7$ and $C_6$ terms receive dominant exchange
contributions from  the W, charged  Higgs  and the charginos. The gluino
and neutralino exchange terms can also contribute. The gluino exchange contributions have also been computed \cite{Everett:2001yy}.
However, it turn out that in the minimal flavor violation (MFV) scenario, the contributions from the gluino and neutralino exchanges
 are indeed relatively small.  The analyses of $b\to s+\gamma$ in beyond the MFA scenario where generational
mixings are taken into account have been carried out in the work of
 \cite{Hahn:2005qi,Foster:2004vp,Foster:2005wb}.
The most complete analyses of $B\rightarrow X_s \gamma$ in SUSY with the inclusion of NLO' effects is given in
\cite{Degrassi:2006eh,Gomez:2006uv,Buras:2002vd}. Specifically in the analysis of
\cite{Gomez:2006uv,Gomez:2005nr} it is shown that the $\epsilon$'s as well as the decay
$B\rightarrow X_s \gamma$ are sensitive to the CP phases.

\section{CP phases in $\nu$ physics and leptogenesis }\label{n}

Recent experiments discussed later in this section show that neutrinos are not
massless. In general neutrinos could have either a Dirac mass, a Majorana mass
or perhaps a mixture of the two. For a neutrino to have a Dirac mass  there must 
be a corresponding right handed neutrino to give a mass term of the type 
$m_D \bar \nu_L\nu_R +H.c.$. On the other hand, one can generate  a Majorana
mass term from just the left handed neutrinos, i.e., a mass term of the form
$\nu_L^T C^{-1} m_L \nu_L + H.c.$, where $C$ is the charge conjugation matrix. 
For the case  of three neutrino species 
the Majorana mass matrix is in general a symmetric mass matrix of
the from \cite{Mohapatra:2004vr,Mohapatra:2005wg,Nunokawa:2007qh}
\beq {\cal
M}_{\nu}= \left(
\begin{array}{ccc}
m_{ee} & m_{e\mu}  & m_{e\tau}\\
  & m_{\mu\mu} & m_{\mu\tau}\\
    &   & m_{\tau\tau}\end{array}\right)
\eeq
The Majorana neutrino mass matrix can be diagonalized by an orthogonal transformation so that
\beq
V^T{\cal M}_{\nu} V=        {\cal M}_{\nu}^D
     \eeq
     where $V$ can be written as  $V=U K$ and where the matrix $U$ is similar to the
     CKM matrix and K is a diagonal matrix with two independent Majorana phases.
     For U one can use the parametrization

          \beq
          U=    \left(
\begin{array}{ccc}
c_1 c_3 & c_3 s_1  & s_3 e^{-i\delta}\\
- s_1 c_2 -c_1 s_2 s_3 e^{i\delta} & c_1 c_2 -s_1 s_2 s_3 e^{i\delta} & s_2 c_3\\
- s_1 s_2 -c_1 c_2 s_3 e^{i\delta} & c_1 c_2 -s_1 c_2 s_3 e^{i\delta} & c_2 c_3\end{array}\right)
     \eeq
where $c_1=\cos \theta_{12}$, $c_2=\cos \theta_{23}$, $c_3=\cos \theta_{13}$ and
similarly for $s_1, s_2$ and $s_3$, where $\theta_{ij}$ and $\delta$ are constrained
so that $0\leq \theta_{ij} \leq \pi/2$ and $0\leq \delta \leq 2\pi$.  The matrix $K$ is diagonal
and can be taken to be

           \beq
          K=    \left(
\begin{array}{ccc}
1 & 0 & 0\\
0 & e^{i\phi_1}  & 0 \\
0 & 0&  e^{i\phi_2}
\end{array}\right)
     \eeq
Thus we have three  diagonal masses, three mixing angles and three phases which together
exhaust the full 9 parameter set of the Majorana neutrino mass matrix.
The Majorana CP phases do not enter in the neutrino oscillations, and only the Dirac phase
$\delta$ does.
Thus the oscillation probability from flavor $\nu_{\alpha}$ to $\nu_{\beta}$ is
given by
\beqn
P(\nu_{\alpha}-\nu_{\beta})= \delta_{\alpha\beta}  -4\sum_{i>j} U_{\alpha i} U_{\beta j} U^*_{\alpha j} U^*_{\beta i}
 \nonumber\\
\times \sin^2(\Delta m_{ij}^2L/4E_{\nu})~~
\label{deltaphase}
\eeqn
where $\Delta m_{ij}^2= |m_i^2-m_j^2|$.
From the  solar neutrino and the atmospheric
neutrino data~\cite{Hampel:1998xg,Abdurashitov:1999zd,Altmann:2000ft,Ahmad:2002jz,Ahmad:2002ka,Ambrosio:2001je,Fukuda:2000np}
one finds that the neutrino
$mass^2$ differences are given by
\beqn
\Delta m^2_{sol} =(5.4-9.5)\times 10^{-5}~ eV^2,\nonumber\\
\Delta m^2_{atm} =(1.4-3.7)\times 10^{-3}~ eV^2.
\eeqn
A fit to the  solar and atmospheric data using the three neutrino-generations
gives constraints only on the  neutrino mass  differences and on the mixing angles.
One has
\beqn
\Delta m^2_{sol}=||m_2|^2-|m_1|^2|,\nonumber\\
\Delta m^2_{atm}=||m_3|^2-|m_2|^2|,\nonumber\\
\sin^2\theta_{12}=(0.23-0.39), ~~\sin^2\theta_{23}=(0.31-0.72),\nonumber\\
\sin^2\theta_{13}<0.054.~~ \label{nu-mixing} \eeqn An interesting
aspect of Eq.(\ref{nu-mixing}) is that  the mixing angles
$\theta_{12}$ and $\theta_{23}$ are  large with $\theta_{23}$ being
close to maximal while $\theta_{13}$ is small. This feature was
rather unexpected and quite in contrast to the case of the quarks
where the  mixings are small.  An important point to note is that
the neutrino oscillation experiments do not give us any information
on the absolute  value of the neutrino masses. Other experiments are
necessary to provide information on the absolute values such as from
cosmology and from the neutrinoless double beta decay. Thus from
cosmology one has  the following upper bound on each species of
neutrino masses
\cite{Spergel:2006hy,Hannestad:2003xv,Hannestad:2003jt,Hannestad:2003ye}
\beqn \sum_i|m_{\nu_i}|< (0.7-1) eV \eeqn Similarly the neutrinoless
double beta decay gives  the following upper bound on the effective
neutrino mass $|m_{ee}|$ 
 \cite{Klapdor-Kleingrothaus:2000sn,Bilenky:2004hw}
\beqn
|m_{ee}|<(0.2-0.5) eV
\eeqn
where \cite{Mohapatra:2005wg}

\beqn
|m_{\nu}^{ee}|= |\cos^2\theta_{13} (|m_1|\cos^2\theta_{12} +\nonumber\\
 |m_2| \sin^2\theta_{12} e^{2i\phi_1})
+ \sin^2\theta_{13} |m_3|e^{2i\phi_2}|
\label{majphases}
\eeqn
Several scenarios for the neutrino mass patterns have  been discussed in order to explain the data.
One possibility considered is that the third generation mass is much larger than the neutrino masses for the
first two.
Among these are the following: (i) $|m_{\nu_3}|$ $>>$ $ |m_{\nu_1,\nu_2}|$, (ii)
 $|m_{\nu_1}$ $ |\sim |m_{\nu_2}|$,  $|m_{\nu_1,\nu_2}|$ $>>$ $|m_{\nu_3}|$, (iii)
 $|m_{\nu_1}$ $ |\sim |m_{\nu_2}$ $ |\sim  |m_{\nu_3}|$, $|m_{\nu_1,\nu_2,\nu_3}|$ $ >> $
 $||m_{\nu_i}| -|m_{\nu_j}| |$.
Neutrino oscillations are sensitive to $\delta$ but not to the
Majorana phases\cite{Barger:2001yr,Barger:1980jm}. As  is clear from
Eq.(\ref{majphases}), Majorana phases do enter in the neutrinoless
double beta decay, but an actual determination of CP violation in
$0\nu\beta\beta$  appears difficult\cite{Barger:2002vy}.\\

We discuss now briefly the possible determination of $\delta$ in the next generation of neutrino experiments
such as NO$\nu$A\cite{Ayres:2004js,Ayres:2002nm}
 and T2KK \cite{Hagiwara:2006vn}. 
  We begin by noting that under the condition that CPT is conserved, 
the conservation of CP would require $P(\nu_{\alpha}\to \nu_{\beta})-P(\bar \nu_{\alpha}\to \bar \nu_{\beta})=0$.
In the presence of CP violation this difference is non-vanishing. Thus specifically
one has\cite{Nunokawa:2007qh,Barger:2007jq}
\begin{eqnarray}
P(\nu_{\mu}\to \nu_{e})-P(\bar \nu_{\mu}\to \bar \nu_{e}) =\nonumber\\
-16 J \sin(\frac{\Delta m_{12}^2 L}{4E})  \sin(\frac{\Delta m_{13}^2 L}{4E})
 \sin(\frac{\Delta m_{23}^2 L}{4E}),
 \label{nu1}
\end{eqnarray}
where $E$ is the neutrino beam energy, L is the oscillation length and 
 $J$ is the Jarlskog invariant for the neutrino mass matrix similar to the one for the quark 
mass matrix 
\begin{eqnarray}
J= s_{12}c_{12}s_{23}c_{23}s_{13}c_{13}^2 \sin\delta.
\label{nu2}
\end{eqnarray}
We note  that $J$ depends on $\theta_{13}$ and $\delta$ both of which are currently unknown
and thus one has only an upper limit for $J$ so that $J\leq 0.04$. 
Thus the observation of a CP violation via Eq.(\ref{nu1}) depends  on other factors. 
For example, $J$ vanishes if  $\theta_{13}$ vanishes and thus the effect of CP violation
via  Eq.(\ref{nu1}) would be unobservable. Similarly, if there  was  a degeneracy in the neutrino
masses, for example if  $|m_{\nu_1}|\sim |m_{\nu_2}|$, then again the  observation of CP violation
via  Eq.(\ref{nu1}) would be difficult.  However, aside from these extreme situations the process
Eq.(\ref{nu1}) holds the strong possibility that long baseline experiments should allow one 
to observe  CP violation due to $\delta$  in the neutrino sector. Two experiments are ideally suited
for this observation. One of these  is  NO$\nu$A \cite{Ayres:2004js,Ayres:2002nm}
which will be  25kton liquid scintillator detector  
 placed 810 km away from the NuMI neutrino beam in Fermilab (see Sec.XIV).  The configuration
 will allow runs in the neutrino as  well as  in anti-neutrino mode.   
 The second possibility  is the T2KK detector \cite{Hagiwara:2006vn} which 
 is discussed in Sec.XIV.

\subsection{CP violation  and leptogenesis}\label{nA}
As already mentioned in  Sec.1,  achieving baryon asymmetry in the
universe requires three conditions: violation of baryon number,
violation of C and of CP, and departure from thermal equilibrium.
Quantitative analyses show that the Standard Model falls short of
fulfilling these conditions.  Specifically,  the amount of CP
violation is found not sufficient.  Thus in the framework of the
electroweak baryogenesis
 the effective CP suppression factor that enters is $f_{CP}$ where \cite{Shaposhnikov:1986jp,Farrar:1993sp}
\beqn f_{CP} = T_C^{-12} (m_t^2-m_c^2)   (m_t^2-m_c^2)
(m_t^2-m_u^2)\nonumber\\(m_b^2-m_s^2) (m_b^2-m_d^2)( m_s^2-m_d^2)
s_{12}s_{23}s_{31} \sin\delta~~ \eeqn where $s_{ij}=\sin\theta_{ij}$
and $\theta_{ij}$ are the three mixing angles, and $\delta$ is the
CKM phase, and $T_C$ is the temperature of the electroweak phase
transition (EWPT).  The EWPT is supposed to  occur at values
$T_C\sim 100$ GeV, which leads to $\delta_{CP} \sim
(10^{-18}-10^{-20})$. A rough estimate of  baryon asymmetry in EWPT
is
 $B \simeq  10^{-8} f_{CP}$
and the Standard Model in this case leads to $B\simeq
(10^{-26}-10^{-28})$  which is far  too small compared to the
desired value of $B\sim 10^{-10}$.  Additionally there are stringent
constraints on the Higgs mass which are already in violation of the
current limits. Analysis of baryogenesis in MSSM relieves some of
the  tension both because there are  new sources of CP violation,
and also because the Higgs mass limits are significantly larger,
e.g., $m_h \leq 120$ GeV. However,
the analysis  requires a significant fine tuning of parameters.\\

  An attractive alternative to conventional baryogenesis (for reviews see \cite{Cohen:1993nk,Riotto:1999yt})
  is baryogenesis via leptogenesis  (\cite{Fukugita:1986hr}. For recent reviews see
  \cite{Nardi:2005hs,Buchmuller:2004nz,Chen:2007fv,Nir:2007zq}).
  The essential idea here is that if one  can generate
  enough lepton asymmetry, then it  can be converted into baryon asymmetry via sphleron interactions which
  violate  $B+L$ but preserve $B-L$.   Leptogenesis   is a natural
   consequence of the See-Saw mechanism \cite{Minkowski:1977sc,Mohapatra:1979ia,Yanagida:1979as,Glashow:1979,Gell-Mann:1980}
     which
  is a popular  mechanism for  the generation of small neutrino masses
(see also \cite{Schechter:1980gr,Schechter:1981cv,Valle:2006vb} for early work on the See-Saw phenomenology).
    To generate a See-Saw one needs
  heavy Majorana neutrinos and one can characterize the Lagrangian for the Majoranas  by
  \beqn
  L_{N}= M_i N_i N_i  + \lambda_{i\alpha} N_i L_{\alpha} \phi
  \eeqn
where $N_i$ are the Majorana fields, and
 $\lambda$ are  in general complex and thus the $\lambda$ terms violate CP. Further, $L_N$ violates
lepton number and  $B-L$.  Thus the Lagrangian of the above type has
the general characteristics that  might lead to the  generation of
baryon asymmetry via leptogenesis.
The CP violation occurs in the decay of the Majoranas because of the overlap of the tree and the loop. \\

 \begin{figure}
\includegraphics*[angle=0, scale=0.40]{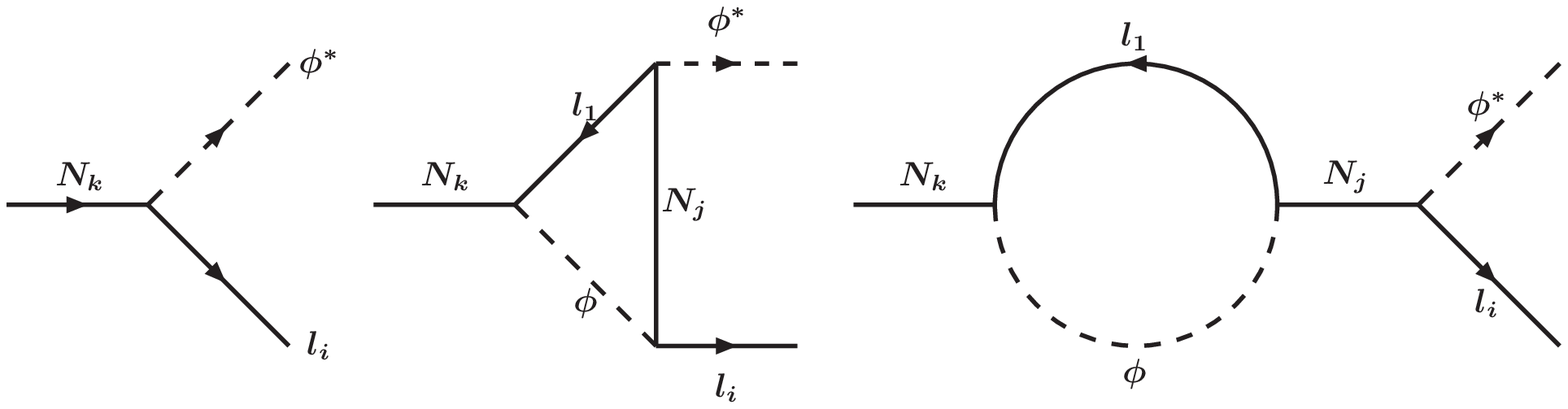}
\vspace{-6.5cm}
\caption{Generation of lepton number asymmetry via decay of the right handed  neutrino (N)
by interference between the tree,  the vertex,  and the  self energy
loop diagrams.}
\label{lepto1}
\end{figure}

One can define a CP asymmetry parameter so that
\beqn
\epsilon_{1}= \frac{  \sum_{\alpha} [\Gamma(N_i\to  l_{\alpha} \phi ) -  \Gamma(N_i\to \bar l_{\alpha} \phi^{\dagger})] }
{\sum_{\alpha} [ \Gamma(N_i\to  l_{\alpha} \phi )  + \Gamma(N_i\to \bar l_{\alpha} \phi^{\dagger})] }
\eeqn
For the case of just two Majorana neutrinos the analysis of $\epsilon_1$ gives
\beqn
\epsilon_{1}= C(\frac{M_2^2}{M_1^2}) \frac{Im(\lambda\lambda^{\dagger})^2_{12}}{(\lambda\lambda^{\dagger})_{11}}
\eeqn
where $C(z) =C_1(z)+C_2(z)$  where \cite{Covi:1996wh}

 \beqn
 C_1(z) = (8\pi)^{-1} {\sqrt z} [1-(1+z) ln(\frac{1+z}{z}),\nonumber\\
 C_2(z) =  (8\pi)^{-1} \frac{\sqrt z} {1-z}
 \eeqn
 For the case of two singlets  and $M_1< M_2$ one has

 \beqn
\epsilon_1= -\frac{3}{8\pi} (\frac{M_1}{M_2})  \frac{Im(\lambda\lambda^{\dagger})^2_{12}}{(\lambda\lambda^{\dagger})_{11}}
 \eeqn
 Next consider the case when the initial temperature $T_i$ is larger than the mass of the lightest singlet neutralino $N_1$. In this
 case  neglecting the effect of the decays of the heavier neutralinos, one can write  the Boltzman equations that govern
 the number densities $n_{N_1}$ and $n_{B-L}$ so that \cite{Buchmuller:2001sr,Buchmuller:2002rq,Buchmuller:2004nz}

 \beqn
 \frac{dn_{N_1}}{dx}= -(D+S) (n_{N_1}-n_{N_1}^{eq})\nonumber\\
  \frac{dn_{B-L}}{dx}= -  \epsilon_1 D (n_{N_1}-n_{N_1}^{eq}) - W n_{B-L}
 \eeqn
 Here $x\equiv M_1/T$, and $W=\Gamma_W/(Hx)$ is the washout term. The processes contributing to the
Boltzman equations  are the decays, inverse  decays, scattering
processes with $\Delta L=1$ and processes with $\Delta L=2$, where
$D= \Gamma_D/(Hx)$ includes decays and inverse decays and
$S=\Gamma_S/(Hx)$ includes $\Delta L=1$ scattering. Two parameters
that enter prominently in the analysis are the effective mass
$\tilde m_1$ which is defined by \beqn \tilde
m_1=\frac{(\lambda\lambda^{\dagger})_{11} |<\phi>|^2}{M_1} \eeqn

and
the equilibrium neutrino mass $m*$ defined by
 \beqn
 m*= \frac{16 \pi^{5/2} \sqrt {g*}}{3\sqrt 5} \frac{ |<\phi>|^2}{M_{Pl}}
 \eeqn
 where $g*$ is the total number of degrees of freedom ($g*=106.75$ for SM). Numerically $m*\sim 10^{-3} eV$.

 The ratio $\tilde m_1/m*$ controls whether or not $N_1$ decays are out of equilibrium. When $\tilde m_1 <m*$
 (the weak washout  region), $N_1$ decay is slower than the Hubble expansion and leptogenesis can occur efficiently.
 For the case  $\tilde m_1 > m*$ (the strong washout region)  the back reactions that  tend to washout are fast and leptogenesis
 is rather slow. However, even for $\tilde m_1/m* >>1$, a sufficient amount of lepton asymmetry can be generated.
 The solution to $n_{B-L}$ can be obtained in the form

 \beqn
 n_{B-L}(x) = n_{B-L}^f exp(-\int_{x_1}^x dx' W(x')) -\frac{3}{4} \epsilon_1 \kappa(x)
  \eeqn
where $n_{B-L}^f=n_{B-L}(x=\infty)$, and  where $\kappa(x)$ is given by

 \beqn
 \kappa(x)=-\frac{4}{3} \int_{x_1}^x \frac{D}{D+S} \frac{dn_{N_1}}{dx'} exp(-\int_{x'}^x  dx" W(x")).
 \eeqn
 The $(B-L)$ asymmetry is converted into baryon asymmetry by spheleron processes so that
 \beqn
 \eta_{B}= \frac{a_{sph}}{f} N_{B_L}^f-\frac{3}{4} \frac{a_{sph}}{f} \epsilon_1 \kappa_f
  \eeqn
  where $a_{sph}$ is the spheleron conversion factor ($a_{sph}= 28/79$), and f is a dilution factor
  $f=n_{\gamma}^{rec}/n_{\gamma}^*$ which depends  on the photon production from the beginning
  of leptogenesis till the point of recombination, and numerically   f= 2387/86. One then has
  \beqn
  \eta_B \simeq 10^{-2} \epsilon_1 \kappa_f
   \eeqn
 Now an upper limit on $\epsilon_1$ can be obtained assuming that $N_1$ decay dominates the asymmetry as assumed
 above  with a hierarchical pattern of heavy neutrino masses, and assuming that the decay of $N_1$ occurs for  $T\geq 10^{12}$ GeV.
 In this case one can deduce, under the assumption $M_1/M_2<< 1$, the result  \cite{Davidson:2002qv}
  \beqn
 |\epsilon_1 \leq \frac{3}{16\pi} \frac{M_1(m_3-m_2)}{|<\phi>|^2}
 \eeqn
 With $|m_3-m_2|\leq \sqrt{\Delta m_{32}^2} \sim .05$ eV, one finds a lower bound on $M_1$ so that

 \beqn
 M_1\geq 2\times 10^9 {~~\rm GeV}
 \eeqn
 This result  implies a lower  bound on the reheating temperature, and this bound
 appears to be  in conflict with the upper  bound on the reheating temperature  to control the  gravitino
 overproduction for the supersymmetric case. Consequently several  variants of leptogenesis have been
 studied such as resonant leptogenesis \cite{Pilaftsis:2005rv,Pilaftsis:2003gt,Pilaftsis:1997jf},
  soft leptogenesis \cite{Grossman:2003jv,Boubekeur:2004ez,Grossman:2004dz},
 and  non-thermal leptogenesis \cite{Fujii:2002jw}.
 The type of CP violation that occurs in leptogenesis involves  neutrinos  which are Standard  Model singlets,
 and hence have no direct gauge interactions with the normal particles, and in addition are very heavy.   Thus a direct
 observation of CP violation that enters leptogenesis would be  essentially
 impossible  in laboratory experiments.  However, in unified models CP phases could be inter-related  across different
 sectors and thus indirect constraints on such phases could arise in such models.

\subsection{Observability of Majorana phases}\label{nB}
In the previous subsection we found that  the  leptogenesis does
depend crucially on the Majorana phases (for a review of Majorana
particles and their phases, see
e.g.,\cite{Kayser:1984ge,Kayser:1984xc}). However, these phases
arise from heavy Majoranas and are not the  same as the
 Majorana phases that arise in light neutrino mass sector.
It was noted in our discussion of the neutrino masses that Majorana phases do not
enter in neutrino oscillations which depend only on the Dirac phase.
The Majorana phases do enter in the neutrinoless double beta decay.
However,  they do so only  in a CP even fashion and further their  observation
 in the $0\nu \beta \beta$ appears difficult.
The question one might ask is in what  processes  the Majorana phases can enter
in a manifestly CP odd fashion.  It is known that one such process is neutrino-anti-neutrino
($\nu \to \bar \nu$)  oscillations  \cite{Schechter:1980gk,Bernabeu:1982vi,deGouvea:2002gf}.
The analysis of \cite{deGouvea:2002gf} sets out some simple criteria for  their appearance
in scattering phenomena.   Thus consider  the amplitude for  the process  $X$ where
\beqn
A_X=    e^{i\xi_X} (A_1 +A_2 e^{i(\delta+ \phi)})
\eeqn
where we have pulled out a common phase  factor $e^{i\xi_X}$   so  $A_1$ has no phase dependent
factor multiplying it,  $\delta$ is a  CP even phase  and $\phi$ is a  CP odd phase.  Then the mirror
process $\bar X$ has the following  amplitude

\beqn
A_{\bar X}=    e^{i\xi_{\bar X}} (A_1 +A_2 e^{i(\delta -\phi)})
\eeqn
where $A_{1,2}$ are assumed not to contain any CP violating effects and are the same
in processes $X$ and $\bar X$.
The difference $\Delta \Gamma_{CP} =|A_{\bar X}|^2-|A_{X}|^2$ is then given by
\beqn
\Delta \Gamma_{CP} = 4 A_1 A_2 \sin(\delta)\sin(\phi)
\eeqn
The above simple analysis points  to the following three conditions necessary for  CP odd effects
to arise in the process $X$ vs its mirror process $\bar X$. These are  (a) the existence of
two distinct contribution to the amplitude, (b) the two contributions must have  a non-vanishing
relative CP odd phase, and (c)  they must also have a non-vanishing relative CP even phase.
The analysis of  \cite{deGouvea:2002gf}
 considers  application to the process
\beqn
l_{\alpha}^+ W^-\to \nu  \to \l_{\beta}^- W^+
\eeqn
for which one has the   amplitude
\beqn
A_{X}= \sum_{i} (\lambda_i U_{\alpha i} U_{\beta i}) \frac{m_i}{E} e^{-i \frac{m_i^2 L}{2E}} S
\eeqn
where  $E$ is the energy of the intermediate state which propagates a microscopic  distance L,
U is the mixing matrix,
and S depends on the initial and the final states  and on kinematical factors.
For the CP-conjugate  process $l_{\alpha}^-W^+\to  l^+_{\beta} W^-$ one has
\beqn
A_{\bar X}= \sum_{i} (\lambda_i U_{\alpha i} U_{\beta i})^* \frac{m_i}{E} e^{-i \frac{m_i^2 L}{2E}} \bar S
\eeqn
where the combination $(\lambda_i U_{\alpha i} U_{\beta i})$ is free of the 
phase-convention \cite{Kayser:1984ge,Bilenky:1984fg,Nieves:1987pp,Nieves:2002vq}.
Limiting the analysis to the case of two generations we write
\beqn
U=  \left(
\begin{array}{cc}
\cos\theta  & \sin\theta\\
-\sin\theta & \cos\theta\end{array}\right)
\left(
\begin{array}{cc}
1 & 0\\
0 & e^{i\phi}\end{array}\right)
\eeqn
Under the  approximation $\lambda_1=1=\lambda_2$, $|\bar S|=|S|$, $\alpha =e$ and $\beta =\mu$
this leads  to
\beqn
\Delta \Gamma_{CP} = |A_{\bar X}|^2- |A_{X}|^2= \frac{m_1m_2}{4 E^2} |S|^2 \sin^2 2\theta\nonumber\\
\sin\left(\frac{(m_2^2-m_1^2)L}{2E}\right) \sin\phi
\eeqn
The above  example satisfies all the criterion set  forth earlier for a CP odd effect to appear.
CP odd effects can also appear in lepton number violating meson processes such as
$K^{\pm}\to \pi^{\mp} \mu^{\pm}\mu^{\pm}$.  Thus, for example, if we  write
\beqn
A_{K^+}= e^{i\xi_{K^+}} (A_{1K} +A_{2K} e^{i(\delta_K+ \phi_K) }),\nonumber\\
A_{K^-}= e^{i\xi_{K^-}} (A_{1K} +A_{2K}  e^{i(\delta_K-\phi_K) })
\eeqn
one will have $\Delta \Gamma^{K}_{CP} = |A_{K^-}|^2-  |A_{K^+}|^2$  given by
\beqn
\Delta \Gamma^{K}_{CP} \propto  4A_{1K} A_{2K} \sin(\delta_K)\sin(\phi_K)
\eeqn
 $\Delta L=2$ contributions  do arise  with R parity violation  in supersymmetry and
contribute to $\Delta \Gamma^{K}_{CP}$.  However, the effect turns  out to be extremely small.
Some possible cosmological effects of CP violation in neutrino oscillations  are considered 
 in \cite{Khlopov:1981nq}.

\section{Future prospects}\label{o}
\subsection{Improved EDM experiments}\label{oA}
There are good prospects of improving the EDM
 bounds significantly. Thus future  experiments  may improve the sensitivity of EDM experiments by an
order of magnitude or more
~\cite{Kawall:2003ga,Kozlov:2006kv,Dzuba:2002kg} and in some cases
by a significantly larger
factor~\cite{Semertzidis:2003iq,Lamoreaux:2001hb,Semertzidis:2004uu}.
A recent review on the current experimental situation and future prospects regarding
the electron electric dipole moment can be found  in \cite{commins}.
Regarding  the neutron EDM a sensitivity at the level of $1.7\times 10^{-28}$ ecm
could be achieved \cite{Balashov:2007zj} and even a  sensitivity of $10^{-29}$ ecm is 
possible\cite{Harris:2007fz}. Regarding the EDM of $^{199}Hg$ improved measurements
are in progress  and a factor of 3-4 improvement over the next year or so is possible. 
Beyond that there are various projects aimed at  improving the limit with diamagnetic atoms, using Xe-129, radioactive Ra or Rn. However, all of them are still in the development phase, 
so when one may expect better limits from these experiments is unclear.  
Regarding the muon EDM  one proposed experiment\cite{Semertzidis:2003iq,Semertzidis:2004uu}  
feasible  at  JPARC (Japan Proton Accelerator Reseach Complex)
could extend the sensitivity to as much as $10^{-24}$ ecm. However, it appears that the
earliest muons may become available at JPARC is 2016. However,  recently another proposal
for muon EDM  has been made where the existing muon beam $\mu$EI at PSI could
be used.  It is claimed that the muon EDM with a sensitivity of better than $d_{\mu} \sim 5 \times 10^{-23}$
ecm within one year of data taking is feasible \cite{Adelmann:2006ab}.
Currently there is also an exploration underway regarding the possible determination of the
deutron EDM using techniques similar to the ones used  for muon EDM with the goal of 
reaching a sensitivity of $10^{-29}$ecm\cite{Semertzidis:2003iq,Yannis}.

\subsection{B physics at the LHCb}\label{oB}
LHCb is one of the four detectors at the LHC, the other three being ATLAS, CMS, and
ELLIS. Of the these ATLAS  and CMS are the main particle physics detectors
dedicated to the search for new physics such as supersymmetry or extra  dimensions.
While the ATLAS and CMS can also study B physics their capabilities in this respect
are rather limited. On the other hand LHCb is an experiment which is specifically
dedicated to the study of B physics.
Thus the B mesons produced in collisions at the LHC are likely to lie in angles close to
the beam directions and a detector ideal for the study of B physics should be able
to detect such particles.  This is precisely what the LHCb  is designed to do. Specifically
the detection of charged particle will be  accomplished by  its Ring-Imaging Cherenkov
(RICH) detector.
The precise identification of the interaction region utilizes a vertex locator or VELO
which  can  be used for  B-tagging, and more generally for the separation of primary
and secondary vertices.  The number of B mesons produced at LHCb will be enormous.
Even  a luminosity of $10^{32} cm^{-2}s^{-1}$ will lead to a number of $b\bar  b$ events
at the rate of
$O(10^{11-12}) $ per year.
Thus the LHCb will have an unprecedented opportunity to study B physics  in great depth.
\footnote{See, e.g., http://www-pnp.physics.ox.ac.uk/~lhcb/}

\subsection{Super Belle proposal}\label{oC}
The B factories  are an ideal instrument for  the study of elements of the CKM matrix including the CP phase
$\delta_{CKM}$.  The analyses  provided by the B factories  at SLAC (BaBar) and at KEK (Belle) have given 
a wealth of data and have improved the measurements of the CKM elements.  Specifically they have been able to
measure time dependent  CP asymmetries with good precision.
Further improvements in the
measurements of these elements will come only with significantly greater  luminosity.  The Super Belle proposal
aims at achieving that by an upgrade of the KEKB collider to a luminosity of $10^{35-36} cm^{-2} s^{-1}$.
Such an improvement will also require an upgrade of the vertex detector for the Super Belle and specific proposals
are under study\cite{Kawasaki:2006sg}.

\subsection{Superbeams, $\nu$ physics, and CP}\label{oD}
The answer to the  question  whether or not $CP$ phases appear  in neutrino physics is of crucial relevance
to our understanding of fundamental interactions. The observation of such phases in the light
neutrino sector
is possible  using long baseline experiments and intense beams \cite{Diwan:2006qf,Marciano:2006uc,Marciano:2001tz}
and its observation
 will give greater credence to the hypothesis that such phases also appear in
the heavy neutrino sector which enter in leptogenesis.
Thus the AIP-2004  study recommends "{\it as a high priority, a comprehensive U.S. program to
complete our understanding of neutrino mixing, to determine the character of the neutrino mass
spectrum, and to search for CP violation among neutrinos}"\cite{FreEDMan:2004rt}.
  Such high priority efforts  could include
improved $0\nu\beta\beta$ experiments, and super beams to study
neutrino oscillations and detect CP phases.  Specifically the study
recommends "{\it a proton driver in the megawatt class or above and
neutrino superbeam with an appropriate very large  detector capable
of observing CP violation and measuring the neutrino mass-squared
differences and mixing parameters with high precision}". One such
proposal is an upgraded Fermilab Proton Driver (FPD). Such an
upgrade will improve the study of $\nu_{\mu}\to \nu_{e}$
oscillations  by a significant factor  \cite{Geer:2005sk}. Thus  the
current  Fermilab NuMI proton beam has $10^{13}$ protons at 120 GeV
(a beam power of .2 megawatts).  A secondary beam of charged pions
is generated from the proton beam, and the pions then decay
producing a beam of tertiary $\nu_{\mu}$ as they propagate along a
long corridor to the target  $735$ km downstream.  With .2 megawatt
of proton beam power one generates only $10^{-5}$ interactions  in a
1kt detector at the far end.  Thus an upgrade of the proton beam to
deliver several megawatt of proton beam power coupled with an
upgrade of the detector to 10 kt will significantly enhance the
sensitivity of the detector to observe possible CP effects. A
similar idea being discussed is   T2KK where the far detector is put
on the east coast of Korea  along the Tokai to Kamioka (T2K)
neutrino beam line \cite{Hagiwara:2006vn}.

\section{Conclusions}\label{p}
We have attempted here to give  a broad overview of CP violation and the effect of CP phases
arising from physics beyond the Standard Model. We know that CP violation beyond what is
allowed in the Standard Model must exist in order that one generate the desired amount of baryon
asymmetry in the universe.   We have examined the origin of such CP violation in some of the
leading candidates for physics beyond the Standard Model. These include models based on
extra dimensions, supersymmetric models with soft breaking, and string models.
Specifically supersymmetric models and string models generate
 a plethora of new CP phases
and one problem one encounters is that such phases lead to  EDMs for
the electron and for the neutron in excess of the current limits.
One  way to limit to these is to fine tune the phases to be small
which, however, is not satisfactory from the point of generation of
baryon asymmetry. What one needs is a mechanism which allows at least
some of the phases to be large while suppressing their contribution
to the EDMs.  One possibility is suppression of the EDMs  by having
a heavy sparticle spectrum. However, this possibility puts the
sparticle masses at least for the first two generations  in the
several TeV  range and thus outside the reach of the LHC. An
alternative possibility of controlling the EDMs is the cancelation
mechanism which allows for large phases consistent with the
stringent limits on the EDMs from experiment.
 If the cancelation mechanism is valid, then the effect of CP phases will show up at colliders
 in a variety of  supersymmetric phenomena.  We have discussed some of these phenomena
 in this report.  One important such phenomenon is CP even -CP odd Higgs mixing which
 would lead to discernible signals at Hadron colliders and at a future International Linear 
 Collider (ILC).  Effects of CP could
 also be visible in $B^0_s\to \mu^+\mu^-$, in Higgs decays $h^0\to b\bar b, \tau \bar \tau$ and
 in  sparticle decays.  Dark matter analyses are also affected, specifically the detection cross section
 for neutralino-nucleon scattering.  \\

 The future proposed experiments will investigate CP phenomena with vastly increased data.
 Chief among these is the LHCb experiment which is dedicated to the study of the B mesons.
 The proposed Super Belle will further add to these efforts.  These  will pin down
 the CKM matrix elements to a much greater precision than BaBar and Belle, and may shed
 light on the possibility whether or not new sources of CP violation are visible. However,  if
 the sparticles are indeed observed, as one expects they will be, then a study of their
 branching ratios is likely to put significant limits on CP phases from sparticle  decays.

\begin{acknowledgments}
We  thank Norman F Ramsey for communications regarding the early history of CP 
violation. 
Discussions and communications with a number of other colleagues are also acknowledged.
These include Giuseppe Degrassi,  Damien Easson, 
Yasaman Farzan,  Mario Gomez,  Maxim Khlopov, 
Olaf Kittel,   Alberto Lerda,
Apostolos Pilaftsis,  and Jose Valle. 
Cumunications with Eugene Commins regarding the electron EDM experiment,  
with Philip Harris regarding the neutron EDM experiment,  
with Yannis K Semertzidis regarding the muon EDM and the EDM of the deutron, and 
with Michael Romalis  on the EDM of  $^{199}Hg$   are acknowledged.
 Assistance by Daniel Feldman and Zuowei Liu during preparation of the
manuscript is acknowledged.  
The  work was supported in part by the NSF grant PHY-0546568.

\end{acknowledgments}

\section{Appendices}\label{q}

\subsection{Chargino and neutralino  mass matrices with phases}\label{qA}

Here we give some details on the diagonalization of the
chargino and neutralino mass matrices which are in general complex.
These appear in the analysis  of Secs. (\ref{i}) and (\ref{j}).
We consider the chargino mass matrix first. Here  we have
\beqn
   M_C= \left(
\begin{array}{cc}
\m2 & \r2 m_W \sb \\
    \r2 m_W \cb & \mu
         \end{array}\right)
\label{ch1}
\eeqn
The chargino matrix $M_C$ is
 not hermitian, is not symmetric and is not real
since $\mu$ and $\tilde{m_2}$ are complex. For simplicity
we analyze its diagonalization for real $\tilde{m_2}$ and complex
$\mu$. Generalization for complex $\tilde{m_2}$ and $\mu$ is
straightforward.
$M_C$ can be  diagonalized by
using the following  biunitary transformation
\beq
U'^{*} M_C V^{-1}=M_D
\label{MsubD}
\eeq
Here $U'$ and $V$ are hermitian matrices  and $M_D$ is a diagonal matrix
which, however, is not yet real. $U'$ and V satisfy the relation
\beqn
V (M_C^{\dagger} M_C) V^{-1}={\rm diag}(|\mc1|^2, |\mc2|^2)\nonumber\\
=U'^{*} (M_C M_C^{\dagger}) (U'^*)^{-1} \eeqn We may parameterize
$U'$ so that \beqn U'= \left(
\begin{array}{cc}
\cos \frac{\theta_1}{2}
           & \sin \frac{\theta_1}{2} e^{i\phi_{1}} \cr
           -\sin \frac{\theta_1}{2} e^{-i\phi_{1}}
                &\cos \frac{\theta_1}{2}
         \end{array}\right)
\eeqn
where
\beqn
\tan\theta_1=2\r2m_W
(\m2^2-|\mu|^2
-2m_{W}^2\cos2\beta)^{-1}\nonumber\\
 \left(\m2^2\cos^{2}\beta+|\mu|^2\sin^2\beta
+|\mu|\m2\sin2\beta \cos\theta_{\mu}\right) ^{\frac{1}{2}} \eeqn and
\beq \tan\phi_1= |\mu|\sin\theta_{\mu}\sin\beta \left(\m2 \cos\beta
+|\mu| \cos\theta_{\mu} \sin\beta\right)^{-1} \eeq Similarly we
parameterize V so that \beqn V= \left(
\begin{array}{cc}
\cos \frac{\theta_2}{2}
           & \sin \frac{\theta_2}{2} e^{-i\phi_{2}} \\
           -\sin \frac{\theta_2}{2} e^{i\phi_{2}}
                &\cos \frac{\theta_2}{2}
 \end{array}\right)
\eeqn
where
\beqn
\tan\theta_2=2\r2m_W
(\m2^2-|\mu|^2
+2m_{W}^2\cos2\beta)^{-1}\nonumber\\
 \left(\m2^2\sin^{2}\beta+|\mu|^2\cos^2\beta
+|\mu|\m2\sin2\beta \cos\theta_{\mu}\right) ^{\frac{1}{2}}
\eeqn
and
\beq
\tan\phi_2= -|\mu|\sin\theta_{\mu}\cos\beta
\left(\m2 \sin\beta +|\mu|
\cos\theta_{\mu} \cos\beta\right)^{-1}
\eeq
We wish to choose the phases of $U'$ and $V$ so that the elements of $M_D$
will be positive. Thus we define  $U=H \times U'$ where
\beqn
H= \left(
e^{i\gamma_1},  e^{i\gamma_2} \right)
\eeqn
where $\gamma_1, \gamma_2$ are the phases of the diagonal elements of
$M_D$ in 
Eq.(\ref{MsubD}).  With  the above choice of phases one has
\beq
U^{*} M_{C} V^{-1}= diag(\mc1,\mc2)
\eeq
  Our
 choice of the signs  and the roots is such that
\beqn
M_{(\mc1)(\mc2)}^2 & &={\frac{1}{2} [\m2^2+|\mu|^2+2m_{W}^2](+)(-)}\nonumber\\
& &
{\frac{1}{2}[(\m2^2-|\mu|^2)^2+4m_{W}^4 \cos^{2}2\beta+4m_{W}^2}\nonumber\\
& &
(\m2^2+|\mu|^2+2\m2 |\mu| \cos\theta_{\mu} \sin2\beta)]^\frac{1}{2}\nonumber\\
\eeqn
where the  sign  chosen is such that  $\mc1<\mc2$ if
\beq
\m2^2<|\mu|^2+2m_{W}^2 \cos2\beta.
\eeq

For the  neutralino mass matrix   $M_{{\tilde \chi}^0}$  one has
\beqn
\left(
\begin{array}{cccc}
\tilde m_1 & 0 & -M_Z s_W  c_\beta    & M_Z    s_W  s_\beta \\
                 0  & \tilde m_2 & M_Z c_W   c_\beta & -M_Z c_W  s_\beta \\
         -M_Z s_W  c_\beta &  M_Z c_W c_\beta & 0 & -\mu \\
         M_Z s_W s_\beta  & -M_Z  c_W s_\beta & -\mu & 0\end{array}\right)\nonumber\\
\eeqn
In the above  $s_W=\sin \theta_W$, $s_\beta= \sin\beta$ where $\theta_W$ is the weak angle,
and $c_\beta = \cos \beta$, and $s_\beta=\sin \beta$.
The matrix $M_{\chi^0}$ is a complex non hermitian and symmetric matrix,
which can be diagonalized by a unitary transformation such that
\beq
X^T M_{\chi^0} X={\rm diag}( m_{\chi^0_1}, m_{\chi^0_2}, m_{\chi^0_3}, m_{\chi^0_4})
\eeq

\subsection{Squark and slepton mass$^2$ matrices  with phases}\label{qB}
 ~~
 In this appendix we give details on the diagonalization of the squark and slepton mass
matrices that appear in Secs.(\ref{i}) and (\ref{j}).
We consider the  squark $(mass)^2$ matrix
\beqn
{ \cal{M}}_{\tilde q}^2=
 \left(
\begin{array}{cc}
M_{\tilde{q}11}^2 & M_{\tilde{q}12}^2  \cr
             M_{\tilde{q}21}^2 & M_{\tilde{q}22}^2
 \end{array}\right)
 \label{sq1}
\eeqn
For the up squark case  one has

\beqn
M_{\tilde{u}11}^2=
M_{\tilde{Q}}^2+m_u^2 + M_{Z}^2(\frac{1}{2}-Q_u
s^2_W)\cos2\beta\nonumber\\
M_{\tilde{u}12}^2=
m_u(A_{u}^{*} - \mu \cot\beta)\nonumber\\
M_{\tilde{u}21}^2=
                m_u(A_{u} - \mu^{*} \cot\beta)\nonumber\\
M_{\tilde{u}22}^2=
m_{\tilde{U}}^2+m{_u}^2 + M_{Z}^2 Q_u s^2_W \cos2\beta
\eeqn
Thus the squark mass$^2$ matrix is hermitian and  can be diagonalized by the unitary transformation
\beq
D_{u}^\dagger M_{\tilde{u}}^2 D_u={\rm diag}(M_{\tilde{u}1}^2,
              M_{\tilde{u}2}^2)
\label{sq2} \eeq where one parameterizes $D_u$ so that \beqn D_u=
 \left(
 \begin{array}{cc}
  \cos \frac{\theta_u}{2}
           & -\sin \frac{\theta_u}{2} e^{-i\phi_{u}} \\
       \sin \frac{\theta_u}{2} e^{i\phi_{u}}
        &\cos \frac{\theta_u}{2}
 \end{array}\right)
\label{sq3}
\eeqn
Here $ M_{\tilde{u}21}^2=|M_{\tilde{u}21}^2| e^{i\phi_{u}}$
and we choose the range of $\theta_u$ so that
${{-\pi}\over {2}} \leq  \theta_u \leq {{\pi}\over {2}}$ where
$\tan \theta_u=
\frac{2|M_{\tilde{u}21}^2|}{M_{\tilde{u}11}^2-M_{\tilde{u}22}^2}$.
    The eigenvalues $M_{\tilde{u}1}^2$ and $M_{\tilde{u}2}^2$
can be determined directly from Eq.(\ref{sq1}) so that
\beqn
M_{\tilde{u}(1)(2)}^2=\frac{1}{2} (M_{\tilde{u}11}^2+M_{\tilde{u}22}^2)
    (+)(-)\nonumber\\
    \frac{1}{2}[(M_{\tilde{u}11}^2-M_{\tilde{u}22}^2)^2 +
        4|M_{\tilde{u}21}^2|^2]^{\frac{1}{2}}.
\label{sq4}
\eeqn
The (+) in Eq.(\ref{sq4}) corresponds to the case so that for
$M_{\tilde{u}11}^2>M_{\tilde{u}22}^2$
 one has $M_{\tilde{u}1}^2>M_{\tilde{u}2}^2$
    and vice versa. For our choice of the $\theta_u$ range one has
\beq
\tan \theta_u=\frac{2 m_{u} |A_u m_0 - \mu^*
\cot\beta|}{M_{\tilde{u}11}^2-M_{\tilde{u}22}^2}
\label{sq5}
\eeq
and
\beq
\sin\phi_u=\frac{m_0 |A_u| \sin \alpha_u+ |\mu| \sin \theta_{\mu} R_u}
        {|m_0 A_u  - \mu^* \cot\beta|}.
\label{sq6}
\eeq
where $R_u=\cot\beta$. 
The analysis for the down squark case proceeds in a similar fashion with
the following changes

\beqn
M_{\tilde{d}11}^2=
M_{\tilde{Q}}^2+m{_d}^2-M_{Z}^2(\frac{1}{2}+Q_d
s^2_W)\cos2\beta\nonumber\\
M_{\tilde{d}12}^2=
 m_d(A_{d}^{*} - \mu \tan\beta)\nonumber\\
M_{\tilde{d}21}^2=
                        m_d(A_{d} - \mu^{*} \tan\beta)\nonumber\\
M_{\tilde{d}22}^2=
                          m_{\tilde{D}}^2+m{_d}^2+M_{Z}^2 Q_d s^2_W \cos2\beta\nonumber\\
\eeqn
The other changes are the modification of expressions for $\theta_d$ and $\phi_d$. They read
\beq
\tan \theta_d=\frac{2 m_{d} |A_u m_0 - \mu^*
\tan\beta|}{M_{\tilde{d}11}^2-M_{\tilde{d}22}^2}
\eeq
and
\beq
\sin\phi_d=\frac{m_0 |A_d| \sin \alpha_d+ |\mu| \sin \theta_{\mu} R_d}
        {|m_0 A_d  - \mu^* \tan\beta|}.
\eeq
where $R_d=\tan\beta$. 
Finally for the case of the slectrons 
\beqn
M_{\tilde{e}11}^2=
 M_{\tilde L}^2+m_{e}^2 -M_Z^2(\frac{1}{2}
- s^2_W)\cos 2\beta\nonumber\\
M_{\tilde{e}12}^2=
  m_{e}(A_{e}^* - \mu\tan\beta)\nonumber\\
M_{\tilde{e}12}^2=
m_{e}(A_{e} - \mu^*\tan\beta)\nonumber\\
M_{\tilde{e}22}^2=
  m_{\tilde E}^2 + m_{e}^2 -M_Z^2 s^2_W
    \cos 2\beta
\eeqn
Expressions for $\theta_e$ and $\phi_e$ are 
 identical to the case of the
down quark with the replacement of d by e. \\

\subsection{RG evolution of electric dipole, color dipole and purely gluonic  operators}\label{qC}

In this Appendix we discuss the renormatization group (RG) evolution of the EDMs discussed
in  Sec.(\ref{j}).
As discussed in the text, there are three competing operators  that contribute to the
EDM of the neutron. These are 
\beqn
{\cal O}_{E}= -\frac{i}{2} \bar q \sigma_{\mu\nu} \gamma_5 q F^{\mu\nu}\nonumber\\
{\cal O}_{qC}= -\frac{i}{2} \bar q \sigma_{\mu\nu} \gamma_5 T^a q G^{\mu\nu a}\nonumber\\
{\cal O}_G= -\frac{1}{6} f^{abc} G_a G_b\tilde  G_c
\eeqn
The one loop RG evolution of the electric dipole and of the color dipole operators
can be easily obtained using their anomalous dimensions since these operators are
eigenstates under the renormalization group. Evolving these operators from a high scale
$Q=M_Z$ to a low scale $\mu$ one finds
\beqn
{\cal O}_i(\mu)= \Gamma^{-\gamma_i/\beta} {\cal O}_i(Q)
\eeqn
where
\beqn
\Gamma =\frac{g_s(\mu)}{g_s(Q)}, ~~\gamma_C=(29-2N_f)/3,\nonumber\\
\gamma_E=  8/3, ~~\beta = (33-2N_f)/3, \eeqn where $N_f$ is the
number of light quarks at the scale $\mu$. Regarding the purely
gluonic dimension six operator it obeys the following
renormalization group
equation\cite{Braaten:1990gq,Braaten:1990zt,Dai:1990xh,Weinberg:1989dx,Boyd:1990bx}
\beqn \mu
\frac{\partial}{\partial \mu} {\cal O}_G =\frac{\alpha_s(\mu)}{4\pi}
(\gamma_G {\cal O}_G -6 \sum_q m_q(\mu) {\cal O}_{qC}), \eeqn where
$\gamma_G= -3-2N_f$. The gauge coupling $\alpha_s$ and the running
quark mass satisfy the RG equations \beqn \mu
\frac{\partial}{\partial \mu} g_s(\mu) = - \beta
\frac{\alpha_s(\mu)}{4\pi} g_s(\mu), \eeqn and \beqn \mu
\frac{\partial}{\partial \mu} m_q(\mu) =
\gamma_m\frac{\alpha_s(\mu)}{4\pi} m_q(\mu). \eeqn where
$\gamma_m=-8$. The above operators contribute to the CP violating
Lagrangian multiplied by coefficients which must cancel their $\mu$
dependence. This allows one to obtain for the coefficients the
following relations 
\beqn d^{(E,C,G)}(\mu) \simeq
\Gamma^{\gamma_{(E,C,G)}/\beta} d^{(E,C,G)}(Q), \eeqn 
where $Q$ is the
high scale. In implementing the RG evolution one uses the matching
conditions due to crossing the heavy thresholds for q=b,c. Thus, for
example, \beqn d^G(m_q^-) =d^G(m_q^+) +d^C(m_q) \frac{1}{8\pi}
\frac{\alpha_s(m_q)}{m_q}. \eeqn Using this technique one can evolve
the EDMs from the electroweak scale $Q=M_Z$ down to the hadronic
scale $\mu$.
A more  up-to-date discussion of the RG evolution of operators  including the mixings 
between the electric and the chromoelectric operators  is given in \cite{Degrassi:2005zd}.

\subsection{\label{qD} Satisfaction of the EDM constraints in the cancelation mechanism}
Here we  give some examples of the parameter points where the
cancelation mechanism discussed in Sec.(\ref{jD}) 
works to produce $d_e$, $d_n$  and $d_{Hg}$
consistent with the current limits.  Table 1 gives three sets of
points (a), (b) and (c) for which the corresponding EDMs $d_e$,
$d_n$ and $C_{Hg}$ are listed in Table 2 where
$C_{Hg}$ is related to the $\tilde {d^C}_d, \tilde{d^C}_u, \tilde{d^C}_s$ by 
$C_{Hg}=|\tilde{d^C}_d - \tilde{d^C}_u -0.012 \tilde{d^C}_s|$. Using the experimental 
constraints on $d_{Hg}$ one obtains the following constraint on $C_{Hg}$
\beqn
C_{Hg} <3.0 \times 10^{-26} {\rm cm}.
\eeqn
The values of $C_{Hg}$ listed in Table 2 are  consistent with the above  experimental
constraint. 
\begin{center}
\begin{tabular}{|c|c|c|}
\multicolumn{3}{c}{Table 1. Three  parameter sets with $A_0$  in units of $m_0$.} \\
\hline
\hline
case& $m_0$, $m_{\frac{1}{2}}$, $|A_0|$ &
$\alpha_A$, $\xi_1$, $\xi_2$, $\xi_3$ \\
\hline
\hline
(a) & $200, 200, 4$ & $1, .5, .659, .633$ \\
\hline
(b) & $370, 370, 4$ & $2, .6, .653, .672$ \\
\hline
(c) & $320, 320, 3$ & $.8, .4, .668, .6$ \\
\hline
\hline
\end{tabular}
\label{dataset}
\end{center}

\begin{center}
\begin{tabular}{|c|c|c|c|}
\multicolumn{4}{c}{Table 2.  Electron, neutron   and $H_g$ EDMs.} \\
\hline
\hline
case   &$d_e$ (ecm unit)  & $d_n$ (ecm unit) & $C_{H_g}$ (cm unit)\\
\hline
\hline
(a) &  $1.45\times 10^{-27}$
& $9.2\times  10^{-27}$ &   $7.2\times  10^{-27}$ \\
\hline
(b) &  $-1.14\times 10^{-27}$ &
$-7.9\times 10^{-27}$ & $2.87\times 10^{-26}$ \\
\hline
(c) & $-3.5\times 10^{-27}$ &
$7.1\times 10^{-27}$ &  $2.9\times 10^{-26}$ \\
\hline
\hline
\end{tabular}
\end{center}

\subsection{Combination of CP phases in SUSY processes}\label{qE}
The  various  phenomena discussed in  Secs.(\ref{i}) and (\ref{j}) involve  several
specific  combinations of  CP phases.  Below we exhibit these combinations. 
\begin{center}
\begin{tabular}{|c|c|}
\multicolumn{2}{c}{Table 3:  Examples of CP phases in  SUSY phenomena } \\
\hline
SUSY Quantity  & Combinations of CP phases \\
\hline
$p\rightarrow \bar{\nu_i}K^{+}$  & $\xi_{1,2,3}+\theta_{1}$,
$\alpha_{A_{t,b}}+\theta_{1}$  \\
\hline
$b\rightarrow s+\gamma$ & $\alpha_{A_{t,s,b}}+\theta_{1}$,
  $\xi_{1,2,3}+\theta_{1}$\\
\hline
$H^0_i$ mixing and spectrum & $\alpha_{A_{t,b,{\tau}}}+\theta_{1}$,
  $\xi_{1,2}+\theta_{1}$\\
\hline
$H^{+}\rightarrow \chi^0\chi^+$ & $\alpha_{A_{t,b}}+\theta_{1}$,
$\xi_{1,2}+\theta_{1}$\\
\hline
$g_{\mu}-2$ & $\xi_{1,2}+\theta_{1}$,
$\alpha_{A_{\mu}}+\theta_{1}$\\
\hline
$\tilde{q}\rightarrow q\chi$ & $\alpha_{A_q}+\theta_{1}$, $\xi_{1,2,3}+\theta_{1}$  \\
\hline
  Dark matter & $\alpha_{A_q}+\theta_{1}$, $\xi_1+\theta_{1}$\\
\hline
$H^0\rightarrow \chi^+\chi^-$ & $\xi_2+\theta_{1}$, $\alpha_{A_{b,t}}+\theta_{1}$, $\xi_1+\theta_{1}$\\
 \hline
 $d_e$ ($d_{\mu}$)  &   $\xi_{1,2}+\theta_{1}$,
 $\alpha_{A_e}+\theta_{1}$($\alpha_{A_{\mu}}+\theta_{1}$ )\\
 \hline
 $d_n$ & $\xi_{1,2,3}+\theta_{1}$,
 $\alpha_{A_{ui}}+\theta_{1}$,$\alpha_{A_{di}}+\theta_{1}$ \\
 \hline
\end{tabular}
\end{center}
In the Table (3)  $\theta_1$ is defined so that $\theta_1=\theta_{\mu}+\theta_H$ and the rest of phases are defined  as in Eqs.(\ref{soft1}) and (\ref{soft3}).

\subsection{Details of $g_{\mu}-2$ analysis in SUSY with CP  Phases}\label{qF}
 Here we  give further details of the analysis  of $a_{\mu}$ discussed in 
 Sec.(\ref{kA})
  but limiting ourselves to the case when the muon mass can be neglected relative
 to other masses.  The chargino exchange contribution is given by

 \beq
a^{\chi^{-}}_{\mu}=a^{21}_{\mu}+a^{22}_{\mu},
\eeq
where for $a^{21}_{\mu}$ and $a^{22}_{\mu}$ we consider now
   the limit where $I_3(\alpha,\beta)$ and
$I_4(\alpha,\beta)$ that appear in Eq.(\ref{i1234}) 
have their first arguments set to zero.   In this case one has
\beq
I_3(0,x)= -\frac{1}{2}F_3(x),~I_4(0,x)=-\frac{1}{6}F_4(x)
\eeq
where
\beqn
F_3(x)=\frac{1}{(x-1)^3}(3x^2-4x+1-2x^2 ln x)\nonumber\\
F_4(x)=\frac{1}{(x-1)^4}(2x^3+3x^2-6x+1-6x^2 ln x).
\eeqn
In the limit considered above  one has the following explicit  expressions for the chargino contributions
\beqn
a^{21}_{\mu}=\frac{m_{\mu}\alpha_{EM}}{4\pi\sin^2\theta_W}
\sum_{i=1}^{2}\frac{1}{M_{\chi_i^+}}Re(\kappa_{\mu} U^*_{i2}V^*_{i1})
F_3(\frac{M^2_{\tilde{\nu}}}{M^2_{\chi_i^+}}),
\eeqn
and
\beq
a^{22}_{\mu}=\frac{m^2_{\mu}\alpha_{EM}}{24\pi\sin^2\theta_W}
\sum_{i=1}^{2}\frac{1}{M^2_{\chi_i^+}}
(|\kappa_{\mu} U^{*}_{i2}|^2+|V_{i1}|^2)
F_4(\frac{M^2_{\tilde{\nu}}}{M^2_{\chi_i^+}}),
\eeq
where
\beq
 \kappa_{\mu}=\frac{m_{\mu}}{\r2 M_W \cos\beta}
 \eeq
Next we discuss the neutralino exchange contributions to $a_{\mu}$. These are given by
\beq
a^{\chi^0}_{\mu}=a^{11}_{\mu}+a^{12}_{\mu},
\eeq
where
\beq
a^{11}_{\mu}=\frac{m_{\mu}\alpha_{EM}}{2\pi\sin^2\theta_W}
\sum_{j=1}^{4}\sum_{k=1}^{2}\frac{1}{M_{\chi_j^0}} Re(\eta^k_{\mu j})
I_1(\frac{m_{\mu}^2}{M^2_{\chi_j^0}},
\frac{M^2_{\tilde{\mu_k}}}{M^2_{\chi_j^0}}),
\eeq
and
\beq
a^{12}_{\mu}=\frac{m^2_{\mu}\alpha_{EM}}{4\pi\sin^2\theta_W}
\sum_{j=1}^{4}\sum_{k=1}^{2}\frac{1}{M^2_{\chi_j^0}}X^k_{\mu j}
I_2(\frac{m_{\mu}^2}{M^2_{\chi_j^0}},
\frac{M^2_{\tilde{\mu_k}}}{M^2_{\chi_j^0}}),
\eeq
Here  $\eta^k_{\mu j}$ is defined by
\beqn
\eta^k_{\mu j} & &=-
(\frac{1}{\sqrt 2}[\tan\theta_W X_{1j}+ X_{2j}]D_{1k}^*
-\kappa_{\mu} X_{3j} D_{2k}^*)\nonumber\\
&&
(\sqrt 2\tan\theta_W X_{1j}D_{2k} +\kappa_{\mu} X_{3j} D_{1k})
\eeqn
and  $X^k_{\mu j}$ is defined by
\beqn
X^k_{\mu j}&&=\frac{m^2_{\mu}}{2 M^2_W \cos^2\beta}|X_{3j}|^2 \nonumber\\
&&
+\frac{1}{2}\tan^2\theta_W |X_{1j}|^2
(|D_{1k}|^2+4|D_{2k}|^2)\nonumber\\
&&+\frac{1}{2} |X_{2j}|^2|D_{1k}|^2
+\tan\theta_W |D_{1k}|^2 Re(X_{1j}X_{2j}^*)\nonumber\\
&&
+\frac{m_{\mu}\tan\theta_W}{M_W \cos\beta}
Re(X_{3j}X_{1j}^*D_{1k}D_{2k}^*)\nonumber\\
&&
-\frac{m_{\mu}}{M_W\cos\beta}Re(X_{3j}X_{2j}^*D_{1k}D_{2k}^*).
\eeqn In the limit when the  muon mass is neglected relative to
other masses and the first argument in the double integral is taken
to  be zero one finds a simplification of the form factors so that
\beq I_1(0,x)=\frac{1}{2}F_1(x), I_2(0,x)=\frac{1}{6}F_2(x), \eeq
where \beq F_1(x)=\frac{1}{(x-1)^3}(1-x^2+2x ln x), \eeq and \beq
F_2(x)=\frac{1}{(x-1)^4}(-x^3+6x^2-3x-2-6x ln x). \eeq

\subsection{\label{qG}
Stop  exchange contributions to  Higgs mass$^2$ matrix. } 
For
completeness we give here an analysis of the one loop contributions
from the stop
 sector  with inclusion of CP violating effects in the analysis of CP even-CP odd 
 Higgs  mixings discussed  in Sec.(\ref{kB}).
The contribution to the one loop effective potential from
the stop and top exchanges is given by

\beqn
\Delta V(\tilde t, t)=\frac{1}{64\pi^2}
(\sum_{a=1,2}6M_{\tilde t_a}^4(log \frac{M_{\tilde t_a}^2}{Q^2}-\frac{3}{2})\nonumber\\
-12 m_t^4 (log\frac{m_t^2}{Q^2}-\frac{3}{2}))
\eeqn
Using the above potential our analysis for $\Delta_{ij\tilde t}$ gives

\beq
\Delta_{11\tilde t}=-2\beta_{h_t}m_t^2 |\mu|^2
\frac{(|A_t|\cos\gamma_t -|\mu|cot\beta)^2}
{(m_{\tilde t_1}^2-m_{\tilde t_2}^2)^2}
f_2(m_{\tilde t_1}^2, m_{\tilde t_2}^2)
\eeq

\beqn
\Delta_{22\tilde t}=-2\beta_{h_t} m_t^2
\frac{|A_t|^2[|A_t| -|\mu|cot\beta\cos\gamma_t]^2}
{(m_{\tilde t_1}^2-m_{\tilde t_2}^2)^2}\nonumber\\
f_2(m_{\tilde t_1}^2, m_{\tilde t_2}^2)
+2\beta_{h_t} m_t^2
ln(\frac{m_{\tilde t_1}^2 m_{\tilde t_2}^2}{m_t^4})\nonumber\\
+ 4\beta_{h_t}m_t^2
\frac{|A_t|[|A_t| -|\mu|cot\beta\cos\gamma_t]}
{(m_{\tilde t_1}^2-m_{\tilde t_2}^2)}
ln(\frac{m_{\tilde t_1}^2}{m_{\tilde t_2}^2})~
\eeqn

\beqn
\Delta_{12\tilde t}=
-2\beta_{h_t} m_t^2
\frac{|\mu|[|A_t|\cos\gamma_t -|\mu|cot\beta]}
{(m_{\tilde t_1}^2-m_{\tilde t_2}^2)}
ln(\frac{m_{\tilde t_1}^2}{ m_{\tilde t_2}^2})\nonumber\\
+ 2\beta_{h_t} m_t^2  f_2(m_{\tilde t_1}^2, m_{\tilde t_2}^2)\nonumber\\
\frac{|\mu||A_t|[|A_t|\cos\gamma_t -|\mu|cot\beta]
[|A_t| -|\mu|cot\beta\cos\gamma_t]}
{(m_{\tilde t_1}^2-m_{\tilde t_2}^2)^2}~
\eeqn

\beqn
\Delta_{13\tilde t}=
-2\beta_{h_t} m_t^2
\frac{|\mu|^2|A_t|\sin\gamma_t [|\mu|cot\beta -|A_t|\cos\gamma_t ]}
{\sin\beta(m_{\tilde t_1}^2-m_{\tilde t_2}^2)^2}\nonumber\\
f_2(m_{\tilde t_1}^2, m_{\tilde t_2}^2)~~~
\eeqn

\beqn
\Delta_{23\tilde t}=
-2\beta_{h_t} m_t^2 |\mu||A_t|^2
\frac{\sin\gamma_t (|A_t| - |\mu|cot\beta \cos\gamma_t)}
{\sin\beta(m_{\tilde t_1}^2-m_{\tilde t_2}^2)^2}\nonumber\\
f_2(m_{\tilde t_1}^2, m_{\tilde t_2}^2)
+2\beta_{h_t}\frac{ m_t^2 |\mu||A_t|\sin\gamma_t}
{\sin\beta(m_{\tilde t_1}^2-m_{\tilde t_2}^2)}
ln(\frac{m_{\tilde t_1}^2}{ m_{\tilde t_2}^2})~~
\eeqn
and
\beq
\Delta_{33\tilde t}=
-2\beta_{h_t} \frac{m_t^2|\mu|^2|A_t|^2\sin^2\gamma_t}
{\sin^2\beta(m_{\tilde t_1}^2-m_{\tilde t_2}^2)^2}
f_2(m_{\tilde t_1}^2, m_{\tilde t_2}^2).
\eeq

In the  analysis above the  D terms of the squark
(mass)$^2$ matrices are ignored to   obtain approximate independence of the
renormalization scale Q  similar to  the analysis of \cite{Carena:2000yi,Demir:1999hj}.

\subsection{Fierz rearrangement relations involving Majoranas}\label{qH}
Fierz rearrangements are known to be very useful when manipulating
interactions involving four fermions. Specifically such Fierz rearrangements 
are needed  in the analysis of Sec.(\ref{kG}). 
Here we give these relations
for the case  when two of the fermions are Majoranas (such as
neutralinos) and the other two are quarks.  Thus any four Fermi
interaction with two Majoranas and two quarks can be written as
involving the following combinations

\beqn
\bar{\chi}\chi\bar{q}q,   ~\bar{\chi}\gamma_5\chi\bar{q}\gamma_5q,
~\bar{\chi}\gamma^{\mu}\gamma_5\chi\bar{q}\gamma_{\mu}q,\nonumber\\
\bar{\chi}\gamma^{\mu}\gamma_5\chi\bar{q}\gamma_{\mu}\gamma_5q,
~\bar{\chi}\gamma_5\chi\bar{q}q,  ~\bar{\chi}\chi\bar{q}\gamma_5q.
\eeqn For convenience define the 16   gamma matrices as follows \beq
\Gamma^A=\{1,\gamma^0,i\gamma^i,i\gamma^0\gamma_5,
\gamma^i\gamma_5,\gamma_5,i\sigma^{0i},\sigma^{ij}\}: ~~i,j=1-3 \eeq
with the normalization \beq tr(\Gamma^A\Gamma^B)=4\delta^{AB} \eeq
The Fierz rearrangement formula with the above definitions and
normalizations is \beq
(\bar{u_1}\Gamma^Au_2)(\bar{u_3}\Gamma^Bu_4)=\sum_{C,D}F^{AB}_{CD}
(\bar{u_1}\Gamma^Cu_4)(\bar{u_3}\Gamma^Du_2) \eeq where $u_j$ are
Dirac or Majorana spinors and \beq F^{AB}_{CD}=
-(+)\frac{1}{16}tr(\Gamma^C\Gamma^A\Gamma^D\Gamma^B) \eeq and where
the +ve sign is for commuting u spinors and the -ve sign is for the
anticommuting u fields. In our case we have to use the -ve sign
since we are dealing with quantum Majorana and Dirac fields in the
Lagrangian.
 We give below the Fierz rearrangement  for  four combinations that appear
commonly in neutralino-quark scattering. These are
\beqn
\bar{\chi}q\bar{q}\chi=-\frac{1}{4}\bar{\chi}\chi\bar{q}q-\frac{1}{4}
\bar{\chi}\gamma_5\chi\bar{q}\gamma_5q+\frac{1}{4}\bar{\chi}\gamma^{\mu}
\gamma_5\chi\bar{q}\gamma_{\mu}\gamma_5q\nonumber\\
\bar{\chi}\gamma_5
q\bar{q}\chi=\frac{1}{4}\bar{\chi}\gamma^{\mu}\gamma_5
\chi\bar{q}\gamma_{\mu}
q-\frac{1}{4}
\bar{\chi}\chi\bar{q}\gamma_5q-\frac{1}{4}\bar{\chi}\gamma_5
\chi\bar{q}q\nonumber\\
\bar{\chi}
q\bar{q}\gamma_5
\chi=-\frac{1}{4}\bar{\chi}\gamma^{\mu}\gamma_5
\chi\bar{q}\gamma_{\mu}
q-\frac{1}{4}
\bar{\chi}\chi\bar{q}\gamma_5q-\frac{1}{4}\bar{\chi}\gamma_5
\chi\bar{q}q\nonumber\\
\bar{\chi}\gamma_5
q\bar{q}\gamma_5
\chi=-\frac{1}{4}\bar{\chi}\chi\bar{q}q-\frac{1}{4}
\bar{\chi}\gamma_5\chi\bar{q}\gamma_5q-\frac{1}{4}\bar{\chi}\gamma^{\mu}
\gamma_5\chi\bar{q}\gamma_{\mu}\gamma_5q\nonumber\\
\eeqn
The metric used above  is $\eta_{\mu\nu}=(1,-1,-1,-1)$,
 and since $\chi$'s are Majoranas we have used the properties
$\bar{\chi}\gamma_{\mu}\chi=0$ and $\bar{\chi}\sigma_{\mu\nu}\chi=0$.

\subsection{Effective four-Fermi interaction for dark matter detection
with inclusion of CP phases}\label{qI}
       In  this appendix we give a derivation  of  the four fermi
neutralino-quark effective Lagrangian with CP violating phases given
in Sec.(\ref{kG}). We begin by discussing   the squark exchange
contribution. From the fundamental  supergravity  Lagrangian of
quark-squark-neutralino interactions \beq {-\cal
L}=\bar{q}[C_{qL}P_L+C_{qR}P_R]\chi\tilde{q_1}+
\bar{q}[C^{'}_{qL}P_L+C^{'}_{qR}P_R]\chi\tilde{q_2}+H.c. \eeq the
effective lagrangian for $q-\chi$ scattering via the exchange of
squarks is given by \cite{Chattopadhyay:1998wb,Falk:1998xj}. 
\beqn
{\cal L}_{eff}=\frac{1}{M^{2}_{\tilde{q_1}}-M^{2}_{\chi}}
\bar{\chi}[C^{*}_{qL}P_R+C^{*}_{qR}P_L]q\nonumber\\
\bar{q}[C_{qL}P_L+C_{qR}P_R]
\chi\nonumber\\
+\frac{1}{M^{2}_{\tilde{q_2}}-M^{2}_{\chi}}
\bar{\chi}[C^{*'}_{qL}P_R+C^{*'}_{qR}P_L]q
\bar{q}[C^{'}_{qL}P_L+C^{'}_{qR}P_R]
\chi\nonumber\\
\eeqn
where
\beqn
C_{qL}=\sqrt{2} (\alpha_{q0} D_{q11} -\gamma_{q0} D_{q21}),\nonumber\\
C_{qR}=\sqrt{2} (\beta_{q0} D_{q11} -\delta_{q0} D_{q21}),\nonumber\\
C^{'}_{qL}=\sqrt{2} (\alpha_{q0} D_{q12} -\gamma_{q0} D_{q22}),\nonumber\\
C^{'}_{qR}=\sqrt{2} (\beta_{q0} D_{q12} -\delta_{q0} D_{q22}),
\eeqn
and where $\alpha$, $\beta$, $\gamma$, and $\delta$ are given by

\beqn
\alpha_{u(d)j}=\frac{g m_{u(d)}X_{4(3)j}}{2 m_W\sin\beta(\cos\beta)},\nonumber\\
\beta_{u(d)j}=eQ_{u(d)j}X'^{*}_{1j}+\frac{g}{cos\theta_W}X'^{*}_{2j}\nonumber\\
  \times (T_{3u(d)}-Q_{u(d)}\sin^2\theta_W),\nonumber\\
  \gamma_{u(d)j}=eQ_{u(d)j}X'_{1j}-
\frac{gQ_{u(d)}\sin^2\theta_W}{cos\theta_W}X'_{2j},\nonumber\\
\delta_{u(d)j}= \frac{- g m_{u(d)}X^*_{4(3)j}}{2 m_W\sin\beta(\cos\beta)}.
\eeqn
Here g is the $SU(2)_L$ gauge coupling and
\beqn
X'_{1j}=X_{1j}\cos\theta_W+X_{2j}\sin\theta_W,\nonumber\\
X'_{2j}=-X_{1j}\sin\theta_W+X_{2j}\cos\theta_W.
\eeqn
The effect of the CP violating phases enter via the neutralino
eigenvector components $X_{ij}$
 and via the matrix $D_{qij}$
that diagonalizes the squark mass$^2$ matrix.

Using the Fierz rearrangement  one can obtain now the coefficients
  A, B, C, D, E and F that appear in Eq.(\ref{4fermi})  in a
 straightforward fashion \cite{Chattopadhyay:1998wb,Falk:1998xj}.
 The first two terms ($A$, $B$) are spin-dependent interactions and arise
from the $Z$ boson  and the sfermion exchanges. For these one has
\beqn
A=\frac{g^2}{4 M^2_W}[|X_{30}|^2-|X_{40}|^2][T_{3q}-
e_q sin^2\theta_W]\nonumber\\
-\frac{|C_{qR}|^2}{4(M^2_{\tilde{q1}}-M^2_{\chi})}
-\frac{|C^{'}_{qR}|^2}{4(M^2_{\tilde{q2}}-M^2_{\chi})}
\eeqn
\beqn
B=-\frac{g^2}{4 M^2_W}[|X_{30}|^2-|X_{40}|^2]
e_q sin^2\theta_W +\nonumber\\
\frac{|C_{qL}|^2}{4(M^2_{\tilde{q1}}-M^2_{\chi})}
+\frac{|C^{'}_{qL}|^2}{4(M^2_{\tilde{q2}}-M^2_{\chi})}
\eeqn
The terms $C$, $D$, $E$ and $F$ receive contributions from sfermions
and from neutral Higgs and can be calculated using similar techniques. 

\subsection{Computational tools for SUSY phenomena with CP phases}\label{qJ}
The numerical analysis of supersymmetric phenomena with CP phases is significantly more difficult
than for the case when the phases are absent.   First most numerical integration codes for the  renormalization
group evolution, sparticle spectra and for the analysis of sparticle decays and cross sections are not
equipped to handle phases. Second any physically meaningful set of parameters which include phases
must necessarily satisfy  the stringent EDM constraints which  also require care.  A significant progress
has been in this direction by   the so called CPsuperH\cite{Lee:2003nt},
which is a Fortran code that calculates the mass
spectrum and decay widths of the neutral and charged Higgs  bosons in MSSM with CP
phases.  Obviously there is significant room for further progress in this  area.

\bibliography{cprefs14}

\end{document}